\documentclass[journal, transmag, onecolumn]{IEEEtran}
\usepackage{graphicx}    
\usepackage{float}
\usepackage{amssymb}
\usepackage{cite}
\usepackage{tabularx}
\usepackage{hyperref}
\usepackage{xcolor}
\usepackage{multirow} 
\usepackage[version=4]{mhchem}

\usepackage{tikz}
\def\checkmark{\tikz\fill[scale=0.4](0,.35) -- (.25,0) -- (1,.7) -- (.25,.15) -- cycle;} 

\usepackage{pifont}
\let\oldding\ding
\renewcommand{\ding}[2][1]{\scalebox{#1}{\oldding{#2}}}

\begin{document}
\title{Cell Balancing Paradigms: Advanced Types, Algorithms, and Optimization Frameworks}


\author{\IEEEauthorblockN{
Anupama R Itagi\IEEEauthorrefmark{1},
Rakhee Kallimani\IEEEauthorrefmark{2},
Krishna Pai\IEEEauthorrefmark{3},
Sridhar Iyer\IEEEauthorrefmark{4}, 
Onel L. A. López\IEEEauthorrefmark{5}, and
Sushant Mutagekar\IEEEauthorrefmark{6}
}

\IEEEauthorblockA{\IEEEauthorrefmark{1}Department of Electrical and Electronics Engineering,\\
KLE Technological University, 580032, Hubballi, Karnataka, India\\
Email: anupama\_itagi@kletech.ac.in
}
\IEEEauthorblockA{\IEEEauthorrefmark{2}Department of Electrical and Electronics Engineering, \\
KLE Technological University Dr. M S Sheshgiri Campus, 590008, Belagavi, Karnataka, India\\
Email: rakhee.kallimani@klescet.ac.in
}
\IEEEauthorblockA{\IEEEauthorrefmark{3}Independent Researcher,\\ 560094, Bengaluru, Karnataka, India\\
Email: krishnapai271999@gmail.com
}
\IEEEauthorblockA{\IEEEauthorrefmark{4}Department of Electronics and Communication Engineering, \\
KLE Technological University Dr. M S Sheshgiri Campus, 590008, Belagavi, Karnataka, India\\
Email: sridhariyer1983@klescet.ac.in
}
\IEEEauthorblockA{\IEEEauthorrefmark{5}Faculty of Information Technology and Electrical Engineering,\\
University of Oulu, 90014, Oulu, Finland\\
Email: onel.alcarazlopez@oulu.fi
}
\IEEEauthorblockA{\IEEEauthorrefmark{6}Independent Researcher,\\ Belagavi, Karnataka, India\\
Email: sushantmm.iitm@gmail.com
}

% <-this % stops an unwanted space
%\thanks{%Manuscript received December 1, 2012; revised August 26, 2015. 
%Corresponding author: ABC (email: ABC@gmail.com).}
}

\IEEEtitleabstractindextext{%
\begin{abstract}
%\textcolor{orange}{Start date:20-4-2024}

The operation efficiency of the electric transportation, energy storage, and grids mainly depends on the fundamental characteristics of the employed batteries. Fundamental variables like voltage, current, temperature, and estimated parameters, like the State of Charge (SoC) of the battery pack, influence the functionality of the system. This motivates the implementation of a Battery Management System (BMS), critical for managing and maintaining the health, safety, and performance of a battery pack. This is ensured by measuring parameters like temperature, cell voltage, and pack current. It also involves monitoring insulation levels and fire hazards, while assessing the prevailing useful life of the batteries and estimating the SoC and State of Health (SoH). Additionally, the system manages and controls key activities like cell balancing and charge/discharge processes. Thus functioning of the battery can be optimised, by guaranteeing the vital parameters to be well within the prescribed range. This article discusses the several cell balancing schemes, and focuses on the intricacies of cell balancing algorithms and optimisation methods for cell balancing. We begin surveying recent cell balancing algorithms and then provide selection guidelines taking into account their advantages, disadvantages, and applications. Finally, we discuss various optimization algorithms and outline the essential parameters involved in the cell balancing process.

\end{abstract}

}

\maketitle
\IEEEdisplaynontitleabstractindextext
\IEEEpeerreviewmaketitle

%=================================================

\section{Introduction}
\label{Section:Introduction}

Batteries are gaining importance due to the large demand for efficient energy storage solutions, addressing both Electric Vehicle (EV) needs and grid application challenges. To achieve higher power output, many cells are inter connected in series and shunt configuration, making them ideal for various applications like electric bicycles, scooters, and buses. Lithium-ion (Li-ion) batteries are preferred due to their low self-discharge rate, high energy density, and long lifespan \cite{article11, article12}. While Li-ion batteries offer numerous benefits, they also contribute to certain safety risks. In fact, they can catch fire or even explode if mishandled or stored under unsafe conditions. To address problems related to Li-ion batteries, a Battery Management System (BMS) plays a critical role, managing various parameters and ensuring optimal battery performance. In addition to protecting against over or under-voltage, BMS plays a critical role in balancing cell charge, which is essential for maintaining battery health, optimising performance, by maintaining its operation well within the safety limits. Further, cell balancing serves as a fundamental component for ensuring safety and extending battery life \cite{article13}. Cell balancing aids to maintain uniform voltage or State of Charge (SoC) across all cells in a battery pack \cite{article16} to maintain the health, safety, and overall functionality of the pack. As batteries age or undergo repeated charge/discharge cycles, individual cells can develop imbalances due to variations in capacity, internal resistance, etc. \cite{article6, article7}. These imbalances, if left unchecked, can lead to decreased battery capacity, reduced lifespan, and safety risks such as overheating or overcharging \cite{article8, article9}. 

Two major circuit structures commonly used for cell balancing are passive balancing and active balancing \cite{article5}. Passive balancing utilises shunt resistors to dissipate surplus charge from cells with elevated SoC by converting it to heat, and aligning them with lower SoC cells \cite{article4, article5}. This technique is user friendly and cost-effective; but it is slow and inefficient due to energy loss. On the contrary, active balancing transfers charge from maximum SoC cells to minimum SoC cells using capacitors, windings, transformers, or converters \cite{article4}. This method achieves faster balancing but demands more intricate circuitry, making it typically expensive \cite{article5}. The choice between passive and active balancing depends on factors like cost, balancing speed, and energy efficiency requirements. Table~\ref{tab:Acronyms} describes the main acronyms used in the study.

Integration of charging–discharging control and cell balancing algorithms addresses the complex challenges of modern battery management, ensuring that battery systems can meet the demanding requirements of contemporary applications. The integration of charging–discharging control and cell balancing algorithms within BMS is highly significant for several reasons, which include:

\begin{enumerate}
\item \textit{Performance Optimisation:} Efficient control of charging and discharging allows individual cell in a battery pack to function at its optimal capacity \cite{ article19}. By integrating these control algorithms with cell balancing algorithms, the system can effectively manage the SoC across all cells, leading to improved overall performance\cite{article94}.

\item \textit{Safety Enhancement:} Safety is a paramount concern in batteries. Over-charging or over-discharging individual cells can results in dangerous conditions, like thermal runaway in case of over-charging and degradation of capacity in case of over-discharge \cite{article95}. 
Integrated cell balancing algorithms aid to prevent these issues by keeping all cells within safe operating limits, thereby enhancing protective features and efficiency of the battery pack \cite{article82}.

\item \textit{Durability and Longevity:} The lifespan of a battery pack can be significantly extended by maintaining balanced charge and discharge cycles \cite{article95}. Cells that are kept within their optimal voltage ranges degrade slowly \cite{article82}. Therefore, integrated  (combination of charging/discharging and cell balancing) algorithms that manage both charging/discharging and balancing enhance the durability of the battery system.

\item \textit{Accurate SoC Estimation:} For accurate SoC estimation, it is necessary to have precise control over the charging and discharging processes \cite{article96}. This can be obtained by accurate measurements of voltage and current. Integrated algorithms ensure that data used for SoC calculations reflect the true state of each cell, leading to better management and estimation of battery performance.

\item \textit{Thermal Management:} Charging and discharging processes generate heat \cite{article56}. Integrated algorithms help in managing heat generation by balancing the load across the cells, thus preventing hot-spots and ensuring uniform temperature distribution, which is crucial for battery health \cite{article97}.

\item \textit{Protection:} Integrated control algorithms can provide robust protection mechanisms. These algorithms can detect and respond to anomalies, such as voltage imbalances or temperature spikes, in real-time. This proactive approach prevents potential damage and maintains the integrity of the battery pack \cite{article98}.
\end{enumerate}

\begin{table*}[t!]
\caption{List of Important Acronyms}
\label{tab:Acronyms}
\centering
\resizebox{\textwidth}{!}{
    \begin{tabular}{|m{0.08\textwidth}|m{0.33\textwidth}|m{0.08\textwidth}|m{0.33\textwidth}|} 
    \hline
\textbf{Acronym} & \textbf{Definition} & \textbf{Acronym} & \textbf{Definition} \\ \hline

Ah&Ampere-hour &AKF&Adaptive Kalman Filter\\\hline
BBE&Buck-Boost Equalizers&BMS& Battery Management System  \\\hline
CPC & cell-to-pack-to-cell  &DAH & Dual Active Half-bridge  \\\hline
DP & Dual Polarization &DQN&Deep Q Network\\\hline
DRL&Deep Reinforcement Learning&EIS&Electrochemical Impedance spectroscopy \\\hline
EKF & Extended Kalman Filter &EKPF&Extended Kalman Particle Filter\\\hline
EMF& Electromotive Force &EMF &Electro Motive Force\\\hline
ESS&Energy Storage System   &EVs & Electric Vehicles \\\hline
FHCD & First High SoC Cell Discharge &FLC  &Fuzzy Logic Controller  \\\hline
FLCC& First Low SoC Cell Charge &GA&Genetic Algorithm \\\hline
ICEs & Individual Cell Equalizer &KF & Kalman Filter \\\hline
Li-ion & Lithium-ion&MAE& Mean Absolute Error \\\hline
MOSFET &  Metal–Oxide–Semiconductor Field-Effect Transistor &MPC & Model Predictive Control  \\\hline
MSA& Multi-Agent System   &NN&Neural Network \\\hline
OCV &  Open Circuit Voltage &PF& Particle Filter  \\\hline
PID &  Proportional Integration Differentiation &PLC& Programmable Logic Controller  \\\hline
PNGV &  Partnership for a New Generation of Vehicle &PSO & Particle Swarm Optimization \\\hline
RLS & Recursive Least Squares  &RVM&Relevance Vector Machine\\\hline
SoC & State of Charge  &SoH&State of Health  \\\hline
SVM&Support  Vector Machine&SVSF & Smooth Variable Structure Filter  \\\hline
UKF & Unscented Kalman Filter  &VBL& Time-Varying Smoothing Boundary Layer\\\hline

% &   &  &  \\ \hline
\end{tabular}}
\end{table*}

\subsection{Significance of cell equalisation in BMS}

\begin{figure*}[hbt!]
    \centering    
    \includegraphics[width=0.6\textwidth]{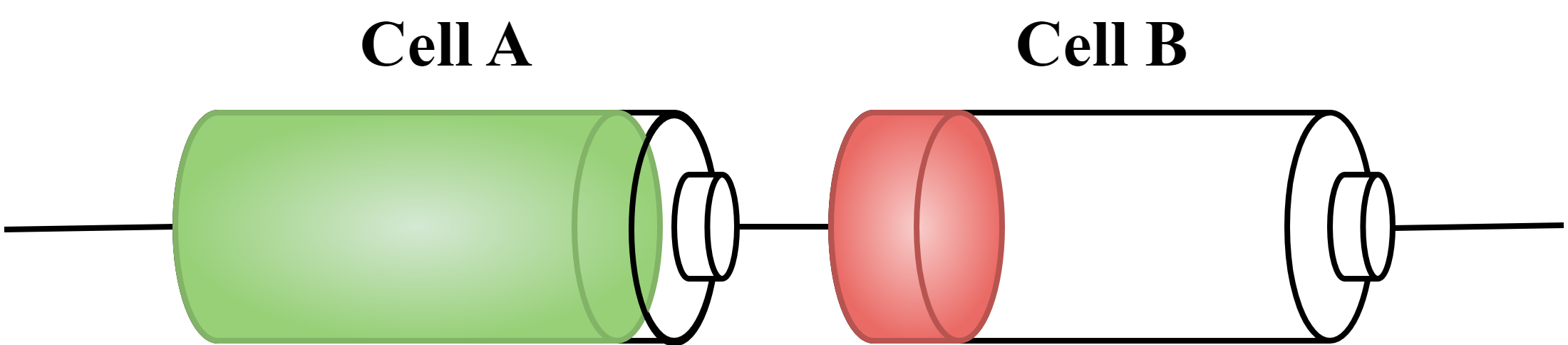}   
    \caption{Illustration of Battery Pack.} 
    \label{fig:Battery Pack}
\end{figure*}

 An unbalanced cell can negatively impact the functionality of the other cells in the battery pack. This lower performance, if continued, may result in potentially making the battery pack useless. For instance,  Fig.~\ref{fig:Battery Pack} shows a battery pack containing two cells, A and B, cell A is in higher SoC state and cell B in lower SoC. This condition puts the battery pack in a lock stage, which implies that the battery can now neither charge further nor can it discharge. In addition, if one attempted to charge forcefully, cell A might be damaged and if it was tried to discharge further, cell B might be damaged. Hence, to avoid such a condition occurring due to cell unbalancing, there is an immediate need to formulate efficient cell balancing algorithms.

The condition of cell imbalance  can occur due to deviations of multiple factors like temperature, internal resistance of the cell and when different cells have distinct Coulombic efficiency \cite{article15}, the condition expressed by (\ref{Eq:Coulombic efficiency}). This leads to i) deviation of one cell’s SoC compared with another cell’s SoC; and ii) different levels of voltage in each cell of a battery caused at the time of charging or discharging \cite{article17}.

\begin{equation}
    z(t) = z(0) - \frac{1}{Q} \int_{0}^{t} \eta(t) i_{\text{net}}(t) dt.
    \label{Eq:Coulombic efficiency}
\end{equation}
Specifically, when all cells have the same SoC $ z(0) $ with the same net current $i_{\text{net}}(t)$ and the same capacity $ Q $; however, due to the distinct cell efficiencies $ \eta $, every cell's SoC starts deviating from each other causing a cell to be unbalanced within the battery pack \cite{article14}.

Another factor contributing to cell unbalancing is variations in the net current $ i_{\text{net}}(t) $ for every cell, which is given by 
\begin{equation}
    i_{\text{net}}(t) = i_{\text{self-discharge}}(t) + i_{\text{leakage}}(t) + i_{\text{pack}}(t).
    \label{Eq:Net Current}
\end{equation}
Note that each cell has its own self-discharge current $ i_{\text{self-discharge}}(t)$, which directly depends on the leakage current $ i_{\text{leakage}}(t)$. 
As shown in  (\ref{Eq:Net Current}), the net current of each cell will be distinct. This is due to dynamic variations in the leakage current, load current $ i_{\text{pack}}(t) $ and self-discharge rate,of the battery pack. Subsequently, the deviation in the SoC at the cell level increases. Differences in the internal resistance of each cells in a battery pack are the chief reason of cell imbalance, leading to unsafe behaviour as well as wasteful energy use. There can be variations in internal resistance of around 15\% amongst cells produced in the same batch. Because the voltage drop across a cell is related to both its internal resistance and the current passing through it, this fluctuation may cause noticeable voltage changes during discharge.

Another factor that may cause cell imbalance is the charge-discharge imbalance. Such imbalances are critical, as overcharging can lead to battery explosions \cite{article17}. Discharging process accelerates degradation in the weaker cells, affecting the entire battery pack. During the discharge cycle, weaker cells deplete their charge more quickly, while stronger cells retain some charge. As a result, the weaker cells experience further degradation due to over-discharge \cite{article17}.
%==============================================================================
\subsection{Recent developments in cell balancing}

A comprehensive overview of recent trends in Unmanned Aerial Vehicle (UAV) BMS is presented in \cite{article95}. This study details battery charging and discharging strategies, battery equalisation, and hybrid energy management. It also delves into system components and addresses various safety concerns. However, the study does not discuss cell balancing algorithms, optimisation techniques, and future research directions. The state-of-the-art battery pack SoC estimation techniques are highlighted in \cite{article117}, highlighting the effect of cell discrepancy on pack functionality and SoC estimation. While it discusses cell balancing methods and categorizes SoC estimation methods, it does not articulate cell balancing algorithms, optimisation techniques, and future research directions. Meanwhile, the authors in \cite{article37} provide an extensive outline of battery balancing arrangements comprising of DC-DC converter, emphasising their crucial role in controlling charge balance and the design of balancing control algorithms. They also address other BMS operations, such as SoC estimation, which impacts balancing functionality, but a comparative analysis of balancing algorithms and optimisation techniques is missing. In \cite{article109}, mainstream equalisation strategies are examined with a focus on efficiency, complexity, and stability through analysis of balancing variables and control algorithms. This work explores variability sources in Li-ion batteries across their life-cycle, detailing classification methods and applications of balance management systems, but it does not detail future research directions. The work in \cite{article120} offers a comprehensive overview of research on battery equalizer circuits (BECs) used in EVs. It includes multiple simulations across various BEC topologies using identical initial conditions and provides quantitative analysis. However, it does not discuss cell-balancing algorithms or related optimisation techniques. Variables, topologies, and modular methods in battery equalizer circuits are analysed in \cite{article121}. Such a work also categorizes dominant cell balancing structures, highlighting their comparative characteristics, and evaluates modular methods using graph theory. However, it does not detail future research directions, not does it discuss cell balancing algorithms and optimisation techniques.

\begin{figure*}[hbt!]
    \centering    
    \includegraphics[width=\textwidth]{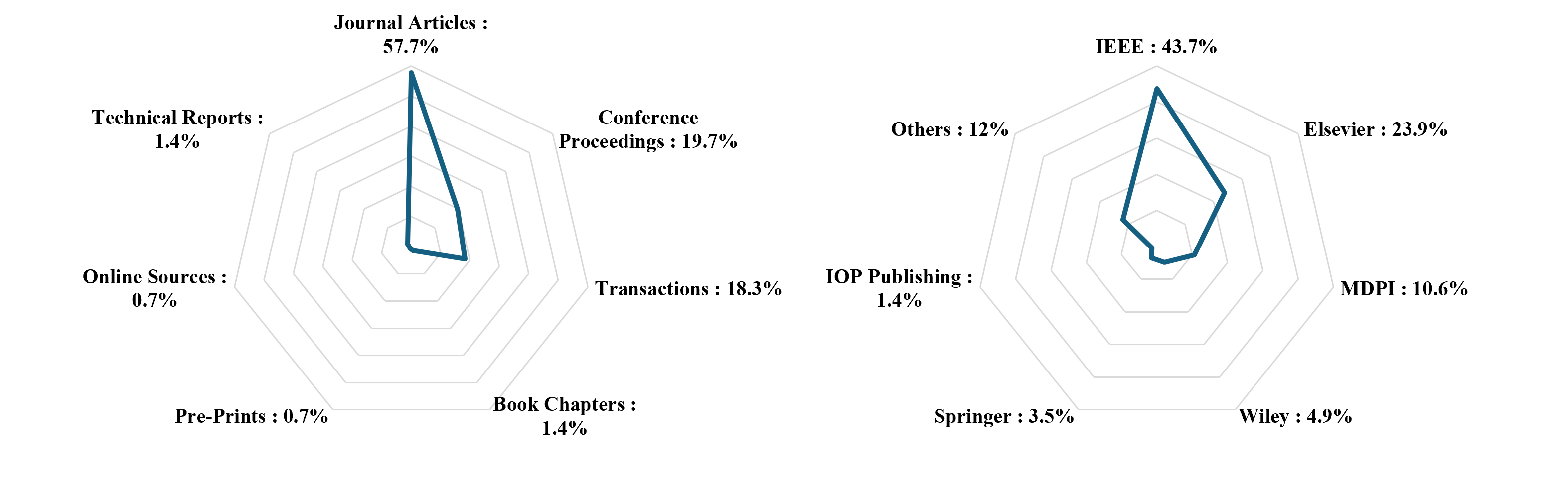}   
    \caption{Types of sources and databases used for study.} 
    \label{fig:Graph}
\end{figure*}

Compared to the previous studies, the current survey article focuses on comparing basic cell balancing algorithms and SoC estimation methods, including direct methods and data-guided methods, highlighting their respective advantages and drawbacks. The summary of the above detailed recent survey articles is presented in Table~\ref{tab:comparison of existing review articles}. Further, to present the detailed survey in this article, we collected 203 articles of which, 164 articles pinned to the keywords. After further scanning  and summarizing, we found 142 articles which matched exactly with the current research. Among these, 108 are primary studies, 30 are proceedings and book chapters, and 4 are technical reports, pre-prints and other sources. For a detailed classification, the reader can connect to Fig.~\ref{fig:Graph}.

\begin{table*}[t!]
\caption{Summary of most recent surveys related to cell balancing}
\label{tab:comparison of existing review articles}
\centering
\resizebox{\textwidth}{!}{
 \begin{tabular}{|m{0.07\textwidth}|m{0.05\textwidth}|m{0.53\textwidth}|m{0.37\textwidth}|} 
    \hline
\textbf{References} &	\textbf{Year} &\textbf{Main Contributions} & \textbf{Constraints related to our survey}\\ \hline

\cite{article95} & 2023	& \ding[0.5]{108}\quad  Overview of recent trends in UAV BMS, encompassing battery charging and discharging strategies, battery equalisation, and hybrid energy management. 

\ding[0.5]{108}\quad Discussion on system components and safety issues.
&  \ding[0.5]{108}\quad Future research directions are not articulated. 
 
 \ding[0.5]{108}\quad No discussion on cell balancing algorithms and optimisation techniques.\\ \hline

\cite{article117} &  2021	& \ding[0.5]{108}\quad  Cutting-edge methods for estimating the SoC in battery packs, emphasizing how cell discrepancy affects pack functionality and SoC estimation.

\ding[0.5]{108}\quad Discussion on cell balancing methods and categorization of SoC estimation methods.
& \ding[0.5]{108}\quad Future research directions are not articulated.
 
 \ding[0.5]{108}\quad No discussion on cell balancing algorithms and optimisation techniques. \\ \hline
 
\cite{article37} &  2020	& \ding[0.5]{108}\quad  An overview of battery balancing systems based on DC-DC converters, highlighting their crucial role in controlling charge balance and the development of control algorithms for the same.

\ding[0.5]{108}\quad  Overview of functions of BMS, such as SoC estimation, which impacts balancing performance.& \ding[0.5]{108}\quad No comparative analysis of balancing algorithms. 

\ding[0.5]{108}\quad No discussion on optimisation techniques.\\ \hline

\cite{article109} &  2020	& \ding[0.5]{108}\quad  Examination of mainstream equalisation strategies, emphasising efficiency, complexity, and stability through examination of balancing variables and control algorithms.

\ding[0.5]{108}\quad Overview of variability in Li-ion batteries across their life-cycle, detailing classification methods and applications of balance management systems.&  \ding[0.5]{108}\quad Future research directions are not articulated.\\ \hline

\cite{article120} &  2020	& \ding[0.5]{108}\quad  Overview of research on battery equalizer circuits (BECs) utilised in EVs.

\ding[0.5]{108}\quad  Simulations across various BEC topologies using identical initial conditions and quantitative analysis are provided. &  \ding[0.5]{108}\quad No discussion on cell balancing algorithms and related optimisation techniques.\\ \hline

\cite{article121} &  2021	& \ding[0.5]{108}\quad  Analysis of variables, topologies, and modular methods in battery equalizer circuits.

\ding[0.5]{108}\quad Categorization of dominant cell balancing structures highlighting their comparative characteristics, and evaluation of modular methods using graph theory.
&  \ding[0.5]{108}\quad Future research directions are not articulated.

 \ding[0.5]{108}\quad No discussion on cell balancing algorithms is available and optimisation techniques. \\ \hline
 
Our survey & 	2024 	& \ding[0.5]{108}\quad  Broad classification of cell balancing algorithms, highlighting their merits and demerits.

\ding[0.5]{108}\quad   Discussion on the strengths and weaknesses of various SoC estimation methods and battery equivalent models.

\ding[0.5]{108}\quad  Provide guidelines for selecting cell balancing algorithms and parameter selection, along with optimization techniques.

\ding[0.5]{108}\quad  Discussion on the performance indicators of batteries after the application of cell balancing algorithms. & - \\ \hline
\end{tabular}}
\end{table*}

%=======================================================

\subsection{Motivation and Key Contributions}
The motivation for conducting this survey is to identify the right choice of cell balancing algorithm and the key aspects that need to be considered in the selection of the algorithm for the desired objective. The existing study highlights key factors contributing to cell imbalance, such as temperature variations within the battery pack and age-related cell degradation. The cell imbalance reduces capacity, efficiency, and battery lifespan simultaneously increasing safety concerns \cite{article111} thereby necessitating the implementation of cell balancing. The design and implementation of cell balancing algorithms include several key aspects to be considered for ensuring effectiveness, efficiency, and safety. The key aspects of algorithm design include minimizing deviations in SoC/voltage \cite{article93, article103} among cells, maintaining safe cell temperatures \cite{article93}, ensuring scalability for various pack sizes and cell chemistry, incorporating safety mechanisms such as over-current and thermal protection \cite{article81}, the right choice between continuous or periodic balancing strategies, and optimizing the overall performance. However, to achieve the optimal and efficient performance of the cell, there is an immediate need to formulate relevant balancing algorithms. The design and formulation of cell balancing algorithms ensures safe, efficient, and reliable battery management which is tailored to the specific requirements of the battery pack and related application(s). Hence, these multiple factors motivate us to explore cell balancing algorithms in detail with an aim to highlight such solutions which can be implemented to enhance cell performance and eventually optimize the battery pack performance.

Considering the above, the key contributions of this survey are:
\begin{itemize}
\item Exploration of the broad classification of cell balancing algorithms, highlighting their merits and demerits.
\item  Scrutiny of the positive and negative attributes of various SoC estimation methods and battery equivalent models in the context of cell balancing applications within BMS.
\item Providing the guidelines for selecting cell balancing algorithms and parameter selection, in addition to the optimization techniques.
\item Examination of the performance indicators of batteries after the application of cell balancing algorithms.
\end{itemize}

Section~\ref{Section:Introduction} provides a crisp overview of the recent advancements and the significance of cell balancing. In Section~\ref{Section: Broad Classification}, various types of cell balancing like SoC, voltage, and capacity, are discussed in detail, along with their classifications. Section~\ref{Section: SoC Classification},~\ref{Section: Volatge Classification} and~\ref{Section: Volatge-SoC Classification} describe the various developments in the SoC, Voltage and  Voltage-SoC dependent cell balancing, respectively. Section~\ref{Section: Optimisation Classification} puts forth the need and developments in optimisation with various performance indicators for cell balancing. Section~\ref{Section: Choice of cell Technique} discusses the factors associated with the selection of cell balancing techniques and optimisation. The article is concluded in Section~\ref{Section: Conclusion}.

\section{Classification of cell equalisation algorithms}
\label{Section: Broad Classification}
Balancing methods vary in their connection structures for transferring energy amongst the cells, allowing for either unidirectional or bidirectional energy flow \cite{article38}. Reducing the considerable energy dissipation is a critical aspect in the case of the passive technique. Several challenges persist with active techniques, such as inaccuracies in cell voltage sensing and balancing, sluggish response times, high voltage stress levels, and increased costs. Accordingly, each balancing method requires a unique control algorithm customized to its specific setup to address these limitations \cite{article25}. Cell-balancing algorithms provide the control logic that dictates the application of cell balancing to ensure the battery pack remains balanced by guiding the occurrences of balancing actions \cite{article10}. The control parameters, which serve as the input for the control scheme, provide insights into the level of disparity among the cells in a battery pack \cite{article37}. These variables include terminal pack voltage, measured cell voltages, Open Circuit Voltage (OCV), SoC, charge capacities, and multi-objective control \cite{article37}, \cite{article38}. The choice of control variable is a crucial factor that directly impacts the precision of balancing approaches. These control algorithms differ according to the design approach and the selection of control variables. The control variables and their respective benefits and drawbacks are examined and tabulated, together with the classification of cell balancing algorithms, in Table~\ref{tab: Control variables} and Table~\ref{tab: Types of cell balancing algorithms}, respectively. The description of the techniques is detailed below.

%******************************************************

\begin{table*}[t!]
\caption{Control variables \cite{article38,article37,article121,article135}}
\label{tab: Control variables}
\centering
\resizebox{\textwidth}{!}{
    \begin{tabular}{|m{0.2\textwidth}|m{0.4\textwidth}|m{0.4\textwidth}|m{0.08\textwidth}|} \hline
    \textbf{Control Variable} & \textbf{Merits} & \textbf{Demerits}  \\\hline

OCV & \ding[0.5]{108}\quad No impact of state estimation inaccuracies.

\ding[0.5]{108}\quad No balancing error due to variations in total charge capacity. & \ding[0.5]{108}\quad Absence of balancing within the flat OCV region.  \\\hline
 Measured cell voltage   &  \ding[0.5]{108}\quad State estimation error has no influence. 
 
 \ding[0.5]{108}\quad Influence of measuring error only.
 
\ding[0.5]{108}\quad No state estimation required. &   \ding[0.5]{108}\quad  Balancing error due to internal resistance variance.

\ding[0.5]{108}\quad  Balancing error due to variation of total charge capacity.

\ding[0.5]{108}\quad  Balancing not required in the flat OCV area. \\\hline
SoC    &   \ding[0.5]{108}\quad Compensation for internal resistance variance is possible.
    
    \ding[0.5]{108}\quad Balancing in the flat OCV section. 
    
    \ding[0.5]{108}\quad Temperature impact on cell parameters is taken into account
    
      &  \ding[0.5]{108}\quad Impact of state estimation error. 
    
    \ding[0.5]{108}\quad Equalisation error due to total charge capacity variance when not taken into account. \\\hline
    
Charge capacity   &\ding[0.5]{108}\quad Compensation for internal resistance variance.

\ding[0.5]{108}\quad Balancing in the flat OCV section. 

\ding[0.5]{108}\quad Effect of temperature is taken into account.

\ding[0.5]{108}\quad Balanced cell energies. &  \ding[0.5]{108}\quad Different error impact (e.g., SoC estimation, measurements) \\\hline

Multi-objective: e.g., SoC and temperature   &   \ding[0.5]{108}\quad  Compensation of internal resistance variance.

\ding[0.5]{108}\quad  Balancing in flat OCV section.

\ding[0.5]{108}\quad  Concern of temperature dependencies.

\ding[0.5]{108}\quad  Control effect on cell ageing.
 &   \ding[0.5]{108}\quad  Various sources of error (e.g., SoC estimation, measurements).
 
\ding[0.5]{108}\quad  Balancing error due to variations in total charge capacity. \\\hline
 State of Health (SoH)   & \ding[0.5]{108}\quad Discrepancies in battery capacity due to ageing is considered.   &  \ding[0.5]{108}\quad Since conventional sensors cannot directly obtain SoH, it is still hard to make an online SoH estimation. 
 
 \ding[0.5]{108}\quad SoH estimation is relatively tough to acquire when batteries are in use.  \\\hline
State of Function (SoF)  & \ding[0.5]{108}\quad  Details the battery's output capacity under specific conditions.  & \ding[0.5]{108}\quad Ambiguous definition.  \\\hline
 % &    &   \\\hline
 % &    &   \\\hline
 % &    &   \\\hline
% & \ding[0.5]{108}\quad   & \ding[0.5]{108}\quad 
\end{tabular}}
\end{table*}

%******************************************************

\begin{table*}[t!]
\caption{Classification of cell balancing algorithms \cite{article1, article2,article3},\cite {article26,article37,article38}}
\label{tab: Types of cell balancing algorithms}
\centering
\resizebox{\textwidth}{!}{
    \begin{tabular}{|m{0.08\textwidth}|m{0.08\textwidth}|m{0.2\textwidth}|m{0.3\textwidth}|m{0.4\textwidth}|} \hline
    \textbf{Type} & \textbf{Control variable} &\textbf{Definition} & \textbf{Merits} & \textbf{Demerits}  \\\hline

SoC-based balancing algorithm &  Cell SoC & Balances cells by controlling SoC to be within a predefined SoC threshold. The SoC is derived from voltage or current measurements.  &   \ding[0.5]{108}\quad Reflects the existing capacity of the battery pack.

\ding[0.5]{108}\quad Potentially more accurate in achieving charge balance.

\ding[0.5]{108}\quad Balancing in flat OCV section is possible.

\ding[0.5]{108}\quad Less influenced by the cell's operating state.

& \ding[0.5]{108}\quad Requires more complex estimation methods.

\ding[0.5]{108}\quad Balancing effectiveness largely dependent on the accuracy of SoC estimation.

\ding[0.5]{108}\quad Can sometimes intensify imbalance in a battery string, resulting in a more uneven state than if no balancing was applied.

\ding[0.5]{108}\quad SoC estimations can be impacted by the battery model, self-discharge rates, and temperature. \\\hline

Voltage dependent balancing algorithm  & Cell voltage & Balances cells by monitoring the voltage deviation between a cell and the mean voltage of the pack. If the difference exceeds a threshold $V^{th}$, balancing is triggered to address voltage disparities.   & \ding[0.5]{108}\quad  Easy to implement as it operates with direct voltage readings.

\ding[0.5]{108}\quad No SoC estimation is needed, eliminating the risk of state estimation errors.

\ding[0.5]{108}\quad Affected only by measurement errors.

   \ding[0.5]{108}\quad Efficient for quick detection of voltage discrepancies. 

   & \ding[0.5]{108}\quad Influenced by external factors like temperature and internal resistance.
   
    \ding[0.5]{108}\quad The balancing process tends to be most effective when the SoC is at either low levels or high levels.
    
\ding[0.5]{108}\quad Balancing errors occur due to variations in total charge capacity.

\ding[0.5]{108}\quad No balancing in areas where the OCV is flat.

\ding[0.5]{108}\quad Balancing errors can occur as a result of internal resistance and capacity loss due to ageing.

\ding[0.5]{108}\quad Challenges in determining correct voltage deviation in-vehicle applications can lead to over-equalization, causing a significant rise in power consumption.

\ding[0.5]{108}\quad Reducing balancing current helps prevent over-equalisation but results in longer balancing times.   \\\hline

Voltage-SoC balancing control scheme & Cell voltage and SoC  &The focus is on assessing the deviation in SoC between two cells instead of measuring the SoC of each individual cell, allowing for identification of the cell requiring balancing.
& \ding[0.5]{108}\quad Simple with improved performance.

\ding[0.5]{108}\quad Offers more accurate balancing compared to strategies based on terminal voltage.

\ding[0.5]{108}\quad Cells can be fully charged or discharged, enabling full use of the pack's energy capacity even when cells have differing total capacities.

\ding[0.5]{108}\quad This method requires less computational load.
 & 
\ding[0.5]{108}\quad Susceptible to noise and external factors. 

\ding[0.5]{108}\quad There is a possibility of occurrence of either overbalance or under-balance of cells.

\ding[0.5]{108}\quad Accuracy relies on the degree of SoC estimation errors.

\ding[0.5]{108}\quad Balancing errors caused by variations in total charge capacity.

\ding[0.5]{108}\quad No balancing capability in areas where the OCV is flat.\\\hline

% &    &  & &\\\hline
\end{tabular}}
\end{table*}

%----------------------------------------
\subsection{SoC-based cell balancing} 
\begin{figure*}[hbt!]
    \centering    
    \includegraphics[width=\textwidth]{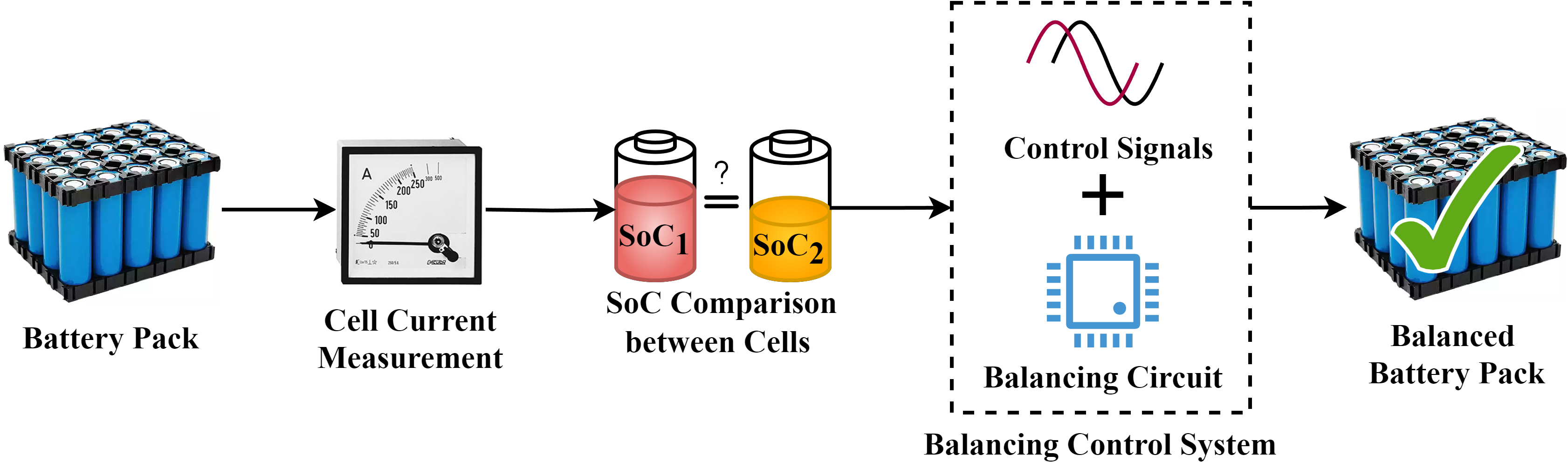}   
    \caption{Control strategy based SoC dependent cell balancing.} 
    \label{fig:SoC Cycle}
\end{figure*}

The constraints of voltage dependent cell balancing can be addressed with SoC-based control \cite{article25}. The latter is a more realistic approach because SoC is the primary factor causing voltage variations. When cells are balanced to have identical SoC, their voltages will naturally align \cite{article1}. The SoC-based cell equalization process adjusts the charge strength of individual cells in a battery pack to maintain balance according to their SoC \cite{article121}. The schematic  representation of SoC-based cell balancing is illustrated in Fig.~\ref{fig:SoC Cycle}. The process involves SoC measurements and in the case of mismatch, a cell balancing algorithm is activated to transfer energy from the cell with elevated SoC to the cell with low SoC to bring the SoC levels in the package to a sensible level during both charging and discharging action \cite{article37}. By redistributing charge among the cells, it ensures a balanced SoC, which improves overall battery performance, extends its lifespan, and enhances safety. This process of accurate estimation of SoC is complex, leading to significant computational burden \cite{article25}.

\begin{figure*}[hbt!]
    \centering
    \includegraphics[width=\textwidth]{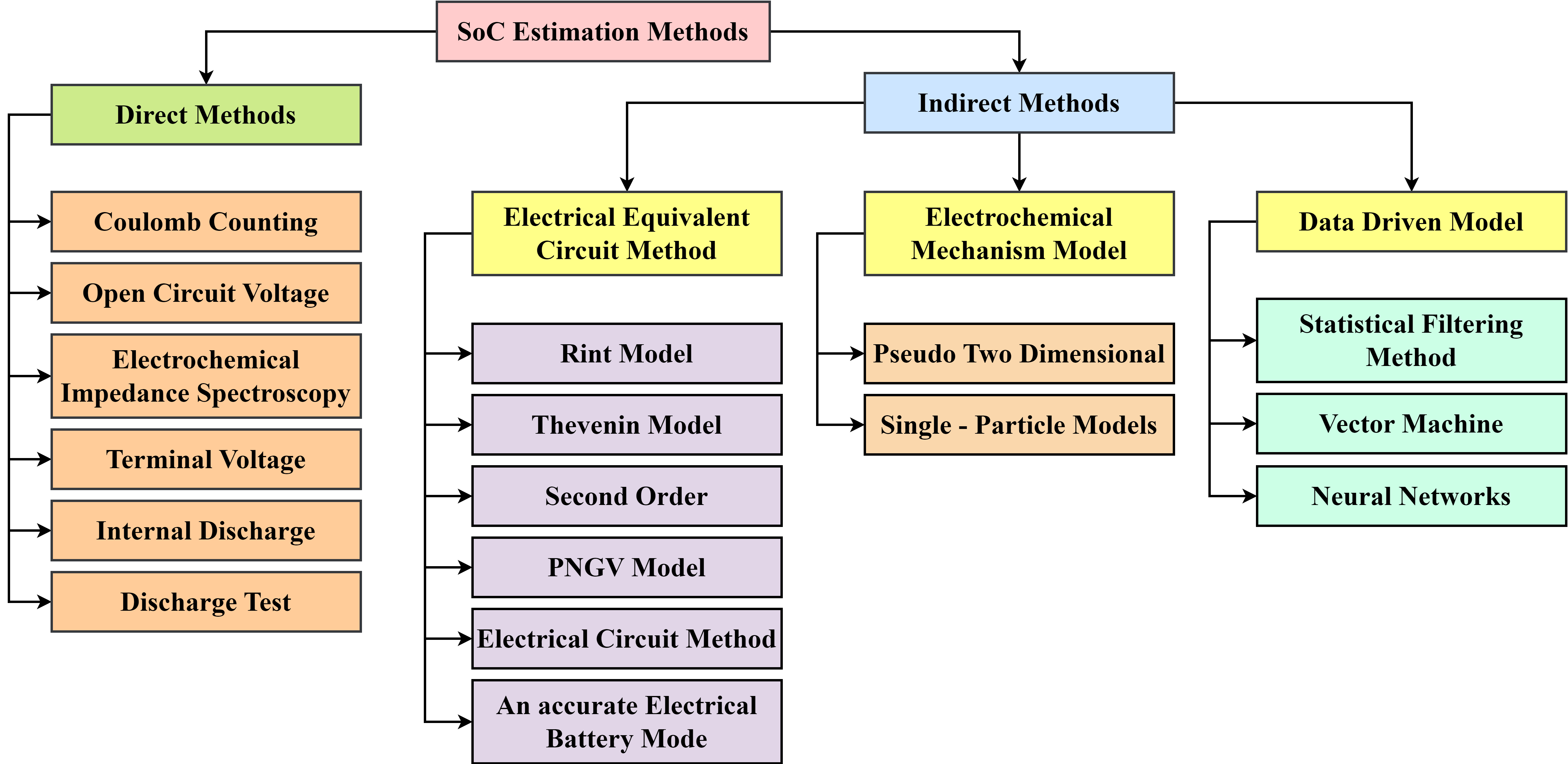}
    \caption{Classification of SoC estimation methods.} 
    \label{fig:SoC methods}
\end{figure*}

In the context of cell balancing, the approaches depicted in Fig.~\ref{fig:SoC methods} offer a means to estimate SoC with sufficient precision. This information helps to determine when and how much balancing is necessary to ensure uniformity among the cells. These methods play a significant role in guiding balancing decisions during cell balancing.

\subsubsection{Direct Methods}
The direct method, which does not rely on a battery model, estimates the battery's SoC dependent on measurable parameters such as voltage,internal resistance, current, impedance, and other reproducible battery parameters comprising a strong connection with the battery's SoC. These parameters are relatively easy to measure in real-world applications \cite{article46}. A key benefit of this system is that it does not require constant connection to the battery, and measurements can be noted once the battery is connected \cite{article52}. Table~\ref{tab:Direct SoC estimation methods} shows the various direct methods of SoC estimation.

\begin{table*}[t!]
\caption{SoC estimation using Direct methods \cite{article46,article52,article53,article54,article55,article100}}
\label{tab:Direct SoC estimation methods}
\centering
\resizebox{\textwidth}{!}{
    \begin{tabular}{|m{0.12\textwidth}|m{0.4\textwidth}|m{0.25\textwidth}|m{0.4\textwidth}|} \hline
    \textbf{Method Type} & \textbf{Operation} &\textbf{Merits} & \textbf{Demerits}   \\\hline
 
Coulomb Counting or Ah integration method  &  Coulomb counting is a method for estimating the SoC of a battery by tracking the flow of electrical charge in and out of the battery over time. It involves measuring the current and integrating it over time to calculate the net charge added to or removed from the battery. & \ding[0.5]{108}\quad Simple and straightforward, making it easy to implement.

\ding[0.5]{108}\quad Minimal requirements for controller hardware and data storage.

\ding[0.5]{108}\quad Serves as the foundation for many other SoC estimation algorithms.
 & \ding[0.5]{108}\quad This approach demands precise current sensors, but their accuracy can be influenced by noise, temperature variations, and other arbitrary disturbances.

\ding[0.5]{108}\quad Over time, the cumulative error in SoC estimation tends to increase.

\ding[0.5]{108}\quad The method lacks an initial convergence mechanism, heavily relying on the initial SoC value, making the accuracy of the starting point crucial. \\\hline

OCV & The relatively stable relation between a battery's OCV and its SoC enables the estimation of the SoC. This relationship can be determined through lookup table or curve fitting.   &  \ding[0.5]{108}\quad Simple.&  \ding[0.5]{108}\quad  SoC estimation relies on the relationship between OCV and SoC, which varies among different batteries.

\ding[0.5]{108}\quad  Since the OCV-SoC curve is affected by temperature and discharge rate, this method accurately estimates SoC only at the start and end of charging and discharging cycles, after the battery has been detached from the load for a prolonged period.

\ding[0.5]{108}\quad The OCV can only be measured after a lengthy battery relaxation period, rendering it unsuitable for real-time applications.\\\hline

Terminal Voltage &   When a battery discharges, its terminal voltage decreases due to internal impedances, making the Electromotive Force (EMF) proportional to the terminal voltage, as the battery's EMF is generally proportional to its SoC, the terminal voltage also tends to correlate with the SoC. & \ding[0.5]{108}\quad Can be utilised to estimate SoC across various discharge currents and temperatures. & \ding[0.5]{108}\quad Can be significantly inaccurate at the end of battery discharge due to the sudden drop in terminal voltage.
\\\hline

Electrochemical Impedance spectroscopy (EIS) &  EIS is a technique used to measure and analyse the impedance of an electrochemical system, such as a battery, across a range of frequencies. It provides insights into the system's electrical and chemical characteristics, allowing the study of reaction mechanisms, internal resistances, and other critical properties.  & \ding[0.5]{108}\quad The offline EIS technique offers highly accurate impedance measurements.

\ding[0.5]{108}\quad It can be adapted for online use to provide a more robust and efficient alternative.
Provides valuable data by using galvanostatic or potentiostatic excitation signals across a wide frequency range to measure battery impedance during charging and discharging.
 &  \ding[0.5]{108}\quad Traditional EIS measurement equipment tends to be large and expensive, making it unsuitable for direct use in vehicles.
 
\ding[0.5]{108}\quad Can only be practically used if implemented as compact, efficient, and cost-effective onboard systems for EVs. \\\hline

Internal resistance method &   The internal resistance method estimates the SoC of a battery by leveraging the monotonic relationship between the battery's internal resistance and SoC, provided the internal resistance is known. & \ding[0.5]{108}\quad  Simple.

\ding[0.5]{108}\quad Precision of estimation is high.
 &  \ding[0.5]{108}\quad The resistance testing device is costly, while the internal resistance is low with a narrow range of variation, and is easily influenced by temperature and cycle count.\\\hline
 
\end{tabular}}
\end{table*}

\subsubsection{Indirect Methods}
Indirect methods are based on battery models. They use equations to represent physicochemical phenomena like diffusion, intercalation, and electrochemical kinetics that take place in a battery \cite{article24}.

\paragraph{Electrical Equivalent Circuit models}

\begin{figure*}[hbt!]
    \centering    
    \includegraphics[width=\textwidth]{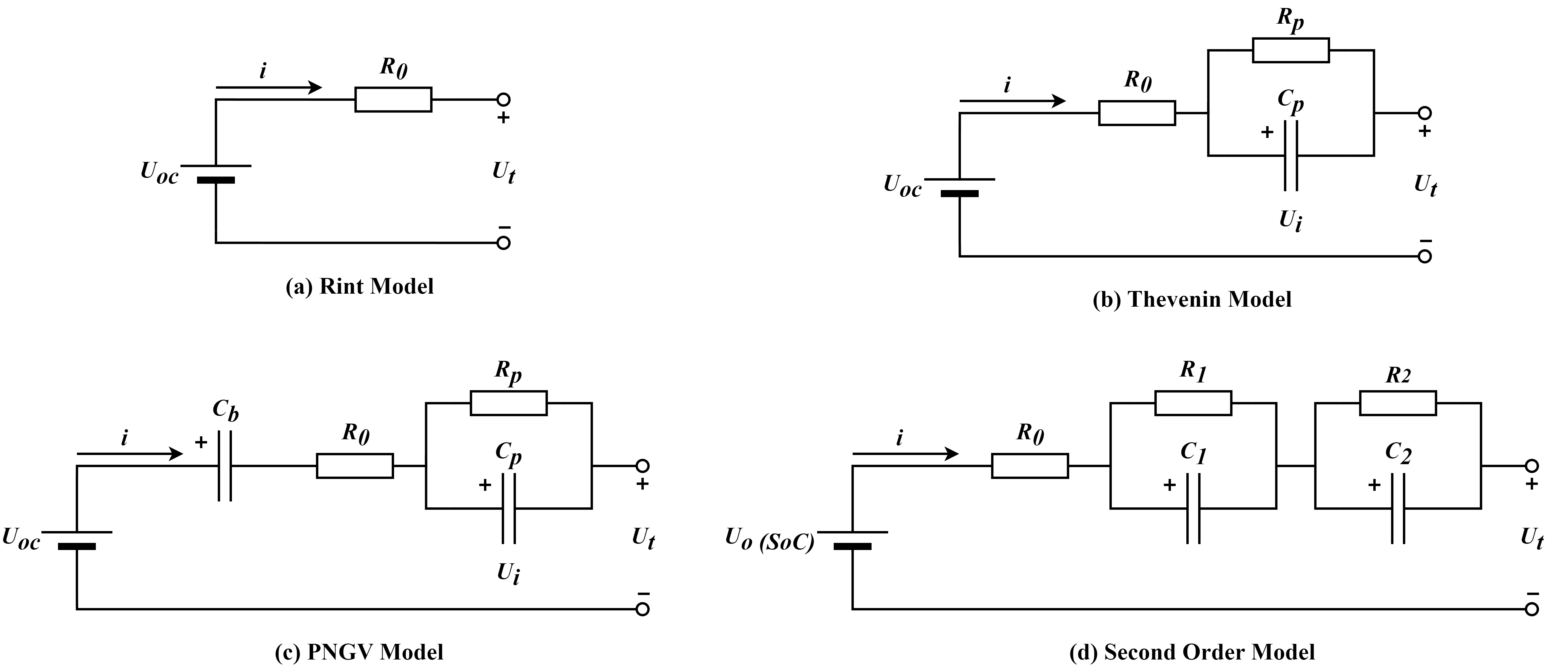} 
    \caption{Battery equivalent circuit models.} 
    \label{fig:Circuit Models}
\end{figure*}

Battery electrical equivalent circuit models allow to understand the model's behaviour with regard to electrical characteristics, charging status, and capacity \cite{article42}. For circuit designers aiming to manage energy consumption in battery-powered systems and predict battery characteristics, accurate information such as charging status, current, and voltage is critical \cite{article43}. The significant aim of BMS is to monitor the condition and state of the traction battery pack, including metrics like SoC and SoH \cite{article136}. Since these parameters cannot be directly measured by any sensor, they ought to be estimated, often employing model dependent algorithms \cite{article47}. This indicates that accurate battery models are essential. These models will assist in comprehending dynamic battery behaviour \cite{article43}, simultaneously playing a role in boosting system performance and efficiency. By utilizing precise and efficient circuit and battery models, circuit designers can anticipate and enhance battery runtime and circuit functionality. \cite{article46}.In addition to imparting precise estimations, it is crucial to balance model complications with precision so that these models can be integrated into microprocessors and deliver reliable results in real time \cite{article47}. The battery equivalent circuit models are given in Fig.~\ref{fig:Circuit Models}. The Rint model shown in Fig.~\ref{fig:Circuit Models}(a) is derived by approximating the external characteristics of the battery as linear \cite{article121}. This model uses an ideal voltage source and the battery's DC internal resistance ($R_{0}$) in series to represent the dynamic characteristics of the power battery. Both $R_{0}$ and the voltage source depend on SoC and temperature \cite{article46}. The Thevenin model given in Fig.~\ref{fig:Circuit Models}(b) is built upon the Rint model by incorporating a parallel RC circuit to simulate the battery's polarization action \cite{ article46,article121}. During charging or discharging, voltage changes at battery terminals exhibit both sudden and gradual variations. In the Thevenin model, $R_0$ represents the abrupt resistance characteristics, while $R_{P}$ and $C_{P}$ simulate the capacitance characteristics responsible for the slow voltage changes \cite{ article46}. Building on the Thevenin model, a capacitor ($C_{b}$) can be connected in series to create a non-linear equivalent circuit model i.e Partnership for a New Generation of Vehicles (PNGV) model \cite{article46}. This capacitor accounts for changes in the battery's open circuit voltage due to current integration over long-term charging and discharging processes. The PNGV model, illustrated in Fig.~\ref{fig:Circuit Models}(c)is widely adopted because it combines high model accuracy with ease of parameter identification \cite{article121}. Fig.~\ref{fig:Circuit Models}(d) gives a second-order model that consists of two capacitors and three resistors. This design considers both the impact of transient and  polarization action of the battery during operation. However, it excludes efficiency inconsistencies during battery charging and discharging \cite{article46}.

Table~\ref{tab:Battery equivalent circuits} gives a comparison of various battery models, indicating their component count, advantages, and disadvantages.

\begin{table*}[t!]
\caption{Battery equivalent circuits \cite{article42,article43,article44,article46}}
\label{tab:Battery equivalent circuits}
\centering
\resizebox{\textwidth}{!}{
    \begin{tabular}{|m{0.15\textwidth}|m{0.3\textwidth}|m{0.4\textwidth}|m{0.4\textwidth}|m{0.4\textwidth}|} \hline
    \textbf{Battery model} & \textbf{Circuit elements} & \textbf{Merits} & \textbf{Demerits} \\\hline
  Rint Model &  Constant ideal voltage source, battery internal resistance  &  \ding[0.5]{108}\quad  Simple.
  
 \ding[0.5]{108}\quad  Basis to other equivalent circuits.
 
 &  \ding[0.5]{108}\quad  The model doesn't consider load transfer polarisation and diffusion polarisation, leading to a narrow application range and inadequate measurement accuracy.
 
\ding[0.5]{108}\quad  In the course of charging and discharging, some of the electrical energy is converted into heat energy.\\\hline
 Thevenin model   &  Battery ohmic internal resistance, polarisation internal resistance, capacitance is the polarisation capacity, constant ideal voltage source.  & \ding[0.5]{108}\quad Takes into account both the battery's heat consumption and its internal polarisation action.

\ding[0.5]{108}\quad Applicable for non-linear circuits.

\ding[0.5]{108}\quad Simple.
& \ding[0.5]{108}\quad SoC estimation accuracy decreases as a consequence of battery ageing. \\\hline
 
Second-order    &   To Thevenin model, additionally parallel RC network is included to consider the non-linear polarisation response. & \ding[0.5]{108}\quad  More accurate.

\ding[0.5]{108}\quad  Offers a more precise depiction of internal ohmic polarisation, electrochemical polarisation, and concentration polarisation.

\ding[0.5]{108}\quad Model align more closely with the actual operating characteristics of the battery.
 & \ding[0.5]{108}\quad Complex structure. \\\hline
 PNGV model   & In addition to circuit elements of second-order, a large capacitor is included to capture the variation in battery OCV.   &  \ding[0.5]{108}\quad It is a non-linear equivalent circuit model, known for its high precision in getting transient responses.

\ding[0.5]{108}\quad It is well-suited for high current, step-type, and more complex charging and discharging conditions.
& \ding[0.5]{108}\quad Complex structure.

\ding[0.5]{108}\quad Involves a substantial amount of calculations.

\ding[0.5]{108}\quad Low real-time performance. \\\hline

% &    &  & \\\hline
\end{tabular}}
\end{table*}

\paragraph{Electrochemical Mechanism model}
The electrochemical mechanism model establishes equations for electrochemical power and transmission based on the internal workings of the battery. It takes into consideration the physical and chemical properties of the positive and negative materials, the internal diffusion process, the electrochemical reaction process, and other relevant factors. Further, it provides a comprehensive and precise description of the internal physical and chemical processes, and also the external characteristics of the battery \cite{article46,article48}. The merits and demerits of various electrochemical mechanism models are given in Table~\ref{tab:Electrochemical Mechanism model}.

\begin{table*}[t!]
\caption{Electrochemical Mechanism model \cite{article46,article47,article51}}
\label{tab:Electrochemical Mechanism model}
\centering
\resizebox{\textwidth}{!}{
    \begin{tabular}{|m{0.15\textwidth}|m{0.35\textwidth}|m{0.55\textwidth}|} \hline
    \textbf{Control Variable} & \textbf{Merits} & \textbf{Demerits}  \\\hline

 Pseudo-Two-Dimensional (P2D) &  \ding[0.5]{108}\quad Known for its rigor and accuracy.  &  \ding[0.5]{108}\quad The partial differential equations within the model lack analytical solutions.
 
\ding[0.5]{108}\quad Typically, the finite difference method is utilised to determine the solution of the equations, which can be time-consuming.

\ding[0.5]{108}\quad Computational complexity and hence unsuitable for real-time SoC estimation.
 \\\hline
 Single-particle models (SP)  & \ding[0.5]{108}\quad It is a simplified variant of the P2D model with only two control equations, decreasing the number of parameters and increasing the simulation speed.
 
\ding[0.5]{108}\quad Improved simulation speed.
  &  \ding[0.5]{108}\quad Due to its simplifications, the model has a lower accuracy.
   
\ding[0.5]{108}\quad Suitable only for low magnifications and constant current situations.

\ding[0.5]{108}\quad Error increases when the current magnification escalates, resulting in significant electrolyte concentration changes.
 \\\hline
 %  &    &   \\\hline
 % &    &   \\\hline
 % &    &   \\\hline
% & \ding[0.5]{108}\quad   & \ding[0.5]{108}\quad 
\end{tabular}}
\end{table*}

\paragraph{Data-Guided methods}
Data-guided models have gained considerable attention for their flexibility and model-free benefits \cite{article135}. They avoid the complexities of modelling and parameter identification associated with model-based estimation methods, allowing for direct analysis of masked information and evolving patterns from the external characteristic parameters of a battery. These methods have found wide applications in the field of battery analysis\cite{article50}. Further, the data-guided models have gained considerable attention for their flexibility and model-free benefits. These methods have also found wide applications in the field of battery analysis \cite{article46}. The choice of a particular SoC estimation method is influenced by different factors. These factors include the battery's electrochemical characteristics, the operational necessities of the determined application, and the expected degree of estimation accuracy \cite{article108}. The merits and demerits of different data-guided models are detailed in Table~\ref{tab:Methods for SoC estimation}.

\begin{table*}[t!]
\caption{SoC estimation using Data Guided models \cite{article30,article31,article39,article40,article41,article46,article48,article49}}
\label{tab:Methods for SoC estimation}
\centering
\resizebox{\textwidth}{!}{
    \begin{tabular}{|m{0.1\textwidth}|m{0.08\textwidth}|m{0.45\textwidth}|m{0.55\textwidth}|} \hline
    \textbf{Approach} & \textbf{Methods} & \textbf{Merits} & \textbf{Demerits} \\\hline
\multirow{5}{0.1\textwidth}{Statistical filtering method} & KF & \ding[0.5]{108}\quad Excels in handling fluctuations in the initial SoC values.

\ding[0.5]{108}\quad Acts as a closed-loop observer, guaranteeing precision and uninterrupted estimation functionality.
& \ding[0.5]{108}\quad Restricted to linear systems. 
                    
\ding[0.5]{108}\quad They are insufficient for addressing the non-linear characteristics commonly observed in batteries. \\\cline{2-4}

  & EKF &  \ding[0.5]{108}\quad  Simple.
 
\ding[0.5]{108}\quad  Linearises non-linear models.
&  \ding[0.5]{108}\quad The complexity of implementing EKF arises from the necessity to linearise non-linear models.

\ding[0.5]{108}\quad Tuning EKF parameters, like noise covariances, demands expertise because of their complexity.

\ding[0.5]{108}\quad EKF's susceptibility to model errors, particularly stemming from linearisation, can compromise both its accuracy and reliability.

\ding[0.5]{108}\quad Its reliability comes into question when dealing with systems that demonstrate only marginal linearity throughout update intervals.  \\\cline{2-4}
 &  UKF  &  \ding[0.5]{108}\quad 	Able to achieve greater estimation performance.

\ding[0.5]{108}\quad Applicable to non-linear systems.

\ding[0.5]{108}\quad Exhibits better convergence characteristics and greater accuracy
&\ding[0.5]{108}\quad The system's robustness is compromised due to factors like abnormal disturbances and uncertainties in initial values, leading to divergence.\\\cline{2-4}
  
 & AKF & \ding[0.5]{108}\quad   Capable of continuously estimating the system condition in real-time and mitigating the noise effects. & \ding[0.5]{108}\quad Complex model training

\ding[0.5]{108}\quad High computing resource necessity.

\ding[0.5]{108}\quad Extensive configuration needed.

\ding[0.5]{108}\quad There's a tendency for the model to over-fit the training data, potentially reducing its generalisation ability. \\\cline{2-4}

   & PF &   \ding[0.5]{108}\quad Unrestricted by the linear and Gaussian conditions of the system model.

\ding[0.5]{108}\quad Minimal restrictions on the probability distribution of state variables.
& \ding[0.5]{108}\quad  Estimation precision tends to be unstable, and there's a high likelihood of particle depletion.\\\hline

Vector machine & SVM, RVM & \ding[0.5]{108}\quad Strong generalisation ability.

\ding[0.5]{108}\quad No dependence on battery model.

\ding[0.5]{108}\quad High precision estimation.

\ding[0.5]{108}\quad Rapid convergence when using a small number of samples. & \ding[0.5]{108}\quad The estimation accuracy relies significantly on a huge volume of specimen data and weight parameters. \\\hline

\multirow{5}{0.12\textwidth}{NN}& Artificial NN, Convolution NN & \ding[0.5]{108}\quad No battery model required.

\ding[0.5]{108}\quad It has a strong capacity to process a wide range of variables.

\ding[0.5]{108}\quad The method can adapt and improve through self-learning.

\ding[0.5]{108}\quad Suitable for Real-Time SoC detection. & \ding[0.5]{108}\quad The quality and quantity of samples significantly influence the training outcomes.

\ding[0.5]{108}\quad Need large number of training samples.\\\cline{2-4}
& Deep learning& \ding[0.5]{108}\quad  Strong generalisation ability.

\ding[0.5]{108}\quad	Parallel processing capability.

\ding[0.5]{108}\quad  High accuracy.

\ding[0.5]{108}\quad  The outcomes maintain consistent performance and reliability.
 & \ding[0.5]{108}\quad  The process of training the model involves significant complexity.
 
\ding[0.5]{108}\quad  Requires high computing resource.

\ding[0.5]{108}\quad  There's a tendency for the model to over-fit the training data, potentially reducing its generalisation ability.\\\hline
% &    &  & \\\hline
\end{tabular}}
\end{table*}

\paragraph{Summary of the indirect approach}
Table~\ref{tab:Summary of indirect approach}  overviews the electrical equivalent circuit, electrochemical mechanism, and data-guided models. It is apparent that each battery model has its unique pros and cons. The precision of SoC estimation is closely linked to the accuracy of the model used. Generally, higher model accuracy requires increased complexity. The electrochemical mechanism model offers a detailed representation of the inherent chemical reactions within the battery, resulting in higher estimation accuracy. However, this model requires significant computational resources due to its complexity. Data-guided models provide flexibility and adaptability, allowing SoC estimation with a data-centric focus. However, they may demand substantial computational resources and rely heavily on algorithm choice and large datasets for training and accuracy.

\begin{table*}[t!]
\caption{Summary of the indirect approach \cite{article46,article47,article51}}
\label{tab:Summary of indirect approach}
\centering
\resizebox{\textwidth}{!}{
    \begin{tabular}{|m{0.1\textwidth}|m{0.07\textwidth}|m{0.1\textwidth}|m{0.16\textwidth}|m{0.07\textwidth}|m{0.3\textwidth}|m{0.4\textwidth}|} \hline
    \textbf{Model type} & \textbf{Accuracy} & \textbf{Computational complexity}  &\textbf{Time}&\textbf{Interpret capability} &\textbf{Merits} &\textbf{Demerits}\\\hline

  Electrochemical mechanism & Very high   &  Very high & Solving control equations is time-consuming. & Low & \ding[0.5]{108}\quad Provide an accurate reflection of battery characteristics, offering a high level of precision. & \ding[0.5]{108}\quad Lacks adaptability to certain operating conditions, resulting in inaccurate estimation outcomes.

\ding[0.5]{108}\quad Models utilise partial differential equations with numerous unknown parameters. This complexity often demands substantial memory and computational resources.

\ding[0.5]{108}\quad Not suitable for actual battery management in EVs.
\\\hline

 Electrical equivalent circuit model   &  Medium  &  Medium to low & Simple and easy to understand, resulting in moderate time consumption. & High & \ding[0.5]{108}\quad  Simple structure.
 
\ding[0.5]{108}\quad  Easy approach to get model parameters.

\ding[0.5]{108}\quad  These are lumped models with a relatively minimal parameters.
& \ding[0.5]{108}\quad Cannot accurately represent the internal characteristics of the battery.\\\hline

 Data-guided models    &  Medium  & Medium & Requires less time since earlier battery knowledge is unessential. & Low & \ding[0.5]{108}\quad  Independent of battery model, avoiding the cumbersome process of physical modelling. 

\ding[0.5]{108}\quad  It allows for a quick evaluation and analysis of the battery's internal state.

\ding[0.5]{108}\quad  Well-designed models are suitable for large-scale applications.

 & \ding[0.5]{108}\quad  The accuracy of the estimation is more dependent sample size, and the convergence speed is slow.
 
\ding[0.5]{108}\quad  With a small sample size and a high value of numerical error, the model may experience overfitting or underfitting.

\ding[0.5]{108}\quad  The system's ability to generalise depends greatly on selecting the appropriate algorithm.

\ding[0.5]{108}\quad  Require extensive training periods, which could delay deployment or adaptation to new conditions.

\ding[0.5]{108}\quad  Choice of critical input parameters is very critical.
\\\hline
  %    &    &  & & & & \\\hline
      
% & \ding[0.5]{108}\quad   & \ding[0.5]{108}\quad 
\end{tabular}}
\end{table*}

%===============================
%********************************************************************
\subsection{Voltage dependent cell balancing}
\begin{figure*}[hbt!]
    \centering    
    \includegraphics[width=\textwidth]{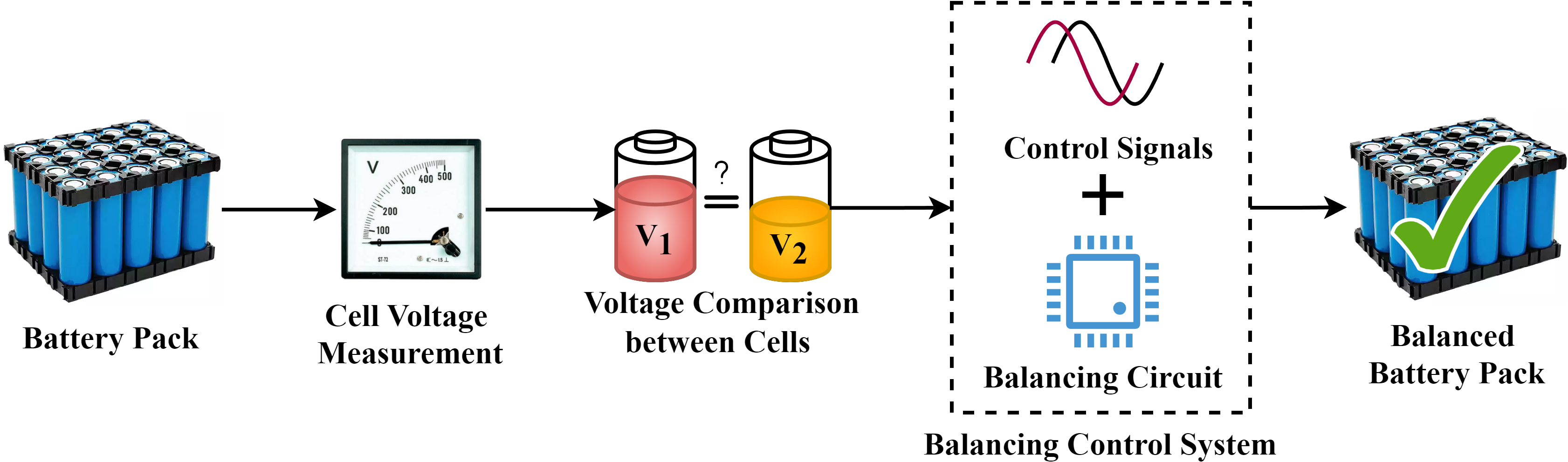}   
    \caption{Control scheme dependent on terminal voltage or Illustration of Voltage Cycle.} 
    \label{fig:Voltage Cycle}
\end{figure*}

The voltages of individual cells connected in series tend to become imbalanced over time due to discrepancies in characteristics like capacity or capacitance, self-discharge rates, and internal impedance\cite{article36}. In the voltage dependent cell balancing technique given in Fig.~\ref{fig:Voltage Cycle}, the balancing mechanism is activated when a cell’s voltage significantly deviates from the mean voltage of all cells in the battery pack. The battery balancing arrangement initiates the balancing activity by comparing the voltage difference between the cells to a pre-set threshold \cite{article1}. If the voltage variation surpasses this threshold, in active balancing, cells with minimum voltage are charged via cells with maximum voltage within the battery pack \cite{article37}. This method is simple to implement because it relies on direct voltage measurements, which are generally straightforward to obtain from a BMS. The system often uses resistive (passive) balancing to discharge cells with higher voltages, equalising them with those having lower voltages. However, voltage might not completely represent the actual characteristics of the batteries \cite{article3}. Also, cell voltage can be affected by factors other than SoC, like temperature deviation or internal resistance. This external impact can lead to less accurate balancing decisions \cite{article1}. For Li-ion batteries, the balancing process tends to be most effective when the SoC is at either low levels or high levels, where the OCV to SoC relationship has a steeper gradient \cite{article2}. This steepness allows for more sensitive and accurate balancing based on voltage measurements, as small changes in SoC are reflected by significant changes in voltage in these sections. However, variations in internal resistance can cause deviation in terminal voltages even when the SoC is the same. This inconsistency means that even if cells are balanced according to their terminal voltages, their usable capacities might not align \cite{article37}. 

\subsection{Voltage-SoC-based cell balancing} 
\begin{figure*}[hbt!]
    \centering    
    \includegraphics[width=\textwidth]{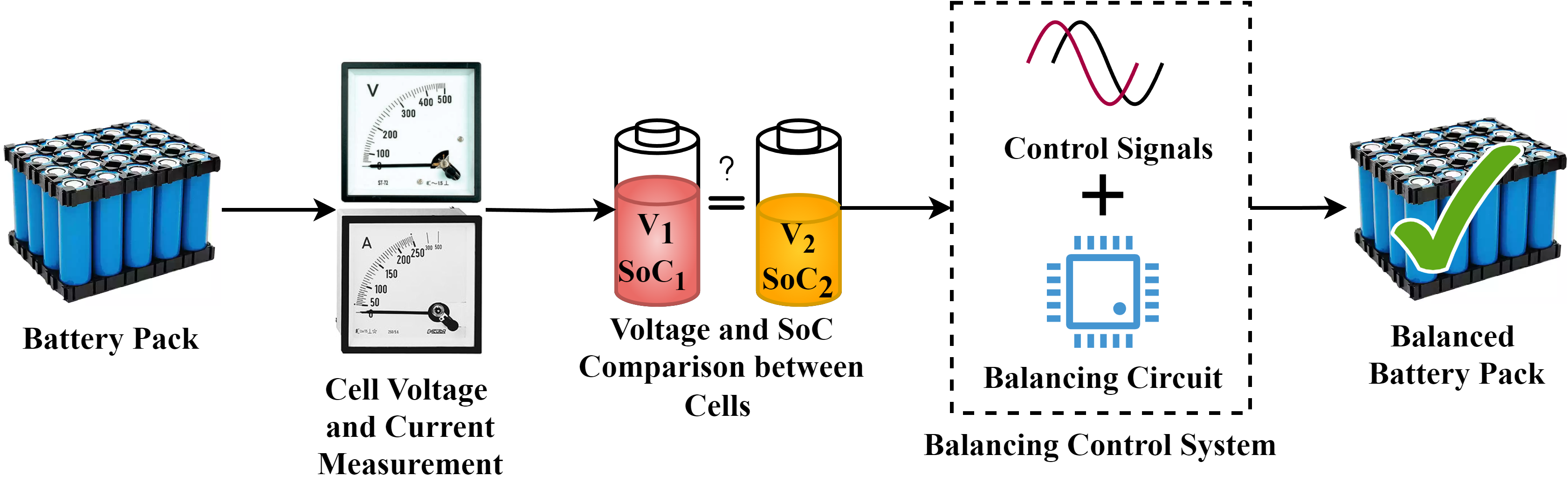} 
    \caption{ Control strategy based on Voltage-SoC Cycle.} 
    \label{fig:Voltage-SoC Cycle}
\end{figure*}

The widely used cell balancing flow for a voltage-SoC-based scheme is shown in Fig.~\ref{fig:Voltage-SoC Cycle}. In this scheme, indirectly deviation among the SoCs of the cell is estimated through the terminal voltage and the SoC of battery pack. This is accomplished through measurements of the terminal voltage and the SoC of the battery pack. By determining the SoC variation amongst the cells, this scheme identifies the cell in need of balancing. This approach enables effective battery pack balancing through Ah counting, determining the moment when both cells achieve balanced SoC levels. As a result, this method facilitates battery pack balancing without requiring explicit knowledge of the SoC of each cell \cite{article26}. Further, this scheme combines simplicity with improved performance while avoiding the drawbacks of high computational demands found in SoC-based balancing schemes and the reduced accuracy of voltage dependent balancing methods \cite{article26}. 

While the Ah counting method operates on an open-loop principle, it remains vulnerable to measurement noise and diverse external factors. As a result, the SoC estimated through Ah-counting may not perfectly correspond with the SoC projected by this method. This discrepancy can lead to situations of overbalance or under-balance within the battery system.

%----------------
\subsection{Capacity-based cell balancing}
This technique uses releasable, rechargeable, and total capacity data as input variables \cite{article37} to balance the existing capacity of the battery \cite{article109}. Although, factors like current and temperature impact the existing capacity of the batteries, posing challenges in accurately determining and maintaining uniform battery capacity  \cite{article109}. In voltage and SoC dependent balancing, the cells in a battery pack may have equal SoC or voltage values, but their releasable capacities can still vary due to differences in inherent resistance and capacity \cite{article37}. These control strategies offer significant advantages \cite{article37}. Firstly, they maximize energy utilization, successively extends on the whole the lifetime of the battery, a critical factor for cost-efficient EVs. Nonetheless, real-time capacity estimation for each cell places a significant computational burden on the BMS hardware \cite{article37}. Additionally, the requirement of SoC details to estimate capacity further increases computational difficulty.

%========================================================================

\section{SoC-based cell balancing algorithms}
\label{Section: SoC Classification}
In this segment, we survey the existing SoC-based cell balancing algorithms. 

A comparative analysis of passive cell balancing algorithms is discussed in \cite{article82}. Balancing among different batteries is achieved by decreasing the differences in SoCs between the cells. A unique Proportional Integral (PI) controller and an NN based controller for passive cell balancing technology are implemented for BMS. The scaled conjugate gradient, Bayesian regularisation, and Levenberg Marquardt algorithms of NN are used individually for control operations. An improved performance is observed in terms of reduced balancing currents, decreased average power dissipation, and reduced average heat dissipation.

Balancing currents, conventionally used to rectify charge or voltage imbalances in Li-ion battery packs and mitigate cell imbalance, often overlook temperature mitigation. Addressing this gap, the study in \cite{article18} explores the trade-off's between charge and temperature imbalances. Specifically, the study utilises a model predictive control approach to optimise balancing currents and validate description across diverse scenarios. That is for a battery pack i) lacking balancing hardware, ii) using balancing currents to rectify charge imbalances, and iii) utilising these currents to expedite temperature heterogeneity removal between cells. The observations reveal that: i) employing an average current through two cells in series can decrease temperature diversity removal time by up to 35\%; ii) the effect of charge balancing on temperature heterogeneity varies with the signs and magnitudes of temperature dynamics; and iii) temperature balancing can affect charge heterogeneity positively or negatively, contingent on the signs and magnitudes of charge dynamics.

For energy storage applications, especially for e-mobility domain, a systematic review is conducted by the authors \cite{article113}. This work reviews cell balancing techniques, four types of converter topologies with optimal design on balancing time and optimised energy storage systems, and related performance evaluation. The study also recommended a prototype for a 4 series connected battery pack. This pack current and cell voltage are measured using a current sensor (TMCS1108B) and a voltage sensor (INA117P), respectively. A microcontroller TMS320F28379D, is used for monitoring the cell voltages, current and SoCs, executing cell balancing algorithms. The key contributions of this comprehensive study are (i) discussion of different cell balancing methods, cell balancing algorithms, the need for battery modelling, equivalent circuit model (ECM), and SoC estimation (ii) Different types of batteries which are effective for EV applications.
and (iii) comparative study on the balancing time with DC-DC converters such as Flyback, Buck-boost, Cuk,  and duty cycle-based converters. Lastly, the study identifies key challenges and highlights future directions.

The author in \cite{article20} proposed a novel cell balancing perspective employing a Multi-Agent System (MSA) with a control engaged in a cell-to-pack to cell-based arrangement of cell balancing \cite{article21}. The proposed system acts as nodes connected to edges called Individual Cell Equalisers (ICE) \cite{article22}. The ICEs function as energy-transferring channels between the cells. Further, the ICE can be of distinct types of circuit designs, and the authors chose the isolated modified bidirectional buck-boost converters type, which comprises two quantities of MOSFETs and resistors along with a transformer. The complete node-edge system, which is in line with the topology graph, operates in a coordinated way with a consensus constituted equalising the SoC algorithm within the cells. An extended version of ICEs, i.e, efficient cell-to-cell-based balancing arrangement \cite{article23}, is employed in this algorithm. The maximum reduction in the cell balancing time can be obtained by placing the ICEs in fixed numbers and optimal positions in between the cells. This cost inflation due to the addition of the ICEs can be also reduced by the preceding optimal positioning. The algorithm test results show that with fewer losses in energy transfer and one ICEs placed in the battery pack, the cell balancing time is lowered and average SoC-dependent balancing in the battery is increased.

The constraints applied during the manufacturing of battery packs are a key reason for inconsistency in the cells. Such hardware issues can be solved using proposed algorithms. The authors in \cite{article45} recommended a novel algorithm based on active balancing for the existing capacity of the cells. Once the balancing process begins, a Particle Filter (PF) method is utilised for SoC estimation. This method helps to remove any drift noises from the current sensor. The proposed methods are first low SoC cell charge (FLCC) and first high SoC cell discharge (FHCD) balancing. Here, a topology based on the bidirectional transformer is also introduced. The PF method demonstrates Mean Absolute Error (MAE) and Root Mean Square Error (RMSE) in SoC estimation of 2\% and 1\%, respectively. After performance testing by setting $\pm$1.5\% and $\pm$0.5\% SoC bounds, the FLCC algorithm shows higher consistency, whereas the FHCD algorithm has a shorter execution time, lesser standard deviations, and is more favourable for an Energy Storage System (ESS) application.

Designing effective BMS for EVs presents multiple challenges. The study in \cite{article69} investigates a 48 V battery pack's performance through simulation and validation, focusing on charging, discharging, and cell balancing techniques using MATLAB/Simulink. A comparative study of passive and active cell balancing techniques is presented by estimating SoC. Results demonstrate enhancements in battery life cycle, , security, drive functionality and power management. The study emphasises the importance of safe charging/discharging to improve battery capacity retention and overall performance.

With the objective to enhance the safety and service life performance of the battery pack, the authors in \cite{article74} put forth a combination of methods namely, adaptive robust UKF with SVSF-VBL. Further, for studying the characteristics and behaviour of the pack, a simulation of the mean difference model of the  Dual Polarisation (DP) model and Rint model is performed as shown in Fig.~\ref{fig:Circuit Models}.(a). The study works in two phases, the first being the application of improved forgetting factor Recursive Least Square (RLS) \cite{article75}. This helps in obtaining DP model parameters. Here, the UKF and adaptive filter provide a robust SoC estimation. The second phase is the application of Smooth Variable Structure Filter with time-Varying smoothing boundary Layer (SVSF-VBL) based on the Rint model for efficiency and accurate SoC difference estimates between each cell. The secondary performance indicator provides a complete capture of the battery imbalance. With MAE of adaptive robust UKF and SVSF-VBL to be less than or equal 3.38\% and 3.69\%, respectively, the proposed methods are shown to be robust in their SoC estimation and cell balancing.

SoC estimation using model-based control solutions can be inaccurate and complex. To address this, the authors in \cite{article80} propose a straightforward, model-free approach viz., a consensus-based cell balancing algorithm. This algorithm utilises SoC estimates from online measurements to produce sequential balancing decisions. It balances the SoC of all cells, even in the presence of noise using SoC a general non-i.i.d. (independent and identically distributed) noise model. Additionally, performance guarantees are provided for the more challenging scenario of balancing during charging or discharging. This model-independent cell balancing method is simple to implement, making it a significant advancement with respect to practical active cell balancing.

Cell balancing with model-based techniques is addressed in \cite{article81}. Moreover, a unique model-independent charging protocol is furnished to optimise the weakest cell's charge, enhancing the battery's usable capacity while maintaining safe voltage limits and accommodating deviations in cell impedance, capacity, and degradation rates. Experiments are conducted to measure cell voltage drift in various Li-ion battery types during extended cycling without regular balancing. These outcomes led to the expansion of a battery management method that removes regular balancing, using a charging algorithm to augment energy throughput while adapting variations in capacity, impedance, and ageing of individual cells. Extensive cycle testing on distinctive battery arrangements showed no evidence of balance loss or reduced lifetime functionality compared to traditional systems with routine dissipative balancing. Laboratory experiments and trials covered a range of battery chemistries, cell types, operating temperatures, charge and discharge rates, and degradation states, all without regular cell voltage balancing post-assembly.

Active balancing for Li-ion battery packs presents challenges with respect to multi-variable control problems. The study in \cite{article83} developed analytical equations to assess the stability of an architecture using low-voltage bypass DC-DC converters. The analysis employs a simple battery model with a linear SoC-based voltage source and an integrator for SoC calculation. The study examines three scenarios: (i) initial unbalance: all cells start unbalanced with the string current (istr) set to zero; (ii) robustness: evaluates the nested loops' response to disturbances and internal resistance imbalances within cells; and (iii) load and capacity variations: analyses how fluctuations in load impedance on the low voltage bus and capacity imbalances in the cell module affect control performance. Results show that under certain mild conditions, stability is inherently ensured.

Meanwhile, \cite{article84} presents a high-efficiency BMS, utilising a bidirectional flyback converter, to achieve rapid equalisation. By combining the modelling of non-ideal elements of the converter with a Genetic Algorithm (GA), optimal parameters are selected to maximise BMS's efficiency. The implemented BMS, designed for high-capacity lithium cells, demonstrates an efficiency of 91.73\% at a balancing current of 5A and a peak efficiency of 94.19\% at 2A. This BMS follows a bidirectional module-to-cell configuration suitable for integration into independent cells. The design methodology presented focuses on developing a high-efficiency, high-equalisation current BMS with simple implementation. A key contribution is the analytical modelling of the converter taking into account non-ideal elements, which in combination with GA, selects optimal converter parameters, allowing accurate estimation of the operating point and energy losses to generate a manufacturable flyback transformer design.

Accurate SoC estimation is essential for ensuring the optimal functionality and longevity of Li-ion battery packs. The method proposed in \cite{article86} is simple and is applied to SAMSUNG ICR18650-22P cells configured in seven series cell modules, with each module accommodating nine cells in parallel. The study addresses the SoC estimation for a Li-ion battery pack with cells connected in series and shunt. It considers voltage noise and current bias in the SoC estimation process. To enhance accuracy, a second-order dynamic model of the cell is used. By employing a sigma point Kalman filter (SPKF), the average SoC of the pack and the current measurement bias are estimated with the first estimator, while the second estimator determines the SoC differences among cell modules from the pack's average SoC. The SoC of the cell modules is derived from the sum of the pack's average SoC and the SoC variations. A comparative analysis between first and second-order models is reported. The SoC estimation is shown to be more accurate using the second-order model.

An active cell balancing technique is implemented in a system comprising seven batteries, utilising a bidirectional flyback circuit converter. An Extended Kalman Filter (EKF)-based SoC control algorithm is developed in \cite{article31} for equalisation. The process begins by measuring the voltage and temperature of individual battery modules utilising an Analog to Digital Converter (ADC) sensor. Then, the EKF algorithm estimates the SoC for each battery. Analysing these SoC values helps identify the battery with the highest and lowest SoC within the pack and calculates the average SoC for the entire pack. Starting with the first cell, the SoC is determined relative to the average, and this difference is utilised to determine if equalisation control is needed. If the variation exceeds a threshold, the micro-controller selects and equalises the cell. If not, it repeats the process for the next cell. Simulation and experimental validation verify the efficacy of both, the hardware arrangement and control strategy for the equalisation system, demonstrating its stable and reliable operation.

Note that manufacturing technique defects and environmental conditions affect the battery balancing performance, thus the power and energy availability in the battery pack. The authors in \cite{article89} proposed an active and novel equalisation circuitry strategy to solve this performance issue. The proposed strategy consists of a bidirectional topology of forward transformers with a switch matrix. Further, aiming to identify various target cells that must be balanced, a novel strategy of clustering analysis is introduced. GA \cite{article90} is employed  to speed up the balancing and optimise the entire classification process. The k-means algorithm \cite{article29} is used for clustering cells into three categories, namely, being charged, no equalisation, and being discharged. Furthermore, GA is used to achieve highly reliable results, setting a fitness function as optimization objective. This proposed algorithm is tested on practical grounds which provides improved results of battery capacity after balancing by 16.84\%, and also decreasing the equalisation time by 23.8\%.

The work in \cite{article93} discusses an active balancing scheme for a series configuration of: Li-ion cells. This scheme majorly works on an active balancing circuit design with a modified dual-layered inductor with a facility to connect the first and last cell group. This allows improving the balancing speed by 18\%. With an intention to upsurge the cell balancing control efficacy, the EKPF algorithm is used for SoC estimation. To estimate precisely the SoC, the authors use the following balanced control strategy:

\begin{equation}
  \begin{array}{rcl}
   \text{Turned on : } SoC_{i} - SoC_{n} \ge 0.05 (i,n = 1,2,.......,i \neq n) \\
    \text{Turned off : } SoC_{i} - SoC_{n} \le 0.01 (i,n = 1,2,.......,i \neq n)
    \end{array}
\end{equation}

The data is collected from the battery pack placed in an environmental chamber. This data is then sent to a battery testing system where the EKPF algorithm is used. Here, the mean and the covariance for each particle are calculated and used to generate a recommended Gaussian density distribution for each particle, which further helps in improving the accuracy. The accuracy estimated by the EKPF algorithm is verified by applying conditions of the urban dynamometer driving schedule (UDDS). This provides the estimation curve and the respective errors for SoC which is completely based on the PF algorithm. This equilibrium algorithm and SoC estimation topology are effectively used for the cell balancing scheme to obtain accurate equalisation. However, the effects of battery aging and temperature are not considered in this scheme. 

The range of EVs, particularly those with aged battery packs, is partially constrained by cell-to-cell imbalances, where the weakest cell commands the overall functionality of the battery pack. Active cell balancing can address these disparities, yet merely balancing voltage or SoC mismatch does not always yield optimal range performance, as it involves complex models. Rather than attempting to model the balancing hardware in detail, the authors in \cite{article99} focused on accurately representing the non-linear battery input-output behaviour to improve voltage estimation precision. Each battery is modelled using a discrete-time non-linear model. Only SoC dependency on the parameters is considered. Fully discharging the battery always results in parametric changes due to low SoC, which is crucial for accurate voltage prediction. The load is represented by a power profile derived from a speed profile employing a longitudinal vehicle model. Voltage bounds are set such that power delivery from the battery pack is halted when the terminal voltage of any cell in the pack exceeds the lower or upper limit. Furthermore, the proposed charge-based balancing approach surpasses prevailing voltage and SoC-based active balancing approaches, particularly close to the End of Life (EoL), by fully utilising the existing balancing power. A 10\% increase in the beneficial lifetime of a battery pack is reported.

Authors in \cite{article102} introduced a novel approach where a reconfigurable double-layer equalisation arrangement is developed to enhance the equalising efficiency and robustness for retired batteries. The inner layer utilises a series reconfigurable equalisation arrangement employing two MOSFETs per cell, chosen for its effectiveness in large-scale energy storage applications. To stabilise the voltage output, buck-boost converters are employed in the outermost layer, addressing the challenge of maintaining stable voltage from a single reconfigurable circuit. During equalisation, cells are deemed sufficiently balanced when their SoC differences are less than the preset threshold value. The process begins by isolating the cell with the lowest SoC, allowing the remaining cells to discharge in series until their SoC matches that of the isolated cell. This cycle repeats until all cells achieve similar SoCs, marking the end of the equalisation process for that group. Subsequently, all cells discharge in series, and multiple converters in series adjust their output voltages based on the SoC hierarchy of battery groups to optimise power distribution. The functionality of this method is validated through MATLAB/Simulink simulations of a 4-series 3-parallel equalisation circuit, demonstrating its practical application and performance under varying battery conditions.

For enhanced functionality and safe operation of a series connected Li-ion battery pack in EVs, effective cell balancing is crucial to have uniform SoC across all cells. However, there are areas for improvement. The traditional  CPC equalisation system is not ideal for large-scale battery packs caused by high voltage burden on the equalizer hardware and challenges in repair and maintenance from a lack of modular design. Additionally, existing literature does not consider cells' current constraints, leading to potential safety and health hazards when the cells' currents exceed their maximum bounds during high charging or discharging currents. The study in \cite{article126} introduces an enhanced module-based cell-to-pack-to-cell (CPC) equalisation system. In this system, CMC and module-level equalizers are used to balance the SoC of cells within each module and across the battery modules, respectively. Compared to conventional CPC balancing systems, this approach offers a simpler modular structure and easier maintainability. A two-layer Model Predictive Control (MPC) scheme is discussed: the top-layer MPC controls the ML equalizers, while the bottom-layer MPC designs the CMC equalising currents in parallel for each module. The proposed equalisation control strategy is mathematically proven to converge using the Lyapunov stability theorem, ensuring reliable performance in practical applications. Extensive results verify the superior performance of this improved module-based CPC equalisation system and the two-layer MPC control approach. The key drawback of this study is that the recommended equalisation control algorithm depends on SoC estimation for each cell, which is not practical for large packs with multiple cells due to high financial and computational costs. Future research must focus on developing a high-performance voltage dependent equalisation strategy.

Table~\ref{tab:Contributions and Applications of SoC} summarises the contributions and applications of various SoC-based cell balancing algorithms.

\begin{table*}[t!]
\caption{Contributions and Applications of various SoC-based cell balancing algorithms}
\label{tab:Contributions and Applications of SoC}
\centering
\resizebox{\textwidth}{!}{
    \begin{tabular}{|m{0.06\textwidth}|m{0.2\textwidth}|m{0.6\textwidth}|m{0.03\textwidth}|m{0.06\textwidth}|} \hline

    \multirow{2}{*}{\textbf{Reference}} & \multirow{2}{*}{\textbf{Control Algorithm}} & \multirow{2}{*}{\textbf{Contribution}} & \multicolumn{2}{l|}{\textbf{Applications}} \\ \cline{4-5} 
    &  &  &EV&ESS/Grid \\ \hline

\cite{article82} & PI controller and NN (scaled conjugate gradient, Bayesian regularisation, and Levenberg Marquardt) & Improved performance in passive cell balancing in terms of reduced balancing currents, decreased average power dissipation, and reduced average heat dissipation.& \checkmark & \\ \hline

\cite{article93} &Modified dual layered inductor + EKPF Algorithm & This scheme improved the balancing speed by 18\%. For accurate SoC estimation, the EKPF algorithm is used. The battery aging and temperature effects are not considered under this scheme.&\checkmark &  \\ \hline

\cite{article81} & Non-routine balancing charging method & Development of a BMS that eliminates routine balancing. This method employs a charging algorithm designed to enhance energy throughput while considering variations in cell capacity, impedance, and aging.& \checkmark & \\ \hline

\cite{article31} & EKF-based SoC control & Development of a battery balancing system that utilizes a bidirectional DC-DC flyback converter combined with an EKF prediction SoC equalization algorithm to obtain rapid cell equalization. & \checkmark & \checkmark\\ \hline

 \cite{article18} & MPC &  A study of charge and temperature balancing among Li-ion in a battery pack, using only current as the input.& \checkmark & \checkmark \\ \hline

 \cite{article20}& MSA + ICEs + Consensus-based equalising algorithm & The proposed system test results show a time reduction in cell balancing, reduced losses in energy transfer, and an increase in average SoC balancing in the battery.&\checkmark &\checkmark \\ \hline
  
\cite{article45} & FLCC / FHCD + SoC PF balancing algorithm& Reduction in drift noises from the current sensor, and performance testing within set SoC bounds using FHCD algorithm demonstrated shorter execution times and lower standard deviations.& \checkmark &\checkmark \\ \hline

\cite{article69} & k-means algorithm, GA & A novel strategy is introduced to accelerate the entire balancing process and to optimise the entire classification process.&\checkmark & \\ \hline

\cite{article74} & Adaptive robust UKF + SVSF-VBL& The combined method produced MAE $\le$ 3\% and revealed to be robust in their SoC estimation and cell balancing.& \checkmark & \\ \hline

\cite{article80} & Consensus-based stochastic control & A model independent and practically implementable cell balancing method is discussed to estimate SoC even in the presence of noise. & \checkmark & \\ \hline

\cite{article83} & Simplified SoC-dependent control algorithm & SOC balancing method for an active BMS using parallel and series output-connected BPMs, enhancing modularity and effectiveness by regulating input currents and utilising a central and module controller. & & \checkmark \\ \hline

\cite{article84} &GA & Development of a high-efficiency BMS that utilizes non-ideal elements for a bidirectional flyback converter to achieve rapid equalization..  & \checkmark & \\ \hline

\cite{article86} & Second-order dynamic model + sigma point Kalman filter & SoC estimation for a Li-ion battery pack with series and parallel cells, accounting for voltage noise and current bias.& \checkmark & \\ \hline

\cite{article89} & Bidirectional topology (forward transformers + switch matrix) + K-means + GA Algorithm & The proposed algorithm was tested on practical grounds, improved results of battery capacity balancing by 16.84\% and decreased the equalisation time by 23.8\%.&\checkmark &\checkmark \\ \hline

\cite{article99} & Discrete-time non-linear model of battery with SoC dependent parameters & Development of charge-based balancing approach, which achieves a 10\% increase in the useful lifetime of a battery pack. & \checkmark & \\ \hline

\cite{article102} & Reconfigurable double-layer equalisation & Balancing of retired batteries based on SoC thresholds, validated through MATLAB/Simulink simulations for practical implementation. & &\checkmark  \\ \hline

\cite{article126} & A two-layer model predictive control strategy  &   The advanced module-based CPC equalisation approach and the two-layer MPC control scheme recommended demonstrate superior performance. It significantly reduces computational complexity compared to centralised MPC, making it more suitable for real-time cell balancing implementation. & \checkmark &  \\ \hline

  % & & \\ \hline
   % & & \\ \hline
% & \ding[0.5]{108}\quad   & \ding[0.5]{108}\quad 
\end{tabular}}
\end{table*}

%========================================================
\section{Voltage dependent cell balancing algorithms}
\label{Section: Volatge Classification}
In this section, we survey voltage and voltage-SoC dependent cell balancing algorithms in detail. 

The article \cite{article105} addresses the challenge of charge equalisation in batteries with voltages up to 1000 V, operating in dynamic modes with varying charge and discharge rates. It recommend a voltage balancing method for battery elements, establishing a reference signal equivalent to the element's reference voltage for its current SoC. This method is applicable to both, passive balancing approach using ballast resistors and active circuits using electromagnetic energy redistribution. The discussed solution calculates the average battery voltage and initiates balancing when any cell's voltage surpasses the permissible variation, typically within several tens of mVs. This approach allows for a straightforward balancing method that not only prevents cell overcharging and equalises voltages in the battery's full charging zone, it also provides balancing throughout the complete charging operation and for cells disconnected from the power supply. Cells charge normally if their SoC is below the average voltage. However, if a cell exceeds the dead zone, it activates balancing, while other cells exceeding only the average voltage continue normal charging. Simulink-based research results demonstrate the proposed method's advantages in speeding up balancing time, enhancing accuracy, and preventing overcharging. The results also demonstrate the advantages of the recommended method such as speeding up balancing time, enhancing accuracy, and preventing overcharging. It is shown that the proposed work is particularly effective for high-power, batteries in dynamic operating modes.

BMS is generally composed of power electronics, electronic circuits, integrated chips, and additional hardware requirements. However, BMSs developed with micro-controllers often have limitations with reference to hardware components and the current levels they can handle. Moreover, using a independent controller for each battery pack can complicate these problems. The study in \cite{article27} presents a Programmable Logic Controller (PLC) based Switched Resistor Passive Balancing (SRPB) technique for Li-ion batteries to address the limitations faced by traditional micro-controller-based BMSs. This study develops algorithms capable of simultaneous operation using two distinct balancing techniques, ensuring both control and protection during cell balancing. The first method entails measuring all cell voltages with a PLC, identifying the cell with the lowest voltage, and calculating voltage differences. Cells with voltages exceeding a reference level undergo balancing, dissipating surplus energy as heat through resistors. This process maintains uniform voltage levels, preventing cell overload during charging and safeguarding the battery pack. This study employs the second method to establish a reference maximum charging voltage for cell balancing. When the BMS detects the usual maximum charging voltage, it switches to protection mode. Meanwhile, the specified protection algorithm operates continuously, and the balancing process from the first method continues simultaneously. Setting a particular reference maximum charging voltage ensures system protection during testing, with the value being lower than the overall normal maximum charging voltage. With the PLC-based design and the newly developed algorithm, a controller can govern many battery packs and operate at elevated current levels.

The current in the fixed duty cycle approach is variable, preventing the operation of the converter at the maximum efficiency point and leading to increased power losses. This limitation is addressed in \cite{article85} which introduces a current-mode controller for switched-inductor arrangement designed to obtain voltage balancing between cells or modules using active cell balancing topology. The proposed work offers several advantages, such as a well-controlled balancing current independent of cell voltages and system parasitic elements. The current mode controller for the switched inductor topology effectively manages and maintains the inductor current within component rating limits without the need for additional over-current shutdown protection. The control algorithm is designed for a battery pack with n cells, incorporating n-1 current control loops. The proposed balancing controller attains open circuit voltage balancing, a more reliable indicator of SoC than terminal voltage. The cell balancing algorithm is summarised below: 
\begin{itemize}
    
\item Charge Transfer: The controller sets initial inductor reference values for each battery. The algorithm monitors the voltage deviations among adjacent cells to determine the direction of charge transfer. If the voltage variation surpasses a threshold value, the controller commands the inductor current to transmit charge from the higher voltage cell to the lower voltage cell. The algorithm dynamically adjusts the direction of the inductor current based on the voltage differences to achieve voltage balancing among the cells. The balancing current reference is chosen to ensure that the balancing circuit works at  efficiency, resulting in lower power losses. A thorough DC analysis of the balancing arrangement is provided. 
\item Series Resistance Estimation: Battery series resistance (Rs) can vary with factors such as temperature, aging, and operating conditions. Real-time estimation enables the system to adapt to these changes dynamically. The estimation process measures the changes in battery voltage and current due to a small inductor current change. Using Ohm's law, the  Rs of each battery is calculated. This process may involve iterative steps to ensure accurate real-time determination of Rs for each cell. 

\item Real-Time Estimation of OCV: The OCV is considered a more precise indicator of SoC compared to terminal voltage, as it considers the voltage drop across the series resistance. The real-time estimation of OCV in the context of battery balancing involves compensating for the voltage drop across the Rs of each battery to accurately determine the open circuit voltage. The estimation process involves adjusting the measured terminal voltage of each battery based on the voltage drop caused by the series resistance. During discharging, the voltage drop (IRs) is added to the terminal voltage, while during charging, it is subtracted. The calculated OCV values are then used to compute the inductor current direction and to conclude on enabling/disabling the converters.

\end{itemize}

Addressing the various problems concerning the cell passive equalisation such as slower speed and inefficient energy is critical. The authors in \cite{article88} proposed an algorithm that configures faster balancing, high energy, and multiple charge transfers for large-scale battery modules in the case of commercial vehicles. The control algorithm establishes and balances a total of 176 cells and their respective sub-modules independently. Each module consists of a switch matrix with DC-DC converter and then finally connected to a bus which consists of other sub-modules. The proposed cell balancing algorithm for a BMS is outlined. The process begins with monitoring the voltage of individual cell. It then determines the minimum, maximum, and average voltages for each sub-module and identifies the cell with the highest and lowest voltage in each module. Next, cells with minimum voltages are sorted in increasing order, and those with maximum voltages are sorted in decreasing order. These sorted cells are then paired. The algorithm checks the voltage variation between the paired cells. If the deviation is less than or equal to 1 mV, the algorithm stops. However, if the difference is greater than 1 mV, the average voltage is used to determine the next step. The cell with the maximum, voltage in module i is discharged if its voltage exceeds the average, and the cell with the lowest voltage in module is charged if its voltage is below the average. This process loops back to sorting the cells and continues until the desired voltage difference condition is met, ensuring effective cell balancing by continuously adjusting the voltages of the cells to maintain them in a specific range of each other. The proposed algorithm reduces the final voltage variation among the cells and also drastically decreases the balancing time for the same when compared with existing algorithms.

The proposed self-re-configurable battery topology introduces a paradigm shift by achieving stable voltage throughout the battery pack cycle without relying on a DC-DC converter, distinguishing it from traditional methods \cite{article103}. This proposal addresses complexities and costs associated with traditional methods while reducing energy losses. Key advantages of this system include the elimination of the DC-DC converter, ensuring voltage stability, improving efficiency, enhancing capacity utilisation, and simplifying control strategies. In this system, the battery pack maintains voltage within a predefined range even as SoC varies from 0 to 100\%. This is achieved by prioritising the discharge of cells with high SoC and voltage that meet operational requirements, while continuously substituting cells with lower SoC to preserve pack uniformity and stabilise voltage. The control scheme focuses on maintaining consistent SoC across cells within the pack while stabilising overall voltage. During discharge, cells with maximum SoC are discharged first, ensuring that those meeting voltage criteria are integrated into the series-connected battery pack, while cells with lower SoC are bypassed. The topology involves strategically configuring parallel cells in series to achieve desired voltage outputs. This arrangement enables voltage stabilisation and SoC consistency without relying on a DC-DC converter, except when essential to match high-load voltage requirements. Overall, the proposed self-re-configurable battery topology presents a significant advancement in battery management technology, offering enhanced operational efficiency, simplified control, and improved voltage stability throughout the battery pack's life-cycle.

The authors in \cite{article106} examined a Li-ion battery equalizer  pertaining to the Dual Active Half-bridge (DAH) architecture. Also, analysis of key parameters related to energy flow among cells within the identical group and cells across dissimilar groups is discussed. Despite the DAH-based equalizer having few switching devices and simple control, the open-loop DAH equalizer exhibits two main drawbacks: 

i) When the voltage deviation between cell groups is negligible, the equalisation speed is slow. In fully charged or discharged states, this slow speed can cause overcharging or over-discharging of some cells, potentially damaging the battery.

ii) At low switching frequencies, achieving a fast balance within the half-bridge group requires large and heavy transformers, contrary to the trend towards miniaturisation and lightweight designs. 

Based on these analyses, a phase shift control scheme is recommended, which solely enables energy flow among cells in the similar and dissimilar groups but also allows the equalizer to operate at high frequencies. While the DAH equalizer draws on the phase-shifting control strategy of the Dual Active Bridge (DAB) used in high-power applications, it has distinct characteristics: 

i) The DAH equalizer consists of battery pack, requiring bi-directional energy flow between cells of the batteries.

ii) Unlike DABs used for high voltage and power, the DAH equalizer manages energy transmit at lower voltages (approximately 2.5–3.6 V) and currents (1–1.5 A), where line impedance cannot be ignored. 

One goal of realising the phase shift control in the DAH equalizer is to improve its operating frequency. A prototype of the DAH equalizer, managing four lithium batteries is developed and tested. Experimental results demonstrate that the proposed phase shift control-based equalizer efficiently balance the batteries. The key findings include: i) the voltage deviation among cells in the group is minimised, and the equalizer operates at a higher switching frequency; ii) the phase shift control strategy enables significant equalising currents between different battery groups, allowing rapid energy equilibrium; iii) the proposed approach decreases the volume and cost of the equalizer, enhancing its competitiveness in the market; and iv) an experimental prototype verified the impact of the phase shift control strategy, with results confirming its efficacy.

A fast equalisation approach based on inductor balancing is discussed in \cite{article8}. The inductor based approach is known to be the fastest when compared with other cell balancing methods. However, there is a need to address the issue of extended energy transfer between the two farthest cells in the process. Here, eight cells are connected in series with set voltage in the range between 3–3.6 V, and the energy transfer for balancing occurs cell by cell. To address this drawback, an alternate topology is presented which mainly consists of (n-1) inductors for every n cells. The MOSFETs function as switches with body diodes with switching frequency, capacitors, and inductors with values set as 100 kHz, 50 mF, and 10 \textmu H respectively. These are controlled via a 50\% duty-cycle of a pair of complementary synchronous patterned triggers. The simulation results demonstrated improved balancing of all eight cells when compared to other methods.

Unlike the aforementioned study where the energy transfer occurs cell by cell, the study proposed in \cite{article115} features a way for direct cell-to-cell energy transfers for faster balancing with the help of high-power density-based planar coupled inductors. The Buck-Boost or fly-back operates as a bridge for the shortest path for cell balancing, resulting in much lesser balancing time. The circuit for the same can be built as: (i) for 2n batteries, coupled inductors, switches and a group of batteries in series are connected as 2n, 6n, and so on, respectively; and (ii) for 2n+1 batteries, coupled inductors, switches and groups of batteries in series are connected as 2n+2, 6n+4 and so on, respectively. During operation, the source battery with the help of Buck-Boost charges the respective target cells at times when a couple of batteries are imbalanced in the inner groups. During inter-group imbalance situations, the source cells with the help of flyback charge the target cell. Here, Continuous Current Mode (CCM) is implemented to reach higher levels of energy transmission during a single switching period with a frequency of 20kHz. The hysteresis logic-based control strategy is used to judge if the obtained voltage differences between the cells are within the limits of the hysteresis. If not, then corresponding action modes are executed in the strategy. 

Authors in \cite{article124} presented the study and design method of a voltage equalizer, which integrates a boost full-bridge input cell and a Symmetrical Voltage Multiplier (Sy-VM).  The proposed voltage equalizer successfully charges batteries, even under varying initial voltage conditions, to a required identical voltage level. The article also discusses an equalizer based on a voltage multiplier and outlines its analysis and design procedure. The described equalizer combines a boost full-bridge inverter with an SVM. The equalizer features an input cell and a voltage multiplier, where the inverter generates a bipolar square wave with variable voltage amplitude, which is then fed into the voltage multiplier to produce multiple same output voltages. The design procedure, guided by analytical models, identifies the optimal parameter configuration for the equalizer. The optimally designed equalizer automatically ensures voltage balance charging with minimised conduction loss. Compared to traditional equalizers that require multiple switches and/or inductors, the circuit complexity and size of the voltage equalizer based on the voltage multiplier are remarkably reduced. Experimental results validate the design technique and demonstrate that the described voltage equalizer achieves excellent voltage equalisation, reducing the standard deviation to merely a few mVs.

When using battery packs, inconsistencies between individual cells (series connected) arise due to factors like internal resistance and self-discharge rates. These inconsistencies decrease the energy utilisation rate and service life of the battery pack and can also compromise the safety of the battery system. To address these issues, \cite{article125} proposed an equalisation approach based on a flyback converter. The method uses the existing energy (power) of a single cell as an index of disparity. The new equalisation approach employs a bidirectional flyback transformer for output isolation, with staggered control switches on both sides to attain bidirectional equalisation during energy transmission. Compared to traditional equalisation topologies, the recommended design decreases the components count and overall size of the equalisation arrangement. Additionally, the primary side of energy transfer requires a single set of control signals, simplifying the control process. A series of experiments for equalisation are conducted using 12 series connected cells  demonstrated the impact of this new method. The equalisation control scheme utilises the voltage of the series-connected cells as a reference and carry out equalisation with regard to the degree of voltage dispersion. The variation between the highest and lowest cell voltages determines when to start and stop equalisation. Equalisation begins when the voltage difference exceeds a set threshold and stops once it falls below this threshold. To address the issue of voltage spikes, the stopping condition requires calculating the terminal voltage of the series connected cells at rest. The design and testing of a 12-cell battery prototype in charge and discharge equalisation experiments further validate the efficiency of the recommended work. The experimental outcomes illustrate that the new voltage equalisation circuit effectively achieves dynamic voltage equalisation and demonstrates excellent functionality, underscoring the rationality of the proposed topology.

A unique cell voltage equalizer utilising a series LC resonant converter is discussed in \cite{article127} for series-connected energy storage devices such as batteries or supercapacitor cells. The series LC resonant converter operates near the LC resonant frequency, where circuit impedance is minimised, allowing efficient and rapid energy transfer. Similar to active multi-winding transformer circuits, the LC resonant tank provides numerous energy transmission paths between the cells. Nevertheless, this design reduces costs and can be manufactured in a miniature package compared to transformer-based schemes. Bidirectional switches for individual cells can be realised with single integrated switch packages, and all switches for each cell are controlled by the same signal, simplifying the control arrangement. For large series-connected systems with tens or hundreds of cells, the circuit can be easily expanded by stacking additional switch sets in parallel rather than adding extra transformer windings. To reduce voltage equalisation time, the switching frequency is selected close to the resonant frequency to allow maximum current flow and energy transfer in each switching cycle. During operation, the voltage detection circuit selects the cells with the maximum and minimum voltages. The switches in the highest voltage cell are turned on to transmit energy to the series LC resonant tank, which temporarily stores the energy before releasing it to the minimal voltage cell, thus enhancing its voltage. This process repeats each cycle until the required voltage balance is obtained. Simulation and experimental results show that the voltage equalisation approach, reducing an initial voltage variation of 527 mV to a final deviation of 10 mV in 900 seconds for three series-connected supercapacitor cells.

The EV batteries face challenges in achieving rapid balancing and maximum efficiency with low circuit and control complexity. The study in \cite{article128} discusses a single LC tank-based cell-to-cell active voltage balancing algorithm for Li-ion batteries in EV. The LC resonant tank transfers energy from the cell with excessive voltage to the cell with the minimal voltage. The proposed method utilises a resonant converter as the first balancing circuit between two cells for various classes of electrochemical batteries and balancing algorithms. The circuit design reduces the usage of bidirectional MOSFET switches and related components. The balancing speed is enhanced by enabling a small path for voltage transfer, ensuring rapid balancing among any two cells in the battery string, and reducing power consumption by operating all switches in zero-current switching conditions. The implemented algorithm operates as follows: Voltage sensors serve as monitoring integrated circuits (ICs) to assess the status of battery cells. These sensor readings are collected by the monitoring IC and then communicated to the microcontroller. The microcontroller collects the IC data, estimates the voltage of each cell, compares the values, identifies the maximum and minimum voltage cells, and activates the associated switch to connect with the LC resonant converter. The DC-DC converter performs voltage balancing according to the recommended algorithm. Based on the voltage readings from the battery IC, the algorithm decides when to operate the balancing converter. The algorithm addresses unbalanced cells, overcharge, and undercharge conditions in . The circuit is tested with 4400 mAh Li-ion  under static, cyclic, and dynamic charging/discharging conditions. Two cells at voltages of 3.93 V and 3.65 V are balanced in 76 minutes, achieving a balancing efficiency of 94.8\%. Results, under dynamic and cyclic conditions, indicate that the balancing circuit is suitable for energy storage and EV applications. However, temperature variation is not accounted in this study.

%=====================================================================================

\section{Voltage-SoC dependent balancing algorithms}
\label{Section: Volatge-SoC Classification}
Recently, many studies have focused on Voltage-SoC dependent balancing algorithms. In what follows, we discuss details of latest studies which have addressed such algorithms.

Equalisation topologies such as single-cell-to-single-cell balancing, battery-pack-to-single-cell balancing, and equalisation among adjacent cell groups suffer from limitations like extended equalisation times and reduced flexibility in selecting equalisation pathways. A reconfigurable circuit with Buck-Boost circuits is proposed in \cite{article28} to enable dynamic cell grouping and reduce equalisation time, offering modes like multi-cell-to-any-cell, any-cell-to-any-cell, and multi-cell-to-multi-cell balancing. The scheme employs an adaptive duty cycle-based control strategy that adjusts based on the relation between duty cycle and voltage deviation. Due to manufacturing variations, batteries from the same batch may differ, with self-discharge rates and SoC values following a normal distribution over a time period. This property aids in identifying and grouping batteries for efficient balancing and maintenance. The k-means algorithm\cite{article29} is proposed to classify batteries into groups based on their capacities. Since the data follows a normal distribution, the algorithm can effectively differentiate between high-energy battery packs and low-energy battery packs. To enhance the initial clustering centre selection, the k-means++ algorithm is used. This step ensures that the random selection of initial cluster centres does not adversely affect the final grouping, providing a reliable and consistent classification results. The silhouette coefficient is used to evaluate the quality of the clustering. This metric helps in finding out the  most favourable number of clusters (k), ensuring that the classification accurately reflects the natural grouping of batteries based on their capacities. This is followed by the SoC estimation. The process begins by entering the SoC values of the two extreme batteries in the cluster, i.e., the higher and the lower SoC in the battery pack. The initial weights, along with the bias, are determined. The framework ensures accurate SoC estimation for battery clusters by dynamically adjusting weights and bias dependent on the SoC values of the extreme batteries and the overall battery pack. This dynamic adjustment helps maintain balanced charging and discharging states, which is critical for efficient battery management and longevity. The algorithm leverages initial SoC values and iterative decision logic to continuously refine and provide a reliable estimate of the battery pack's overall SoC. The article provides a comparative analysis of various equalisation schemes, including AC2AC, MC2MC, and conventional methods, examining key factors such as equalisation time, computational workload, variance, and energy-efficiency.

%------------ repeatation- But it can be in both  algorithm and optimisation section ----------%
One of the major effects on the performance parameters and the safety of Li-ion batteries is due to an extended range of inconsistencies. To solve this issue, the authors in \cite{article33} proposed a fast and flexible equalisation topology called double layer single structured equalisation. An improved and optimised proposed topology based on single and double-layered buck boost equalisation topologies is discussed. MPC \cite{article34} equalisation strategy is also introduced within the topology to further enhance the speed of equalisation and reduce the respective losses. The proposed system majorly works in three phases. The first phase obtains the future prediction of the sequences of the system. The prediction model consists of double-layer single-ring structured equalisation. The second step is to define various system constraints along with the establishment of an optimisation objective function in line with the obtained prediction sequence. The last step is to obtain a control sequence after solving the objective functions. Hence, the first value in the control sequence is applied to the system. The recommended topology proves to reduce the time of equalisation and increase in energy transfer efficiency by 38.15\% and 27.25\%, respectively when compared with and single layered buck-boost converter based equaliser (BBE) coupled with a mean difference algorithm. Other tests are also considered by the authors under different operating conditions such as constant current charging and New York Bus cycle discharging \cite{article35}.

A redundant battery can also be used for active battery balancing. This aspect is experimented by authors in \cite{article78}. Here, the equivalent model for equalisation is majorly based on multiple numbers of cells to be series connected with a redundant battery. At the time of discharge, the switch controller opts for any one cell from the series to be connected in parallel with the redundant cell. The battery parameters such as SoH, current and SoC of the paralleled connection will be taken into consideration by the switching control strategy. This strategy is framed on the deep reinforcement learning (DRL) structure. This structure has majorly relied on greedy and rule algorithms \cite{article79} and assists in attending early equalisation with minimum counts of switching. Additionally, the study also reveals that DQN algorithm, a DRL form, provides 7 times lesser control switching compared to greedy algorithm and 60s faster equalisation than the rule algorithm.

To overcome the long-time consuming cell balancing methods in the conventional approach, the authors in \cite{article91} proposed a prototype with a balancing algorithm that focus on the combined approach of current capacity and voltage difference balance for \ce{LiFePO4} battery pack. The experimental setup is a series combination of 12 Li-ion cells with a voltage of 36V. The battery voltage and SoC are determined at the same time and these characteristics are stored in a well-developed lookup table. Here, the relation between current capacity and the voltage difference can be derived by using linear interpolation of EMF method \cite{article92}. The methodology used for SoC estimation is coulomb counting. During this process if the voltage increases above the set threshold, the cell is drained instantly. Whenever the voltage of the cell reaches the over-voltage threshold set, the battery pack charging stops. This approach reduces the balancing and charging time hence working even when the voltage difference becomes negligibly detectable. The disadvantage is that this approach can only be employed at no load state of the battery pack.

Effective battery pack management hinges on balanced techniques within the BMS. Poor balance can perpetually limit pack performance due to the weakest cell. Although battery SoC serves as a useful balancing indicator, its accuracy can vary, introducing uncertainties in balancing performance. The study in \cite{article100} introduces a unique switchable indicator to address the limitations of using only SoC and voltage. The necessity of SoC and voltage indicators is supported by an analysis of their impacts, comprising of the initial SoC, current noise, battery aging, and Coulomb efficiency. A specially designed state-switching logic table and balancing algorithm are developed to use this switchable indicator. For implementing the two different indicators, cells with the $m^{th}$-lowest SoC/voltage are bypassed during discharging, and those with the $m^{th}$-highest SoC/voltage are bypassed during charging. To finalise the balancing logic, specific design actions are added for each state. This innovative bypass equaliser features a compact topology, high efficiency, and fault tolerance. It automatically toggles between voltage and SoC indicators, facilitating a new balancing scheme tailored to optimise performance. The discussed work is validated through both, simulation and experimental testing on a Li-ion battery pack.

The series multi-cell connected string of batteries is comparatively less efficient and slow in terms of cell balancing. Hence, to overcome this issue, the authors in \cite{article116} designed an automatic switched capacitor-based cell balancing for approximately 16 cells in the series string. As the speed of balancing in this approach does not rely majorly on initial cell voltages or their differences and the number of cells in the series string; the authors propose a modularised balancing architecture using MATLAB/Simulink for 4, 8, and 16 cells. The approach design configuration is as follows: for each cell, two switches of MOSFETs, and for every adjacent cell, one capacitor is connected. Along with this configuration, additional two capacitors are connected to provide a direct pathway for the cell-to-cell-based transfers. Other multiple capacitors are connected for operation during the switching period of the respective operating cycle of charge and discharge. The modularised method consists of 4 cells per module with multiple levels of switching capacitors for voltage balancing within the cells and modules. This approach, when experimented on a Li-ion battery type with a nominal voltage of 3.7V and rated capacity of 1200mAh, demonstrated that the cell balancing incurs less than an hour to complete, irrespective of the initial voltage, SoC, and cell number differences.

Table~\ref{tab:Contributions and Applications of Voltage} summarises the contributions and applications of various Voltage and Voltage-SoC dependent cell balancing algorithms.

\begin{table*}[t!]
\caption{Contributions and Applications of various Voltage and Voltage-SoC dependent cell balancing algorithms}
\label{tab:Contributions and Applications of Voltage}
\centering
\resizebox{\textwidth}{!}{
    \begin{tabular}{|m{0.06\textwidth}|m{0.2\textwidth}|m{0.6\textwidth}|m{0.03\textwidth}|m{0.06\textwidth}|} \hline
\multicolumn{5}{|c|}{\textbf{Voltage based}}     \\ \hline
    \multirow{2}{*}{\textbf{Reference}} & \multirow{2}{*}{\textbf{Control Algorithm}} & \multirow{2}{*}{\textbf{Contribution}} & \multicolumn{2}{l|}{\textbf{Applications}} \\ \cline{4-5} 
    &  &  &EV&ESS/Grid\\ \hline
    
\cite{article105}  & Reference voltage based algorithm.& Speeding up balancing time, enhancing accuracy, and preventing overcharging, particularly effective for high-power, multi-element batteries in dynamic operating modes. & &  \checkmark \\ \hline

\cite{article27}&Algorithm based on the maximum reference charge voltage level using PLC & A  BMS comprising of PLC is discussed for Li-ion batteries to tackle the challenges of microcontroller based BMS.& \checkmark & \\ \hline

\cite{article85} & Switched inductor-based balancing technique featuring phases such as charge transfer control, series resistance estimation, and real-time open circuit voltage estimation. & Open circuit voltage balancing, a  precise indicator of SoC than terminal voltage.& \checkmark &  \\ \hline

\cite{article88}&Fast balancing - simultaneous charge transfer algorithm &The proposed algorithm can reduce the final voltage difference and also drastically decrease the balancing time. & \checkmark & \\ \hline

\cite{article103} & Voltage stabilizing principle + SoC dependent cell equalization & A self-reconfigurable batteries topology is discussed to achieve good equalization with low voltage fluctuations without DC-DC converter.& \checkmark &\checkmark \\\hline

\cite{article106} & DAH Phase Shift Control & The proposed DAHB-PSC algorithm improves Li-ion battery equalisation by enabling efficient energy flow between cells across different groups with high-frequency operation, reducing volume, cost, and enhancing overall system performance. &\checkmark &  \\ \hline

\cite{article8}  & Inductor-based balancing method with MOSFETs& The cells connected in series are balanced faster when compared to other methods explored within the scope of work& \checkmark&  \\ \hline

\cite{article115} & Planar Coupled Inductor + hysteresis logic-based control strategy &This work scheme proved to shorten the path for equalisation and fastens the speed for balancing between the cells & \checkmark& \checkmark  \\ \hline

\cite{article124} &  Voltage equalizer, which consists of a boost full-bridge input
cell and an SVM & Presents the study and design procedure of a voltage equalizer, which integrates a boost full-bridge input cell and an SVM. & \checkmark &\checkmark  \\ \hline

\cite{article125} & Flyback converter + voltage dispersion equalisation algorithm & 
An innovative equalisation method consisting of flyback converter is proposed with reduced component count to overcome inconsistencies caused by factors such as internal resistance and self-discharge rates.& \checkmark & \checkmark  \\ \hline

\cite{article127} & Voltage difference minimisation algorithm + circuit operating at resonant frequency. & A simple, cost-effective series resonant-based cell equalizer uses voltage as a parameter and can be easily modified for higher capacity energy storage devices.& & \checkmark  \\ \hline

\cite{article128} & LC tank voltage equalisation algorithm & A novel single resonant converter designed for balancing circuits between two cells in various types of electrochemical batteries, employing streamlined bi-directional MOSFET switches and associated components to enhance efficiency. & \checkmark & \\ \hline

 %-==============================================================================
  
\multicolumn{5}{|c|}{\textbf{Voltage-SoC dependent}}     \\ \hline
    \multirow{2}{*}{\textbf{Reference}} & \multirow{2}{*}{\textbf{Control Algorithm}} & \multirow{2}{*}{\textbf{Contribution}} & \multicolumn{2}{l|}{\textbf{Applications}} \\ \cline{4-5} 
    &  &  &EV& ESS/Grid \\ \hline

\cite{article28}& k-means clustering + adaptive duty cycle & Accurate SoC estimation for battery clusters and comparative analysis of various equalisation schemes focusing on equalisation time, computational workload, variance, and energy efficiency. Balancing topology optimises series-connected battery grouping to expedite equalisation, promptly isolating failed batteries for safety improvements, while also minimising inter-battery energy transfers to extend battery life.
&\checkmark &  \\ \hline
    
\cite{article33}& Double layer single structured equalisation + MPC& The proposed topology proved to reduce the equalisation time and increase energy transfer efficiency by 38.15\% and 27.25\%, respectively.&\checkmark &\checkmark \\ \hline

\cite{article78}& DQN + Greedy and Rule Algorithm + Redundant cell & The paralleled redundant cell and the DQN-based control switch provides 7 times lesser control switching than that of a greedy algorithm and 60 s faster equalisation than the rule algorithm.&\checkmark &\checkmark  \\ \hline
    
\cite{article91}& A combined approach of current capacity and voltage difference balance&This approach Shortened the balancing and the charging time. This approach even worked during the voltage difference was not detectable. No load states type of battery packs can only be used for this approach. &\checkmark & \\ \hline

\cite{article100}& Switchable Indicator - voltage and SoC & Addresses the limitations of individual indicators and enhances balancing performance under varying conditions such as initial SoC, current noise, battery aging, and Coulomb efficiency.& \checkmark &  \\ \hline

\cite{article116} & Modularised Automatic switched capacitors architecture& With a simple, modularised, and flexible design this approach increased the balancing speed and efficiency &\checkmark &  \\ \hline

 % & & & & &\\ \hline
% & \ding[0.5]{108}\quad   & \ding[0.5]{108}\quad 
\end{tabular}}
\end{table*}

%---------------------------------------------------------------------------

\section{Optimisation techniques and Performance indicators}
\label{Section: Optimisation Classification}
Optimisation is the process of identifying the best solution to a problem from a set of feasible solutions, considering specific constraints and objectives \cite{article56,article57}. It involves maximising or minimising a function, often called an objective function, subject to constraints \cite{article58}. 

Optimisation algorithms typically work iteratively, calculating the objective functions and constraints at each point within the design space. The algorithm then uses these objective values, in addition to related information such as past results, gradients, or constraint values, to determine the manner in which design variables should be adjusted in terms of direction and magnitude \cite{article59}. After each iteration, new objective and constraint values are recalculated, guiding the process towards attaining the optimal solution \cite{article60}. This iterative approach continues until a acceptable design is reached or specific stopping criteria are met \cite{article56}.
The essential steps involved in optimising a given complex problem include \cite{article58,article61, article62}:
\begin{itemize}
\item Identification of the system whose performance needs to be optimised.
\item Identification of the optimisation variables that influence the objective function.
\item Clearly defining the objective function and constraints for a well-posed and solvable problem.
\item Choosing an appropriate optimisation technique based on the problem characteristics.
\item Validating the optimisation results and considering the practical feasibility.

\end{itemize}

Further, the benefits of including optimization in cell balancing are as follows:
\begin{itemize}
\item It reduces energy losses throughout the balancing process, ensuring that more energy is available for the intended application \cite{article19, article64}.

\item It ensures that energy is transferred between cells in an efficient manner thereby, reducing wasted energy \cite{article33}.

\item It ensures that all cells maintain a uniform SoC thereby, preventing overcharging and deep discharging of individual cells \cite{article66, article73}. Further, it manages the heat generated during charging and discharging which otherwise can degrade battery life \cite{article19, article67}. As a consequence, this leads to more reliable and consistent performance.

\item It leads to simpler and cost-effective hardware designs, decreasing the overall cost of the BMS.

\end{itemize}

\subsection{Literature review of optimisation techniques}
In general, traditional charging and discharging controller designs like constant-current and constant-voltage, pulse charging, reflex charging, trickle charging and float charging face challenges such as complexity, prolonged charging times, reduced efficiency, elevated temperatures, and over-charging or self-discharging. The study in \cite{article19} introduces a novel approach using a backtracking search algorithm-based fuzzy optimisation to assess the battery's SoC. This method aims to improve control efficiency by optimising input and output fuzzy membership functions linked to the rate of SoC change and power balance. The controller's performance is evaluated by comparing it with particle swarm optimisation-based (PSO) fuzzy and fuzzy-only controllers. The findings suggest that the optimised fuzzy approach offers effective control over battery charging and discharging processes.

Inconsistent Li-ion batteries can harm series battery packs, reducing capacity and lifespan. To address this issue, a dual-layer ring-structured equalisation method is recommended which provides flexible paths and swift balancing \cite{article33}. A double-layer equalisation topology is proposed for a battery pack with n cells in series, divided into k modules, featuring buck-boost converter based equalisers and bidirectional flyback transformer equalisers (BFTE) in both, inner and outer layers. The BBEs connect adjacent cells and modules, while the BFTEs connect the first and last cells or modules, facilitating energy exchange between adjacent or isolated batteries/modules. A MPC model predicts system behaviour based on current conditions, using the SoC of each battery as the state variable. It also enhances speed and reduces the losses. Continuous recursion and rolling optimisation enable the system to adapt and optimise performance with new data. Simulations conducted via MATLAB/Simulink show that the MPC-based equaliser reduces equalisation time by 38.15\% and enhances the energy transfer efficiency by 27.25\% compared to a single-layer BBE. These results are also validated through experiments across various scenarios.

Battery inconsistency, stemming from manufacturing deviations, temperature disparities, and degradation dynamics, impacts various battery parameters beyond SoC. Addressing this issue, the study in \cite{article64} introduced an efficient SoC observer that estimates OCV using RLS, followed by a Luenberger observer for SoC with guaranteed stability. Additionally, a graph-theoretic battery model facilitates energy transfer in a parallel equalisation circuit, formulated as a path-searching problem, and an A-star search algorithm significantly reduces balancing time and energy loss. Experimental results confirm the observer's steady-state error of less than 2\% and illustrates the A-star algorithm's efficiency, reducing balancing time and energy loss by 9.59\% and 19.5\%, respectively, compared to the existing methods.

Traditional controllers often encounter challenges such as rapid temperature fluctuations, overcharging, over-discharging, and complex operations, which can negatively impact battery lifespan. The study in \cite{article65} proposed Fuzzy Logic Controller (FLC) and its optimised version which effectively mitigate these issues, leading to increased battery longevity. The study aims to enhance grid efficiency and cost reduction by introducing a PSO-based FLC for battery ESSs in micro-grid (MG) applications. Initially, the FLC is deployed for charging and discharging regulation, simplifying operations without complex calculations. Subsequently, the FLC's performance is enhanced by optimising its membership function using PSO, considering parameters such as existing power, load demand, battery temperature, and SoC. Additionally, an optimal Binary PSO (BPSO) scheduling algorithm is introduced to efficiently manage ESS, grid, and distributed sources, resulting in substantial reductions in grid power consumption (42.26\%) and energy usage costs (45.11\%). 

The battery charging process is characterised by non-linear behaviour and hysteresis, posing challenges for conventional Proportional Integration Differentiation (PID) control methods. The effectiveness of PID control hinges on its parameters, $K_{\text{p}}$, $K_{\text{i}}$, and $K_{\text{d}}$, which are traditionally challenging to set optimally, impacting charging efficiency. The study in \cite{article66} employs modified PSO to optimise PID parameters, addressing limitations of the basic PSO such as reduced speed and premature convergence. This helps to achieve better dynamic performance in battery charging systems. The input is a specified charging current or voltage, and the output is the corresponding charging current or voltage of the battery. The loop employs a PID control method to ensure improved stability and dynamic response performance in the charging system. The PID controller parameter tuning involves the implementation of the PSO algorithm to optimise the characteristics of $K_{\text{p}}$, $K_{\text{i}}$, and $K_{\text{d}}$. These three parameters act as basic particles that automatically evolve within the solution space, ultimately converging towards a globally optimal solution. Experimental results demonstrate significant improvements in charging efficiency, with the charging temperature rise reduced by 1$^{\circ}$C.

Accurate SoC estimation is paramount in applications such as EVs. However, direct real-time SoC measurement is impractical, necessitating estimation through measurable parameters such as voltage, current, and battery temperature. Current SoC estimation methods frequently depend on pre-calibrated data, resulting in inaccuracies, particularly as operational conditions change or batteries undergo ageing. 
The study in \cite{article67} introduces a real-time SoC estimation approach for Li-ion batteries using PSO applied to a detailed electrochemical model. A simple and efficient algorithm is proposed for accurate SoC estimation while conserving Central Processing Unit (CPU) and memory resources. It takes into consideration measured voltage and current as inputs, and outputs the estimated SoC minimising the expected error between the measured and estimated voltages. Once optimised states are acquired, they are utilised in the single-cell model for SoC estimation. Though the battery model includes six states, the study focuses on optimising one state i.e. the Li-ion concentration in the negative electrode. The choice of electrode lithium concentration ($C_s$) is relevant in representing SoC as it is directly proportional to the OCV. In the PSO algorithm, $C_s$ values are treated as particles, initialised to values arbitrarily around the previous Cs. The terminal voltage is then estimated and optimised for the measured current. The initial particle values are recommended to be the best known values representing the current battery condition for faster convergence. A linear interpolation function is used where, measured voltage and current are inputs, yielding the stoichiometric value of the concentration. This value is used to calculate Cs at the start of the algorithm. In succeeding time steps, the optimised states, including Cs, from the previous step are fed back into the system. The developed modular electrochemical model and constrained PSO-based SoC estimation algorithm are implemented over MATLAB/Simulink. The study demonstrates PSO's efficacy in SoC estimation across diverse battery operating conditions and validates real-time implementation feasibility on a Raspberry Pi 3 embedded device, showcasing promising performance in estimating SoC for both, healthy and aged battery  under various charge and discharge scenarios.

While existing methods partially address the battery inconsistency, the optimal equalisation strategy and balancing current tracking remain unresolved challenges. The study in \cite{article68} addresses these issues by proposing a modified equalisation system model emphasising topological efficiency and deriving an optimal balancing current solution using an MPC-based scheme. The study evaluates dissipative, unidirectional adjacent, bidirectional adjacent, and bus-based equalisation topologies, concluding that the bus-type system is particularly effective for series cell configurations exceeding 15 cells. A Multi-layer Feed Forward Neural Network (NN)-based passive cell balancing approach is employed in \cite{article70}, with training conducted using the Levenberg-Marquardt back-propagation algorithm through MATLAB's NN Toolbox. An analytical method is proposed for determination of bleed resistor, considering factors such as highest power loss, balancing time, and voltage deviation. A successful equalisation of SoC levels is achieved, with the Artificial NN (ANN)-based model exhibiting smoother cell voltage during balancing compared to the Logic-based State-flow model. The methodology offers advantages including analytical determination of the optimal resistance value, reduced voltage ripple during balancing, and minimised voltage and current ripples with bleed resistor adjustments. The parameters $\alpha$, $\beta$, and $\gamma$ are utilised to calculate an optimal resistance value, resulting in approximately 21$\Omega$. The ANN-based passive balancing demonstrates a remarkable reduction in battery voltage ripples (up to 95\%) and a 1\% decrease in battery current in contrast to the logic-based State-flow model during bleed resistor adjustments.

Consensus-based control methods often necessitate precise energy exchanges between the cells, potentially leading to excess energy exchange than necessary. Consequently, the benefits of distributed design may diminish or even be negated by excessive dissipation during the balancing process. To address this problem, the authors in \cite{article71} introduce a unique distributed control method for active balancing of, enabling a scalable and adaptable battery system design. The recommended algorithm assumes a dynamic system topology, where each battery has its own power electronics and processing unit. The system operates as a MAS, with each cell acting as an agent. Using cooperative game theory, cells autonomously compute and optimise control actions without a central leader. The cells also form coalitions to exchange energy for active balancing, enabling flexibility in system size and avoiding the need for algorithm reprogramming. Each cell makes individual runtime decisions, promoting autonomy and communication between neighbouring cells.

Traditional droop control in micro-grids often results in voltage and frequency variations. However, current methods lack optimisation of battery equalisation time and consideration of battery limits. Addressing this, the study in \cite{article72} proposes two equalisation and energy management methodologies are proposed for cell balancing and minimal grid usage. The initial SoC of batteries considered are 70\%, 65\%, 61\% each of 1Ah, 2Ah, and 3Ah capacity, respectively. Utilising a GA-based optimisation, optimal battery currents are determined. Results indicate that both methods effectively manage energy and battery equalisation. The choice between methods depends on prioritising rapid equalisation or minimising costs for both, energy management and equalisation. Further, both approaches demonstrate effective energy management and equalisation without compromising system operation; one achieves equalisation in 29 seconds with grid consumption, while the other achieves it in 110 seconds with no grid usage.

Current studies often design charging and balancing control strategies independently, overlooking their interplay. The latter is crucial to developing rapid charging and equalisation algorithms concurrently, particularly for scenarios involving high charging currents as it leads to safety constraints. The article \cite{article73} introduces a novel approach to address these challenges via a balancing-aware fast charging control framework exploiting DRL. Initially, a cell-to-pack equalisation arrangement is presented to efficiently distribute energy between cells within the pack. Subsequently, the fast charging problem, with considerations for charging time, consistency, and over-voltage safety constraints, is formulated as a multi-objective optimisation challenge. A DRL framework utilising a deep Q-network is developed to determine the optimal policy for this problem.

In \cite{article55}, battery equalisation is attained to improve both, time efficiency and energy efficiency. The authors propose an Ah-EMF Fuzzy Logic Controller. A battery pack of 10Ah and 3.2V is used. It is built with a weighted combination of the ampere hour counting method and the EMF equivalent model method. The controller is designed to estimate static SoC. It operates with a two-stage DC-DC converter. Dynamic SoC estimation is proposed by employing NN. The choice of Fuzzy logic control with NN is used considering the operating current, temperature, and ageing. GA is used to identify the global point at which the energy consumption is the minimum within the time limit specified. The authors built a testing system with the battery pack for the HEV's or EV's applications. A 96-cell battery pack and PLF08150240 are used in the study, with 3.2V nominal voltage and 10Ah nominal Capacity. The experimental set up includes a system consisting of a data acquisition module, a battery dynamic SoC estimation module, and a battery equalisation unit. For testing the built system, UDDS current and voltage profiles are employed. The study compared the outcomes with both the mean-difference algorithm and GA, demonstrating significant improvements in equalization duration compared to the mean-difference algorithm. GA achieves 93.1\% energy efficiency with an equalization time of 1002.5 seconds, highlighting its novel approach to optimizing battery performance and energy consumption under time constraints, aimed at identifying the global minimum energy consumption point.

For SoC balancing, the authors in \cite{article77} proposed a modified bidirectional Cuk Converter circuit with one MOSFET and one relay to control the transfer of energy between two adjacent cells and an additional relay to cut off when the cells have same SoC level. To determine the optimal SoC Balancing control, authors formulated an optimisation problem to determine the optimal duty using sequential quadratic programming (SQP) method. The MATLAB script is used to code SQP in the study. The study exhibited an experimental set up consisting of Samsung Li-ion battery pack of 7 series connected cells, SoC balancing circuit (Cuk converter), a measurement module and a PC installed with MATLAB and LabVIEW, and a load with a charging source. The measurement module is developed to measure current, voltage, and temperature of the cells. The author's connected a 240 W, 24 V BLDC motor with 295 rpm as load to the battery. Based on a second-order equivalent circuit model, the authors developed a two-point sigma Kalman filter estimation algorithm and balancing control algorithm (SQP) coded in MATLAB script and embedded in the LabVIEW environment to estimate the SoC, current, and voltage of the cells at three different scenarios. The results demonstrated balancing speed improvement and the significance of the algorithm initialization.

Using graph theory for equalising the series connected cells for HEV battery pack as storage system is proposed in \cite{article104}. The optimisation  objective is to maximise the efficiency and obtain optimal equalisation. The study demonstrates two-layer equalisation scheme using graph theory for transferring energy between the cells. In order to maximise energy transfer, the objective function is to minimise energy loss in the balancing process. To obtain balancing scheme, the ANT colony algorithm is employed to a 36 cell series connected battery for testing and each equaliser balances 6 cells, with bottom layer equaliser and top layer equaliser set to 0.85 and 0.9 efficiency, respectively. To validate the proposal simulated results are compared with common equalisation strategy like transferring the energy from highest Value of Characteristic factor (VOC) to lowest VOC, and prioritised equalisation within individual equaliser. The overall loss of the complete balancing activity in terms of  convergence process of the algorithm is reported as 0.334 with respect to VOC. The authors also suggest multi-objective optimisation as the future development for equalisation of series connected cells.

A novel stochastic approach is proposed by the authors in \cite{article122}. The study focuses on achieving better balancing of electrochemical cells by SoC method using Hybrid Genetic BPSO with sparsity regularisation for real time optimal balancing in energy storage systems. The study provides balancing architecture for SoC dependent passive balancing method, balancing algorithm aiming to evaluate the best optimal balancing point at each instant. The dependent variable considered in the study is SoC and is estimated using the traditional Coulomb counting technique. The optimisation problem is formulated to minimize variance of the SoC which also evaluates the optimal configuration of the shunt resistors. However, the study also aims at minimising the energy losses corresponding to the activation of the shunt resistors. Yet, the solution aims at providing the activation of minimum number of MOSFET switches. The experimental setup is validated for the effectiveness on the equivalent circuit model of an electrochemical cells simulated for 10 Li-ion 18650 series connected cells. The developed balancing algorithm is simulated on Advanced Vehicle Simulator (ADVISOR) software. A 85 Ah capacity EV with Federal Test Procedure (FTP) driving cycles is considered for the validation. Square Root UKF (SR-UKF) SoC estimation is implemented with initial SoC set to 50 percent. The impact of balancing using the proposed technique is presented with a set of promising results which highlight the need of cell balancing. Unbiased variance of SoC, unbiased variance of cell voltage, energy lost during the activation of resistors are considered as the performance metrics. The results exhibit that BPSO performs better in terms of SOC and voltage variance; however, it incurs highest energy loss. The energy lost in activation of the shunt resistors is reported as 6.24 Wh for voltage based balancing, 12.89 Wh for baseline BPSO, and 8.55 Wh for regularised BPSO.

An analytical approach is presented in \cite{article123} to address the control problem of cell balancing. The authors attended the major challenge of controlling the MOSFET switches, the frequency and duty cycle of the PWM signal in the balancing circuit of the cell equalizer. The aim of the study focuses on minimising the energy consumption in the equalising process and equalize the cells SoC and proposed Non Linear Model Predictive control (NMPC) method. Simulation is conducted for Li-ion cells, employing coulomb counting to estimate SoC, and Cuk converter as equalisation unit in a cell-to-cell transfer. The adaptive control scheme decides the control variables like switching frequency and the duty cycle, and GA is employed to solve the non-linear function introduced due dynamics of the battery. The study encourages researchers to explore the importance of model predictive control, considering control objectives and constraints specific to the identified problem. The authors introduced optimal control variables like control frequency and duty cycle in the equalization circuit, prompting further investigation into existing numerical solutions for adaptive control schemes in battery management.

The article \cite{article134} addresses active cell balancing for a cell-to-cell topology, and implements cell equalising using bidirectional modified Cuk converter for its numerous benefits and operated on a discontinuous inductor current mode (DICM). The authors achieved dynamic cell balancing with the proposed adaptive quasi-sliding mode observer with equalising current as a constraint for series-connected cells in a battery pack. The results demonstrate the convergence of SoC difference to an acceptable range with the converter’s equalising current as a constraint. Four cells of Li-ion battery 18650 are serially connected with three bidirectional modified Cuk converters are implemented using NTD6416AN-1G MOSFETS driven at 7 kHz switching frequency, set up the experimental platform for validating the proposed approach. Federal Urban Driving Schedule (FUDS) is utilised as driving cycle for the battery test on the MATLAB platform. The maximum cell current is set as 3 A and maximum allowed equalising current for the converter is 1 A. The actual SoC is obtained using the Ah counting method. The experiments are validated on (i) battery pack in standby mode, and (ii) battery pack in discharging mode. The study concludes that (i) the maximum cell equalising current should be set to a higher value to reduce the balancing time of the cell; (ii) SoC estimation errors converge to a small bound irrespective of the initial errors; and (iii) SoC-based cell balancing is the best choice as the cell’s OCV is constant for a typical SoC operating range. The scope for further research is to investigate the cell balancing efficiency with different converter topologies and validate the system scalability.

%>>>>>>>>>>>>>>>>>>>>>>>>>>>>>>>>>>

Table~\ref{tab:Estimation Optimisation Table} summarises various optimisation methodology for SoC and Voltage dependent cell balancing and Table~\ref{tab:Comparision of Optimization techniques} provides a detailed comparison of the optimisation methodology for SoC and Voltage dependent cell balancing

\begin{table*}[t!]
\caption{Optimisation Methodology for SoC and Voltage dependent Cell Balancing \cite{article46,article52,article53,article54,article55}}
\label{tab:Estimation Optimisation Table}
\centering
\resizebox{\textwidth}{!}{
    \begin{tabular}{|m{0.3\textwidth}|m{0.25\textwidth}|m{0.25\textwidth}|m{0.25\textwidth}|m{0.07\textwidth}|} \hline
    \textbf{Optimised parameter} & \textbf{Control strategy} &\textbf{Optimisation technique} & \textbf{Constraints} & \textbf{Reference} \\\hline

 SoC rate of change and power balance. & Optimising fuzzy membership functions & Backtracking search algorithm-based fuzzy optimisation & Balancing power,SoC & \cite{article19} \\ \hline 

Equalisation time and energy transfer efficiency &  Model predictive control  & Model predictive control & SoC &\cite{article33}\\\hline

Balancing time and energy loss.&   Recursive Least Squares to estimate OCV and Luenberger observer for estimating SoC &  A-star search algorithm & SoC & \cite{article64}\\\hline

Charging–discharging and scheduling of battery, grid efficiency and cost &  Fuzzy controller  &  PSO and  Binary PSO &  Temperature & \cite{article65}\\\hline

Charging efficiency of the battery and charging temperature & PID control   & PSO and modified PSO & - &\cite{article66}\\\hline
 
SoC &   K means ++ & PSO & - &\cite{article67}\\\hline

Balancing time and energy loss.& MPC   & MPC &  Temperature & \cite{article68}\\\hline

Bleed resistor &   Levenberg-Marquardt backpropagation algorithm & Analytical method &  Bleed resistor & \cite{article70}\\\hline

Energy transfer &  Bargaining game  & Nash solution & SoC, thermal loss, energy conservation & \cite{article71}\\\hline

Battery equalisation time and energy management in micro-grid.&   GA & GA &  Charging current,SoC,battery capacity & \cite{article72}\\\hline

Charging and cell balancing time& DRL   & Model-free DQN algorithm &  Maximum current  and voltage limitations, e in-pack cells’ SoC &\cite{article73} \\\hline

Battery time and energy equalisation    & Static SoC estimation by Fuzzy Logic and dynamic SoC estimation by Neural Network &  GA  & Equalisation time limit  &\cite{article55} \\\hline

SoC difference of cells and energy loss& Optimal SoC balancing algorithm   & Sequential quadratic programming algorithm & SoC operating range, balancing current, and cell current.  &\cite{article77} \\\hline

Energy efficiency & Graph theory   & ANT colony algorithm as balancing scheme & Voltage balance of cells &\cite{article104}\\\hline

Minimize the variance of SoC, Variance of Voltage and Energy lost in shunt resistors& SoC estimation technique using Square root Unscented Kalman Filter    & Hybrid Genetic Binary  PSO. & Energy loss  &\cite{article122}\\\hline

Equalize the cell SoC & Non-linear model predictive control (NMPC) & GA  & Frequency and duty cycle of PWM signal  &\cite{article123}\\\hline
Cell equalising current & Quasi-sliding Mode Control & Discrete-time quasi-sliding-mode  & Saturated equalising current  &\cite{article134}\\\hline

% &    &  &  &\\\hline
% & \ding[0.5]{108}\quad   & \ding[0.5]{108}\quad 
\end{tabular}}
\end{table*}

\begin{table*}[t!]
\caption{Comparison of Optimization techniques applicable to cell balancing \cite{article19,article67, article138,article139,article140,article141,article142,article143,article144,article145,article146, article147,article148,article149,article150, article151}}
\label{tab:Comparision of Optimization techniques}
\centering
\resizebox{\textwidth}{!}{
    \begin{tabular}{|m{0.16\textwidth}|m{0.4\textwidth}|m{0.5\textwidth}|} \hline
    \textbf{Optimisation technique} & \textbf{Merits} & \textbf{Demerits}  \\\hline
    
PSO &  \ding[0.5]{108}\quad Requires less variables to tune.

\ding[0.5]{108}\quad Is computationally efficient.

\ding[0.5]{108}\quad Exhibits a elevated degree of convergence. &   \ding[0.5]{108}\quad Initialization of velocity vectors decides the accuracy of the results.\\\hline

Binary PSO &  \ding[0.5]{108}\quad  Simple to implement. Effective for Discrete Problems.

\ding[0.5]{108}\quad It utilizes the concept of swarm intelligence, enabling particles to share information and collectively converge toward optimal solutions.  &   \ding[0.5]{108}\quad  Exhibits slow convergence rates on certain optimization problems.

\ding[0.5]{108}\quad  Due to its probabilistic nature, the algorithm may become trapped in local optima, preventing it from reaching the global optimum.

\ding[0.5]{108}\quad The position updating mechanism ignores the previous positions of particles, which can cause divergence from the optimal solution.

\ding[0.5]{108}\quad Using a sigmoid function for position updates can result in a 50\% probability of changing position when near the optimum, potentially hindering convergence. \\\hline

Backtracking Search Algorithm  &  \ding[0.5]{108}\quad High convergence speed. 

\ding[0.5]{108}\quad Effectively explores the search space, aiding in the avoidance of local minima and leading to the discovery of better solutions. 

\ding[0.5]{108}\quad Effective for constraint satisfaction problems.

\ding[0.5]{108}\quad Capable of easily identifying the global optimal solution  & \ding[0.5]{108}\quad Exponential time complexity

\ding[0.5]{108}\quad Higher computation time and cost\\\hline

 GA & \ding[0.5]{108}\quad  Simple to implement. 
 
 \ding[0.5]{108}\quad Exhibits global perspective and inherent parallel processing.   &  \ding[0.5]{108}\quad GA performance can be successful, but it does not ensure that the final generation consists of an optimal solution as it can retard at any point.
 
 \ding[0.5]{108}\quad The stalling situation becomes more frequent as the sequence length increases. 
 
 \ding[0.5]{108}\quad Growing resemblance in the population leads to the growth of twins or same chromosomes, which contributes to stalling. 
 
 \ding[0.5]{108}\quad GAs need a usually more number of iterations to converge compared to other less complex problems.\\\hline
 
ANT Colony  &   \ding[0.5]{108}\quad Demonstrates strong performance in solving discrete optimization problems.  & \ding[0.5]{108}\quad While it offers good stability, its convergence speed and solution accuracy may suffer when handling large datasets.

\ding[0.5]{108}\quad Low efficiency and reduced accuracy during convergence. 

\ding[0.5]{108}\quad Can stuck in local minima due to the premature convergence of pheromone trails, leading to sub-optimal solutions.

\ding[0.5]{108}\quad It is less practical for very large-scale optimization problems.\\\hline

A-star Search &  \ding[0.5]{108}\quad Ease of implementation. 

\ding[0.5]{108}\quad Elevated search efficiency.

 \ding[0.5]{108}\quad Strong functionality.
 
 \ding[0.5]{108}\quad Utilized in static global search because of its directional and heuristic capabilities in the search operation.&    \ding[0.5]{108}\quad Does not ensure the identification of the optimal path in a static environment.\\\hline

Hybrid Genetic Binary PSO  &  \ding[0.5]{108}\quad Mitigates the risk of PSO wasting resources on inferior individuals due to the absence of a selection operator.

\ding[0.5]{108}\quad PSO effectively enhances the ability to search for optimal solutions, while GA contributes to exploring the global search space.  

\ding[0.5]{108}\quad Effective for optimization problems in continuous, multidimensional search spaces.

\ding[0.5]{108}\quad Unlike GA, where unselected chromosomes lose their information, PSO retains memory of previous positions, preserving valuable information.& \ding[0.5]{108}\quad The evolutionary nature of the algorithm results in elongated computing time.

\ding[0.5]{108}\quad Lack of probing the complete solution due to possibilities of getting trapped in local minima.

\ding[0.5]{108}\quad Lack of efficient trade-off's.

\ding[0.5]{108}\quad Low memory capability for re-exploring the previous solutions.  \\\hline

MPC   &  \ding[0.5]{108}\quad  Effectively works within the narrow constraints of real actuators. &  \ding[0.5]{108}\quad  Requires more time to execute as it is complex. 

\ding[0.5]{108}\quad  The performance of MPC majorly relies on the precision of the system model.

\ding[0.5]{108}\quad  Any discrepancy between the system model and the actual plant behaviour can significantly impact the performance.\\\hline

Discrete-Time Quasi-Sliding-Mode Control    & \ding[0.5]{108}\quad The system need not cross the sliding hyperplane at each control step but only needs to stay within a miniature band around it. 

\ding[0.5]{108}\quad The control approach can be linear, which helps to avoid undesirable chattering.

\ding[0.5]{108}\quad Requires the system state to stay within a band all over a time-varying hyperplane that converges to a predetermined location in finite time.

\ding[0.5]{108}\quad Good performance. &  \ding[0.5]{108}\quad Rapid oscillation of the control signal around the sliding surface. \\\hline

Model-Free DQN    & \ding[0.5]{108}\quad Can handle large state spaces without requiring prior knowledge of the dynamic system. 

\ding[0.5]{108}\quad High level of accuracy & \ding[0.5]{108}\quad Increased computation time during the training phase.  \\\hline

Analytical Method      & \ding[0.5]{108}\quad Enhanced precision and accuracy with performance longevity.

\ding[0.5]{108}\quad Uniformity and quick cell balancing.

\ding[0.5]{108}\quad Reduced capacity degradation with an extended battery's lifespan.   & \ding[0.5]{108}\quad Increased complexity with sophisticated algorithm and equipment requirements.

\ding[0.5]{108}\quad Elevated implementation cost.

\ding[0.5]{108}\quad Dependency on sensing and monitoring-based types of equipment.

\ding[0.5]{108}\quad Vulnerable to equipment malfunctions   \\\hline

Sequential Quadratic Programming   &  \ding[0.5]{108}\quad A simple algorithm for enhancing designs in constrained optimization problems. &  \ding[0.5]{108}\quad Lack of robustness. \\\hline
 % &    &   \\\hline
% & \ding[0.5]{108}\quad   & \ding[0.5]{108}\quad 
\end{tabular}}
\end{table*}

\section{Choice of cell balancing techniques and algorithms}
\label{Section: Choice of cell Technique}
Cell balancing techniques and algorithms are crucial for ensuring the secure and efficient functionality of battery packs. The choice of the technique to be implemented depends on user’s determined applications, battery pack size, balancing requirements, and cost constraints \cite{article111}. Table \ref{tab:Summary of} provides a comprehensive summary of the selection criteria for cell balancing techniques, algorithms, and control parameters.

\begin{figure*}[hbt!]
    \centering    
    \includegraphics[width=\textwidth]{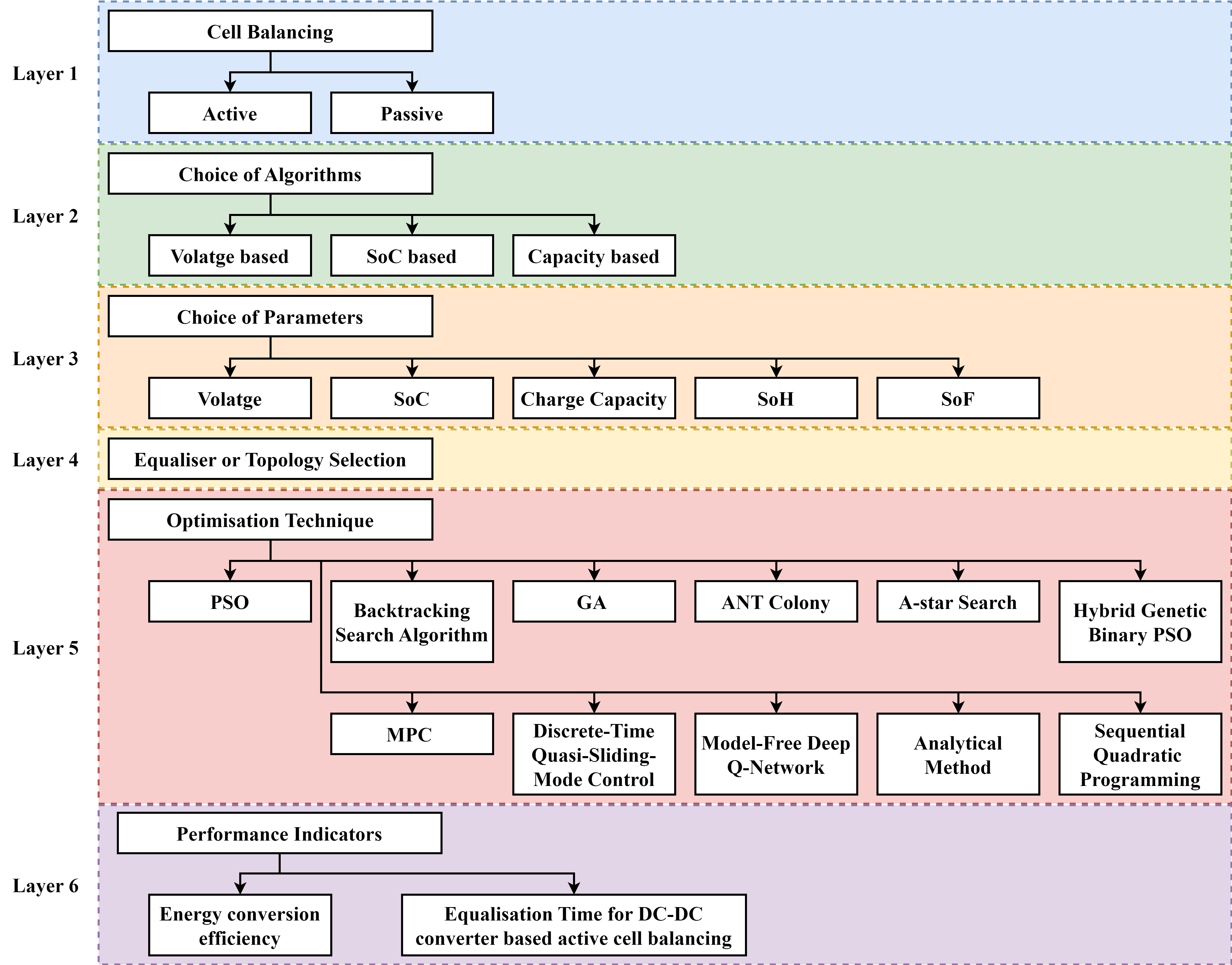}   
    *for Layer 4 (Equaliser or Topology Selection), refer \cite{article111}.
    \caption{A summary for choosing cell balancing algorithms and optimisation schemes.} 
    \label{fig:Method}
\end{figure*}

\begin{table*}[t!]
\caption{Summary of selection criteria for cell balancing techniques, algorithms, and control parameters \cite{article1, article2, article3, article26,article37,article38, article111}}
\label{tab:Summary of}
\centering
\resizebox{\textwidth}{!}{
\begin{tabular}{|m{0.15\textwidth}|m{0.12\textwidth}|m{0.8\textwidth}|} \hline
    \textbf{Key consideration} & \textbf{Key parameters} & \textbf{Applicability} \\\hline
    
\multirow{2}{*}{Cell balancing technique} & Passive & \ding[0.5]{108}\quad Well suited for applications where cost is significant and efficiency is not the highest priority.

        \ding[0.5]{108}\quad Suitable for small, medium and large battery packs.

        \ding[0.5]{108}\quad Applicable for low-power applications.  \\ \cline{2-3} 
                           & Active &  \ding[0.5]{108}\quad Suitable for small, medium, and large battery packs.
                            
        \ding[0.5]{108}\quad Recommended for applications requiring high efficiency and faster balancing. \\ \hline
                            
\multirow{3}{*}{Balancing algorithms} & SoC  & \ding[0.5]{108}\quad  Ideal for applications like EVs, and ESS which demand elevated balancing precision.\\ \cline{2-3} 
        & Voltage  &  \ding[0.5]{108}\quad	Suitable for applications where cost is significant and quick detection of voltage discrepancies is necessary. \\ \cline{2-3} 
        & Charge capacity &  \ding[0.5]{108}\quad	Suitable for applications requiring precise cell balancing, taking into account temperature effects and internal resistance variations.
                            
        \ding[0.5]{108}\quad Effective in systems with significant capacity differences due to manufacturing tolerances or ageing. \\ \hline
                            
\multirow{2}{*}{Control strategies} & High-level & \ding[0.5]{108}\quad Used to analyse, sort, and compare variations in recorded cell variables such as terminal voltage, SoC, and charge capacity to decide the direction of energy flow. \\ \cline{2-3} 
        & Low-level & \ding[0.5]{108}\quad	Recommended to overcome the limitations of fixed duty cycle.
                            
        \ding[0.5]{108}\quad Suitable to control the balancing current in accordance with the voltage changes of the cells. \\ \hline
                            
\multirow{6}{*}{Control parameter} & Terminal Voltage &	\ding[0.5]{108}\quad Useful for real-time balancing as it is a direct indicator of cell imbalance under operational conditions. \\ \cline{2-3}
        & OCV &	\ding[0.5]{108}\quad Suitable when voltage is the chosen parameter, as it offers more accurate balancing compared to terminal voltage dependent strategies. \\ \cline{2-3}
        & SoC &	\ding[0.5]{108}\quad Critical for cell balancing taking into account cell usage. \\ \cline{2-3}
        & Charge capacity &	\ding[0.5]{108}\quad Important for understanding cell health when balancing of cells with different capacities is necessary. \\ \cline{2-3}
        & SoF &	\ding[0.5]{108}\quad Applicable when the cell's ability to deliver power is to be determined. \\ \cline{2-3}
        & SOH &	\ding[0.5]{108}\quad Suitable for cell balancing where, addressing discrepancies in battery capacity due to aging is crucial.\\ \hline
\end{tabular}}
\end{table*}

\begin{enumerate}
\item \textit{Cell balancing technique:} Passive balancing is apt for simple, low-cost applications \cite{article112}; whereas, active balancing techniques are preferred as they offer advantages such as faster balancing, higher efficiency, reduced heat generation, etc. \cite{article111}.

\item \textit{Balancing speed:} Time and speed play a prominent role in the process of cell balancing. Hence if the application is dynamic or time-constrained active cell balancing might be the way. In other cases, passive balancing can be chosen.

\item \textit{Chemistry:} Passive cell balancing can be based on the battery 
chemistry. For example, lead-acid or nickel-based strings of cells perform well with passive cell balancing. During the overcharge condition, the excess available energy is dissipated as heat without major damage to the cell or its chemistry. Conversely, active cell balancing does not depend on the chemistry of the cell. Hence, a wide range of battery chemistries can be utilised under active balancing \cite{article137}.

\item \textit{Control parameter:} The equalisation efficiency and balancing time depends upon chosen variables which should be conscientiously examined based on given conditions \cite{ article108}. Voltage based balancing is preferred in low-cost consumer electronics where, cell balancing is not critical \cite{ article108,article111}; whereas SoC dependent balancing is ideal for applications demanding high balancing precision such as, advanced BMS for EVs, renewable energy storage, etc. \cite{article111}.

\item \textit{Cost and complexity:} The complex structure/algorithm comes with a cost. If the application is not cost-constrained, the active cell balancing can be chosen due to its complex circuitry nature. If a simple alternative is needed, passive-based cell balancing will do the work. The number of components required for realization of active balancing is more in contrast to to passive cell balancing.

\item \textit{System efficiency:} Due to high energy losses, passive cell balancing has an overall lower efficiency when compared with active. The active cell balancing feature of charge redistribution during the charge and discharge cycles greatly helps achieve higher efficiency.

\item \textit{Operational context and Algorithmic Perspective:} The choice of cell balancing algorithm depends on whether balancing is required during charging/discharging or in isolation. Algorithms for dynamic balancing during charging/discharging must handle real-time fluctuations in cell voltages and states SoC \cite{article19} ensuring safety and efficiency without compromising battery performance \cite{ article82}. In contrast, algorithms for static or isolated balancing can focus on achieving long-term equilibrium among cells, optimising energy transfer efficiency and minimising losses over extended periods. Thus, the operational context significantly influences the algorithm's design and effectiveness in maintaining battery health and performance. Environmental and operational conditions necessitate careful consideration as well. Choosing the appropriate cell balancing algorithm requires a thorough evaluation of performance criteria, including efficiency, cycle life, power delivery, safety, adaptability, cost-effectiveness, regulatory compliance, and future scalability.

\item \textit{Selection of optimization parameter and hardware:} The selection of parameters to be optimized is crucial for enhancing system performance through optimization techniques. In hardware implementation, restricting the balancing current to a suitable value can lower the cost of balancing hardware and decrease energy losses throughout the balancing operation. However, consideration of the overall functionality of the battery network, introduces additional constraints such as limited time for balancing and potential further deviations in the SoC of the batteries during operation \cite{article71}.

\item \textit{Control algorithm:} Classic control algorithms (CCAs), including methods like the Extreme Value Equalisation Algorithm (EVEA) and the Average Value Equalisation Algorithm (AVEA), have been commonly used for cell balancing due to their simple structure and low complexity. Despite their advantages, CCAs can suffer from mis-equalisation or over-balancing caused by factors such as noise interference and sampling errors \cite{ article109}. To address these issues, Advanced Control Algorithms (ACAs) such as MPC \cite{article37}, fuzzy logic control, PID control\cite{article37}, and data-guided approaches based on machine learning algorithms are used. These methods enhance the precision of system measurements and improve the learning abilities. \cite{ article109, article110}.
\end{enumerate}

 A generalised overview for choosing cell balancing algorithm and optimisation scheme is shown in Fig.~\ref{fig:Method}. The operating conditions of a battery pack are critical to consider when designing the cell balancing topology. Depending on the equalization structure requirements, the appropriate topology is selected \cite{article111}. The balancing algorithm can be based on SoC, voltage, or a combination of both, each with its own pros and cons, as listed in Table \ref{tab: Types of cell balancing algorithms}. To further enhance the performance of the cell balancing algorithm, optimization techniques can be applied by identifying the variable to be optimized. These variables can include SoC, cell terminal voltage, OCV, cell capacity, SoH, and SoF. The process starts with identifying the constraints, such as temperature, operating voltage, and current of the cell following which, an optimization technique is selected. The process involves continuous measurement of the selected parameter to determine how and when to balance cells based on the monitored data. The algorithm updates the selected variable continuously, repeating this process until optimal balancing is achieved which implies attaining the same SoC or voltage for all series-connected cells.

\subsection{Performance Indicators}
A high-performing equalizer is expected to demonstrate precise equalized SoC/voltages across the cells, minimal energy conversion losses, and rapid balancing capabilities. The functionality of an equalizer is assessed through diverse measurements and characteristic curves \cite{article129}. Next, we detail the key criteria used to validate the effectiveness of an equalizer.

\begin{enumerate}
\item Energy conversion efficiency, losses, and efficiency, cannot be directly determined using measuring instruments. Since there is no characteristic curve to depict exact energy loss, it must be determined mathematically. A cell can be modelled as a capacitor \cite{article129, article131}, and energy stored in the cell can be computed using
\begin{equation}
    E_{\text{cell}} = \frac{1}{2}  C_{\text{cell}} V^{2}_{\text{cell}}.
    \label{Eq:Ecell}
\end{equation}
When cells are connected in series or parallel, (\ref{Eq:Ecell}) can be modified to determine the energy stored in series for n cells in a string, and is given by
\begin{equation}
    E_{\text{string}} = \frac{1}{2} \sum_{\text{i=1}}^{\text{n}} C_{\text{i}} V^{2}_{\text{i}} ,
    \label{Eq:Estring}
\end{equation}
where, $C_{\text{i}}$ and $V_{\text{i}}$ are the capacitance and voltage of the $i^{th}$ cell, respectively. The energy stored in the string before equalising the cell voltages is expressed as
\begin{equation}
    E_{\text{string}} \left( 0 \right)= \frac{1}{2} \sum_{\text{i=1}}^{\text{n}} C_{\text{i}} V_{\text{i}}\left( 0 \right)^{2}.
    \label{Eq:equalisingEstring}
\end{equation}
The energy stored in the string after voltage equalisation of the cells at time $t$ is given as 
\begin{equation}
    E_{\text{string}} \left( t \right)= \frac{1}{2} \sum_{\text{i=1}}^{\text{n}} C_{\text{i}} V_{\text{i}}\left( t \right)^{2}.
    \label{Eq:afterEstring}
\end{equation}
Energy conversion loss is calculated as the difference of (\ref{Eq:equalisingEstring}) and  (\ref{Eq:afterEstring}), i.e., 
\begin{equation}
    E_{\text{loss}} = E \left(0 \right) - E \left(t \right) =\frac{1}{2} \left( \sum_{\text{i=1}}^{\text{n}} C_{\text{i}} V_{\text{i}}\left( 0 \right)^{2} - \sum_{\text{i=1}}^{\text{n}} C_{\text{i}} V_{\text{i}}\left( t \right)^{2} \right),
    \label{Eq:Eloss}
\end{equation}
necessitating mathematical analysis to accurately determine this loss. 

To calculate the energy conversion loss in joules (J), direct voltage monitoring of each cell in a battery pack is necessary. The measured voltages across the cells connected in a string, both pre and post equalisation, along with the cell capacitance (F), are needed for this calculation. The energy efficiency of an equalizer for series-connected cells in a string is the ratio of the total existing energy after equalisation to the overall unbalanced energy before equalisation, i.e., \cite{article130} 
\begin{equation}
    \text{Energy conversion efficiency} \left( \% \right)= \frac{E_{\text{string}}\left( t \right)}{E_{\text{string}}\left( 0 \right)} \times 100b,
    \label{Eq:Conversion}
\end{equation}
where \( E_{\text{string}}(0) \) and \( E_{\text{string}}(t) \) represent the electrostatic energy before and after equalisation, respectively.

\item Equalisation Time for DC-DC converter-based active cell balancing
 refers to the duration required to balance the voltages or SoC of individual cells within a battery pack \cite{article133}. The equalisation time is closely associated with the equalisation current, with higher currents resulting in shorter equalisation times. Selecting the most appropriate current path and maintaining the balancing current at the maximum feasible value within specific limits ensures minimal energy consumption and the shortest balancing time \cite{article37}. Additionally, the equalisation time depends on the degree of charge imbalance, like the difference in the SoC \cite{article132}. Therefore, the equalisation time can be determined by considering the equalisation current and the desired SoC difference.
The equalisation time based on the SoC difference \cite{article132} is given by
\begin{equation}
    \Delta \text{SoC} = \frac{Q_{\text{available}}}{Q_{\text{total}}} = \frac{C \Delta V}{Q_{\text{total}}} ,
    \label{Eq:deltaSOC}
\end{equation}
where, \( Q_{\text{available}} \) and \( Q_{\text{total}} \) represent the available and maximum charge for a Li-ion battery, respectively, and \( C \) denotes the remaining capacitance in the battery. To determine the relation between the equalisation time \(\Delta t_{\text{eq}} \) and the equalisation current \( I_{\text{eq}} \), one has that

%\begin{equation}
\begin{align}
C &= \frac{Q_{\text{available}}}{\Delta V} = \frac{I_{\text{eq}} \Delta t_{\text{eq}}}{\Delta V} ,
\label{Eq:C}\\
%\end{equation}
%
%\begin{equation}
I_{\text{eq}} &= I_{\text{out}} - I_{\text{in}} = I_{\text{out}} \left( 1 - \frac{V_{\text{cell}}}{\eta V_{\text{pack}}} \right) %,
%\label{Eq:Ieq}\\
%\end{equation}
%
%\begin{equation}
%&
\simeq I_{\text{out}} \left( 1 - \frac{V_{\text{cell}}}{\eta  n  V_{\text{cell}}} \right) \simeq I_{\text{out}},
%\label{Eq:Iout} ,
\label{Eq:Ieq}
%\end{equation}
\end{align}
where the equalisation current $I_{\text{eq}}$ is the net current between the input current $I_{\text{in}}$ and the output current $I_{\text{out}}$  of the DC-DC converter. $V_{\text{cell}}$ and $V_{\text{pack}}$ represent the voltages of the battery and the battery pack, respectively. The efficiency of the DC-DC converter, $\eta$, is included in (\ref{Eq:Ieq}). From equation (\ref{Eq:C}), the equalisation current can be illustrateed by the output current of the DC-DC converter for n . By using (\ref{Eq:Eloss}) to (\ref{Eq:C}), the equalisation time can be obtained as 
\begin{equation}
\Delta t_{\text{eq}} = \frac{C}{I_{\text{eq}}}  \Delta V = \frac{Q_{\text{total}}}{I_{\text{out}}}  \Delta \text{SoC}.
\end{equation}
\end{enumerate}

%>>>>>>>>>>>>>>>>>>>>>>>>>>>>>

\section{Conclusion}
\label{Section: Conclusion}

The implementation of BMS is key to monitoring and managing the key parameters of a battery to guarantee its safe and efficient functionality. BMS also ensures appropriate cell charging which is essential to maintain battery health, optimise performance, and maintain fundamental battery characteristics.

This articles surveyed a extensive classification of cell balancing algorithms for series-connected cells, highlighting their advantages and disadvantages. The importance of selecting appropriate control variables is also discussed, along with their merits and demerits. An extensive review on SoC estimation methods, detailing their strengths and weaknesses, is presented. The survey is categorised based on the cell balancing algorithms, and a comparative analysis of optimization techniques applicable to cell balancing techniques is also discussed. Overall, through this survey article, the authors aim to offer a guide in regard to the choice of control parameters, cell balancing algorithms, and optimisation techniques.

\section*{Declarations}

\textbf{Funding:} The authors declare that no funds, grants, or other support were received during the preparation of this manuscript.

\textbf{Conflict of Interests:} The authors have no conflict of interests to disclose.

\textbf{Data Availability:} Data sharing not applicable to this article as no data-sets were generated or analysed during this study. 

\textbf{Code Availability:} Not Applicable

\textbf{Author Contributions:} All the Authors contributed equally to the study and approved the final manuscript.

%=================================================

\bibliographystyle{IEEEtran}
\bibliography{references.bib}

\begin{thebibliography}{100}
\providecommand{\url}[1]{#1}
\csname url@samestyle\endcsname
\providecommand{\newblock}{\relax}
\providecommand{\bibinfo}[2]{#2}
\providecommand{\BIBentrySTDinterwordspacing}{\spaceskip=0pt\relax}
\providecommand{\BIBentryALTinterwordstretchfactor}{4}
\providecommand{\BIBentryALTinterwordspacing}{\spaceskip=\fontdimen2\font plus
\BIBentryALTinterwordstretchfactor\fontdimen3\font minus
  \fontdimen4\font\relax}
\providecommand{\BIBforeignlanguage}[2]{{%
\expandafter\ifx\csname l@#1\endcsname\relax
\typeout{** WARNING: IEEEtran.bst: No hyphenation pattern has been}%
\typeout{** loaded for the language `#1'. Using the pattern for}%
\typeout{** the default language instead.}%
\else
\language=\csname l@#1\endcsname
\fi
#2}}
\providecommand{\BIBdecl}{\relax}
\BIBdecl

\bibitem{article11}
S.~M. Lukic, J.~Cao, R.~C. Bansal, F.~Rodriguez, and A.~Emadi, ``Energy storage
  systems for automotive applications,'' \emph{IEEE Transactions on industrial
  electronics}, vol.~55, no.~6, pp. 2258--2267, 2008.

\bibitem{article12}
J.~Campillo, E.~Dahlquist, D.~L. Danilov, N.~Ghaviha, P.~H. Notten, and
  N.~Zimmerman, ``Battery technologies for transportation applications,''
  \emph{Technologies and applications for smart charging of electric and
  plug-in hybrid vehicles}, pp. 151--206, 2017.

\bibitem{article13}
S.~Ci, N.~Lin, and D.~Wu, ``Reconfigurable battery techniques and systems: A
  survey,'' \emph{IEEE Access}, vol.~4, pp. 1175--1189, 2016.

\bibitem{article16}
\BIBentryALTinterwordspacing
R.~Kallimani, S.~Gulannavar, K.~Pai, and P.~Patil,
  \emph{\BIBforeignlanguage{en}{A Detailed Study on State of Charge Estimation
  Methods}}.\hskip 1em plus 0.5em minus 0.4em\relax Singapore: Springer
  Singapore, 2022, vol. 844, p. 191–207. [Online]. Available:
  \url{https://link.springer.com/10.1007/978-981-16-8862-1_14}
\BIBentrySTDinterwordspacing

\bibitem{article6}
Y.~Li, J.~Xu, X.~Mei, and J.~Wang, ``A unitized multiwinding transformer-based
  equalization method for series-connected battery strings,'' \emph{IEEE
  Transactions on Power Electronics}, vol.~34, no.~12, pp. 11\,981--11\,989,
  2019.

\bibitem{article7}
\BIBentryALTinterwordspacing
N.~Samaddar, N.~Senthil~Kumar, and R.~Jayapragash, ``Passive cell balancing of
  li-ion batteries used for automotive applications,'' \emph{Journal of
  Physics: Conference Series}, vol. 1716, no.~1, p. 012005, Dec. 2020.
  [Online]. Available:
  \url{https://iopscience.iop.org/article/10.1088/1742-6596/1716/1/012005}
\BIBentrySTDinterwordspacing

\bibitem{article8}
A.~F. Moghaddam and A.~Van Den~Bossche, ``An active cell equalization technique
  for lithium ion batteries based on inductor balancing,'' in \emph{2018 9th
  International Conference on Mechanical and Aerospace Engineering (ICMAE)},
  2018, pp. 274--278.

\bibitem{article9}
S.~Khaleghi, M.~S. Hosen, J.~Van~Mierlo, and M.~Berecibar, ``Towards
  machine-learning driven prognostics and health management of li-ion
  batteries. a comprehensive review,'' \emph{Renewable and Sustainable Energy
  Reviews}, vol. 192, p. 114224, 2024.

\bibitem{article5}
S.~Hemavathi, ``Overview of cell balancing methods for li-ion battery
  technology,'' \emph{Energy storage}, vol. 203, pp. 1--12, 2020.

\bibitem{article4}
S.~Karmakar, T.~K. Bera, and A.~K. Bohre, ``Review on cell balancing
  technologies of battery management systems in electric vehicles,'' in
  \emph{2023 IEEE IAS Global Conference on Renewable Energy and Hydrogen
  Technologies (GlobConHT)}.\hskip 1em plus 0.5em minus 0.4em\relax IEEE, 2023,
  pp. 1--5.

\bibitem{article19}
M.~Faisal, M.~Hannan, P.~J. Ker, and M.~N. Uddin, ``Backtracking search
  algorithm based fuzzy charging-discharging controller for battery storage
  system in microgrid applications,'' \emph{IEEE Access}, vol.~7, pp.
  159\,357--159\,368, 2019.

\bibitem{article94}
W.~Han, C.~Zou, C.~Zhou, and L.~Zhang, ``Estimation of cell {SoC} evolution and
  system performance in module-based battery charge equalization systems,''
  \emph{IEEE Transactions on Smart Grid}, vol.~10, no.~5, pp. 4717--4728, 2018.

\bibitem{article95}
S.~Jiao, G.~Zhang, M.~Zhou, and G.~Li, ``A comprehensive review of research
  hotspots on battery management systems for {UAV}s,'' \emph{IEEE Access},
  2023.

\bibitem{article82}
S.~Karmakar, T.~K. Bera, and A.~K. Bohre, ``Novel {PI} controller and ann
  controllers-based passive cell balancing for battery management system,''
  \emph{IEEE Transactions on Industry Applications}, 2023.

\bibitem{article96}
J.~Lee and J.~Won, ``Enhanced coulomb counting method for {SoC} and {SoH}
  estimation based on coulombic efficiency,'' \emph{IEEE Access}, vol.~11, pp.
  15\,449--15\,459, 2023.

\bibitem{article56}
T.~Ebbs-Picken, C.~M. Da~Silva, and C.~H. Amon, ``Design optimization
  methodologies applied to battery thermal management systems: A review,''
  \emph{Journal of Energy Storage}, vol.~67, p. 107460, 2023.

\bibitem{article97}
Q.~Yu, C.~Wang, J.~Li, R.~Xiong, and M.~Pecht, ``Challenges and outlook for
  lithium-ion battery fault diagnosis methods from the laboratory to real world
  applications,'' \emph{ETransportation}, p. 100254, 2023.

\bibitem{article98}
B.~Ragchaa, L.~Wu, and X.~Zhang, ``A design of fault-tolerant battery
  monitoring {IC} for electric vehicles complying with {ISO} 26262,''
  \emph{IEEE Open Journal of Circuits and Systems}, 2024.

\bibitem{article15}
\BIBentryALTinterwordspacing
X.~Liu, Y.~Li, X.~Jiang, and K.~Xu, ``\BIBforeignlanguage{en}{Lifespan
  prediction of li-ion batteries in electrical vehicles by applying coulombic
  efficiency: from anode material to battery cell to vehicle application},''
  \emph{\BIBforeignlanguage{en}{Sustainable Energy \& Fuels}}, vol.~8, no.~3,
  p. 621–630, 2024. [Online]. Available:
  \url{https://xlink.rsc.org/?DOI=D3SE01455J}
\BIBentrySTDinterwordspacing

\bibitem{article17}
\BIBentryALTinterwordspacing
U.~Krishnamoorthy, G.~S. Satheesh~Kumar, S.~Barua, and H.~H. Fayek,
  ``\BIBforeignlanguage{en}{w},'' \emph{\BIBforeignlanguage{en}{IET Power
  Electronics}}, vol.~16, no.~15, p. 2492–2503, Nov. 2023. [Online].
  Available:
  \url{https://ietresearch.onlinelibrary.wiley.com/doi/10.1049/pel2.12575}
\BIBentrySTDinterwordspacing

\bibitem{article14}
\BIBentryALTinterwordspacing
D.~G.~L. Plett., \emph{Cell Balancing}, ser. ECE5720: Battery Management and
  Control.\hskip 1em plus 0.5em minus 0.4em\relax University of Colorado,
  Colorado Springs: University of Colorado, Colorado Springs, 2015. [Online].
  Available: \url{http://mocha-java.uccs.edu/ECE5720/ECE5720-Notes05.pdf}
\BIBentrySTDinterwordspacing

\bibitem{article117}
M.~Naguib, P.~Kollmeyer, and A.~Emadi, ``Lithium-ion battery pack robust state
  of charge estimation, cell inconsistency, and balancing,'' \emph{IEEE
  Access}, vol.~9, pp. 50\,570--50\,582, 2021.

\bibitem{article37}
A.~Turksoy, A.~Teke, and A.~Alkaya, ``A comprehensive overview of the {DC}-
  {DC} converter-based battery charge balancing methods in electric vehicles,''
  \emph{Renewable and Sustainable Energy Reviews}, vol. 133, p. 110274, 2020.

\bibitem{article109}
Y.~Hua, S.~Zhou, H.~Cui, X.~Liu, C.~Zhang, X.~Xu, H.~Ling, and S.~Yang, ``A
  comprehensive review on inconsistency and equalization technology of
  lithium-ion battery for electric vehicles,'' \emph{International Journal of
  Energy Research}, vol.~44, no.~14, pp. 11\,059--11\,087, 2020.

\bibitem{article120}
A.~Alvarez-Diazcomas, A.~A. Est{\'e}vez-B{\'e}n,
  J.~Rodr{\'\i}guez-Res{\'e}ndiz, M.-A. Mart{\'\i}nez-Prado, R.~V.
  Carrillo-Serrano, and S.~Thenozhi, ``A review of battery equalizer circuits
  for electric vehicle applications,'' \emph{Energies}, vol.~13, no.~21, p.
  5688, 2020.

\bibitem{article121}
S.~Dai, F.~Zhang, and X.~Zhao, ``Series-connected battery equalization system:
  a systematic review on variables, topologies, and modular methods,''
  \emph{International Journal of Energy Research}, vol.~45, no.~14, pp.
  19\,709--19\,728, 2021.

\bibitem{article111}
A.~R. Itagi, R.~Kallimani, K.~Pai, S.~Iyer, and O.~L. Lopez, ``Cell balancing
  for the transportation sector: Techniques, challenges, and future research
  directions,'' \emph{arXiv preprint arXiv:2404.13890}, 2024.

\bibitem{article93}
X.~Liu, H.~Pang, Y.~Geng, and L.~Wu, ``Design of an active equalization scheme
  with a modified circuit and {SoC} estimation-based {EKPF},'' in \emph{2022
  4th International Conference on Smart Power and Internet Energy Systems
  (SPIES)}, 2022, pp. 1266--1271.

\bibitem{article103}
F.~Ji, L.~Liao, T.~Wu, C.~Chang, and M.~Wang, ``Self-reconfiguration batteries
  with stable voltage during the full cycle without the {DC}- {DC} converter,''
  \emph{Journal of Energy Storage}, vol.~28, p. 101213, 2020.

\bibitem{article81}
J.~Hardy, J.~Steggall, and P.~Hardy, ``Rethinking lithium-ion battery
  management: eliminating routine cell balancing enhances hazardous fault
  detection,'' \emph{Journal of Energy Storage}, vol.~63, p. 106931, 2023.

\bibitem{article38}
M.~Caspar, T.~Eiler, and S.~Hohmann, ``Systematic comparison of active
  balancing: A model-based quantitative analysis,'' \emph{IEEE Transactions on
  Vehicular Technology}, vol.~67, no.~2, pp. 920--934, 2016.

\bibitem{article25}
M.~Hoque, M.~Hannan, and A.~Mohamed, ``Voltage equalization control algorithm
  for monitoring and balancing of series connected lithium-ion battery,''
  \emph{Journal of Renewable and Sustainable Energy}, vol.~8, no.~2, 2016.

\bibitem{article10}
Q.~Ouyang, J.~Chen, C.~Xu, and H.~Su, ``Cell balancing control for serially
  connected lithium-ion batteries,'' in \emph{2016 American control conference
  (ACC)}.\hskip 1em plus 0.5em minus 0.4em\relax IEEE, 2016, pp. 3095--3100.

\bibitem{article135}
J.~Yu, J.~Yang, Y.~Wu, D.~Tang, and J.~Dai, ``Online state-of-health prediction
  of lithium-ion batteries with limited labeled data,'' \emph{International
  Journal of Energy Research}, vol.~44, no.~14, pp. 11\,345--11\,360, 2020.

\bibitem{article1}
C.~Piao, Z.~Wang, J.~Cao, W.~Zhang, and S.~Lu, ``Lithium-ion battery
  cell-balancing algorithm for battery management system based on real-time
  outlier detection,'' \emph{Mathematical problems in engineering}, vol. 2015,
  2015.

\bibitem{article2}
J.~Xu, S.~Li, C.~Mi, Z.~Chen, and B.~Cao, ``{SoC} based battery cell balancing
  with a novel topology and reduced component count,'' \emph{Energies}, vol.~6,
  no.~6, pp. 2726--2740, 2013.

\bibitem{article3}
M.~Lu, Y.~Fan, and B.~Chong, ``A novel comprehensive {SoC}-voltage control
  scheme for lithium-ion battery equalization,'' in \emph{2020 International
  Conference on Power, Instrumentation, Control and Computing (PICC)}.\hskip
  1em plus 0.5em minus 0.4em\relax IEEE, 2020, pp. 1--6.

\bibitem{article26}
T.~Wu, F.~Ji, L.~Liao, and C.~Chang, ``Voltage-{SoC} balancing control scheme
  for series-connected lithium-ion battery packs,'' \emph{Journal of Energy
  Storage}, vol.~25, p. 100895, 2019.

\bibitem{article46}
W.~Zhou, Y.~Zheng, Z.~Pan, and Q.~Lu, ``Review on the battery model and {SoC}
  estimation method,'' \emph{Processes}, vol.~9, no.~9, p. 1685, 2021.

\bibitem{article52}
V.~Pop, H.~Bergveld, P.~Notten, and P.~P. Regtien, ``State-of-charge indication
  in portable applications,'' in \emph{Proceedings of the IEEE International
  Symposium on Industrial Electronics, 2005. ISIE 2005.}, vol.~3.\hskip 1em
  plus 0.5em minus 0.4em\relax IEEE, 2005, pp. 1007--1012.

\bibitem{article53}
\BIBentryALTinterwordspacing
G.~S.~R. Chowdary and M.~Sharam, ``An overview of {SoC} estimation in li-ion
  batteries with direct measurement methods and coulomb counting,''
  \emph{International Journal of Engineering Applied Sciences and Technology},
  vol.~7, no.~6, p. 237–242, Oct. 2022. [Online]. Available:
  \url{https://www.ijeast.com/papers/237-242,%20Tesma0706,IJEAST.pdf}
\BIBentrySTDinterwordspacing

\bibitem{article54}
W.-Y. Chang, ``The state of charge estimating methods for battery: A review,''
  \emph{International Scholarly Research Notices}, vol. 2013, 2013.

\bibitem{article55}
\BIBentryALTinterwordspacing
S.~Zhang, L.~Yang, X.~Zhao, and J.~Qiang, ``A ga optimization for lithium--ion
  battery equalization based on {SoC} estimation by {NN} and {FLC},''
  \emph{International Journal of Electrical Power \& Energy Systems}, vol.~73,
  pp. 318--328, 2015. [Online]. Available:
  \url{https://www.sciencedirect.com/science/article/pii/S0142061515002239}
\BIBentrySTDinterwordspacing

\bibitem{article100}
Y.~Liu, J.~Meng, F.~Yang, Q.~Peng, J.~Peng, and T.~Liu, ``A switchable
  indicator for active balance of the lithium-ion battery pack using a bypass
  equalizer,'' \emph{Journal of Energy Storage}, vol.~68, p. 107696, 2023.

\bibitem{article24}
A.~Mostafa, M.~A. Gaafar, O.~Abdel-Rahim, and M.~Orabi, ``Comparative study of
  the state of charge ({SoC}) balancing techniques for battery power module
  configurations,'' in \emph{2023 IEEE Conference on Power Electronics and
  Renewable Energy (CPERE)}.\hskip 1em plus 0.5em minus 0.4em\relax IEEE, 2023,
  pp. 1--5.

\bibitem{article42}
Q.~Wang, T.~Gao, and X.~Li, ``{SoC} estimation of lithium-ion battery based on
  equivalent circuit model with variable parameters,'' \emph{Energies},
  vol.~15, no.~16, p. 5829, 2022.

\bibitem{article43}
R.~R. Thakkar, Y.~S. Rao, and R.~R. Sawant, ``Performance analysis of
  electrical equivalent circuit models of lithium-ion battery,'' in \emph{2020
  IEEE Pune Section International Conference (PuneCon)}.\hskip 1em plus 0.5em
  minus 0.4em\relax IEEE, 2020, pp. 103--107.

\bibitem{article136}
R.~Kallimani, K.~Pai, and P.~Patil, ``Comparative analysis of state-of-health
  estimation techniques: A comprehensive survey,'' \emph{A Sustainable Future
  with E-Mobility: Concepts, Challenges, and Implementations}, pp. 53--69,
  2024.

\bibitem{article47}
X.~Hu, S.~Li, and H.~Peng, ``A comparative study of equivalent circuit models
  for li-ion batteries,'' \emph{Journal of Power Sources}, vol. 198, pp.
  359--367, 2012.

\bibitem{article44}
M.~Chen and G.~A. Rincon-Mora, ``Accurate electrical battery model capable of
  predicting runtime and {IV} performance,'' \emph{IEEE transactions on energy
  conversion}, vol.~21, no.~2, pp. 504--511, 2006.

\bibitem{article48}
M.~C. VanDyke, J.~L. Schwartz, C.~D. Hall \emph{et~al.}, ``Unscented {K}alman
  filtering for spacecraft attitude state and parameter estimation,''
  \emph{Advances in the Astronautical Sciences}, vol. 118, no.~1, pp. 217--228,
  2004.

\bibitem{article51}
M.~Doyle, T.~F. Fuller, and J.~Newman, ``Modeling of galvanostatic charge and
  discharge of the lithium/polymer/insertion cell,'' \emph{Journal of the
  Electrochemical society}, vol. 140, no.~6, p. 1526, 1993.

\bibitem{article50}
M.-F. Ng, J.~Zhao, Q.~Yan, G.~J. Conduit, and Z.~W. Seh, ``Predicting the state
  of charge and health of batteries using data-driven machine learning,''
  \emph{Nature Machine Intelligence}, vol.~2, no.~3, pp. 161--170, 2020.

\bibitem{article108}
P.~Dini, A.~Colicelli, and S.~Saponara, ``Review on modeling and {SoC}/{SoH}
  estimation of batteries for automotive applications,'' \emph{Batteries},
  vol.~10, no.~1, p.~34, 2024.

\bibitem{article30}
J.~Humpherys, P.~Redd, and J.~West, ``A fresh look at the {K}alman filter,''
  \emph{SIAM review}, vol.~54, no.~4, pp. 801--823, 2012.

\bibitem{article31}
G.~Sun, Q.~Ma, T.~Liu, and H.~Wen, ``Active battery equalization system based
  on extended {K}alman filter and {DC}-{DC} bidirectional flyback converter,''
  in \emph{2023 IEEE 2nd International Power Electronics and Application
  Symposium (PEAS)}.\hskip 1em plus 0.5em minus 0.4em\relax IEEE, 2023, pp.
  2163--2168.

\bibitem{article39}
P.~Kaniewski, ``Extended {K}alman filter with reduced computational demands for
  systems with non-linear measurement models,'' \emph{Sensors}, vol.~20, no.~6,
  p. 1584, 2020.

\bibitem{article40}
S.~J. Julier and J.~K. Uhlmann, ``New extension of the {K}alman filter to
  nonlinear systems,'' in \emph{Signal processing, sensor fusion, and target
  recognition VI}, vol. 3068.\hskip 1em plus 0.5em minus 0.4em\relax Spie,
  1997, pp. 182--193.

\bibitem{article41}
G.~L. Plett, ``Extended {K}alman filtering for battery management systems of
  {LiPB}-based {HEV} battery packs: Part 1. background,'' \emph{Journal of
  Power sources}, vol. 134, no.~2, pp. 252--261, 2004.

\bibitem{article49}
M.~Song, R.~Astroza, H.~Ebrahimian, B.~Moaveni, and C.~Papadimitriou,
  ``Adaptive {K}alman filters for nonlinear finite element model updating,''
  \emph{Mechanical Systems and Signal Processing}, vol. 143, p. 106837, 2020.

\bibitem{article36}
M.~Uno and A.~Kukita, ``Bidirectional {PWM} converter integrating cell voltage
  equalizer using series-resonant voltage multiplier for series-connected
  energy storage cells,'' \emph{IEEE Transactions on Power Electronics},
  vol.~30, no.~6, pp. 3077--3090, 2014.

\bibitem{article18}
D.~J. Docimo and H.~K. Fathy, ``Analysis and control of charge and temperature
  imbalance within a lithium-ion battery pack,'' \emph{IEEE Transactions on
  Control Systems Technology}, vol.~27, no.~4, pp. 1622--1635, 2018.

\bibitem{article113}
\BIBentryALTinterwordspacing
N.~Khan, C.~A. Ooi, A.~Alturki, M.~Amir, Shreasth, and T.~Alharbi, ``A critical
  review of battery cell balancing techniques, optimal design, converter
  topologies, and performance evaluation for optimizing storage system in
  electric vehicles,'' \emph{Energy Reports}, vol.~11, pp. 4999--5032, 2024.
  [Online]. Available:
  \url{https://www.sciencedirect.com/science/article/pii/S2352484724002506}
\BIBentrySTDinterwordspacing

\bibitem{article20}
Q.~Ouyang, J.~Chen, J.~Zheng, and H.~Fang, ``Optimal cell-to-cell balancing
  topology design for serially connected lithium-ion battery packs,''
  \emph{IEEE Transactions on Sustainable Energy}, vol.~9, no.~1, pp. 350--360,
  2018.

\bibitem{article21}
R.~Ling, Q.~Dan, J.~Zhang, and G.~Chen, ``A distributed equalization control
  approach for series connected battery strings,'' in \emph{The 26th Chinese
  Control and Decision Conference (2014 CCDC)}, 2014, pp. 5102--5106.

\bibitem{article22}
Q.~Ouyang, J.~Chen, H.~Liu, and H.~Fang, ``Improved cell equalizing topology
  for serially connected lithium-ion battery packs,'' in \emph{2016 IEEE 55th
  Conference on Decision and Control (CDC)}, 2016, pp. 6715--6720.

\bibitem{article23}
\BIBentryALTinterwordspacing
F.~Baronti, R.~Roncella, and R.~Saletti, ``\BIBforeignlanguage{en}{Performance
  comparison of active balancing techniques for lithium-ion batteries},''
  \emph{\BIBforeignlanguage{en}{Journal of Power Sources}}, vol. 267, p.
  603–609, Dec. 2014. [Online]. Available:
  \url{https://linkinghub.elsevier.com/retrieve/pii/S0378775314006715}
\BIBentrySTDinterwordspacing

\bibitem{article45}
\BIBentryALTinterwordspacing
Y.~Wang, C.~Zhang, Z.~Chen, J.~Xie, and X.~Zhang, ``\BIBforeignlanguage{en}{A
  novel active equalization method for lithium-ion batteries in electric
  vehicles},'' \emph{\BIBforeignlanguage{en}{Applied Energy}}, vol. 145, p.
  36–42, May 2015. [Online]. Available:
  \url{https://linkinghub.elsevier.com/retrieve/pii/S030626191500166X}
\BIBentrySTDinterwordspacing

\bibitem{article69}
P.~Soni and I.~Karuppasamy, ``Performance analysis of a 48v battery pack using
  {SoC} estimation and cell balancing for electric vehicle,'' in \emph{2023
  IEEE 8th International Conference for Convergence in Technology
  (I2CT)}.\hskip 1em plus 0.5em minus 0.4em\relax IEEE, 2023, pp. 1--6.

\bibitem{article74}
\BIBentryALTinterwordspacing
L.~Wang, J.~Ma, X.~Zhao, X.~Li, and K.~Zhang, ``\BIBforeignlanguage{en}{Online
  estimation of state-of-charge inconsistency for lithium-ion battery based on
  {SVSF-VBL}},'' \emph{\BIBforeignlanguage{en}{Journal of Energy Storage}},
  vol.~67, p. 107657, Sep. 2023. [Online]. Available:
  \url{https://linkinghub.elsevier.com/retrieve/pii/S2352152X2301054X}
\BIBentrySTDinterwordspacing

\bibitem{article75}
\BIBentryALTinterwordspacing
T.~Long, S.~Wang, W.~Cao, H.~Zhou, and C.~Fernandez,
  ``\BIBforeignlanguage{en}{An improved variable forgetting factor recursive
  least square-double extend {K}alman filtering based on global mean particle
  swarm optimization algorithm for collaborative state of energy and state of
  health estimation of lithium-ion batteries},''
  \emph{\BIBforeignlanguage{en}{Electrochimica Acta}}, vol. 450, p. 142270, May
  2023. [Online]. Available:
  \url{https://linkinghub.elsevier.com/retrieve/pii/S0013468623004504}
\BIBentrySTDinterwordspacing

\bibitem{article80}
I.~Bistritz and N.~Bambos, ``Consensus-based stochastic control for model-free
  cell balancing,'' \emph{IEEE Transactions on Control of Network Systems},
  vol.~8, no.~3, pp. 1139--1150, 2021.

\bibitem{article83}
M.~Kamel, V.~Sankaranarayanan, R.~Zane, and D.~Maksimovi{\'c},
  ``State-of-charge balancing with parallel and series output connected battery
  power modules,'' \emph{IEEE Transactions on Power Electronics}, vol.~37,
  no.~6, pp. 6669--6677, 2022.

\bibitem{article84}
N.~T. Milas and E.~C. Tatakis, ``Fast battery cell voltage equaliser based on
  the bi-directional flyback converter,'' \emph{IEEE Transactions on
  Transportation Electrification}, 2022.

\bibitem{article86}
C.~Nguyen~Van and T.~Nguyen~Vinh, ``{SoC} estimation of the lithium-ion battery
  pack using a sigma point {K}alman filter based on a cell’s second order
  dynamic model,'' \emph{Applied Sciences}, vol.~10, no.~5, p. 1896, 2020.

\bibitem{article89}
\BIBentryALTinterwordspacing
S.~Jinlei, L.~Wei, T.~Chuanyu, W.~Tianru, J.~Tao, and T.~Yong, ``A novel active
  equalization method for series-connected battery packs based on clustering
  analysis with genetic algorithm,'' \emph{IEEE Transactions on Power
  Electronics}, vol.~36, no.~7, p. 7853–7865, Jul. 2021. [Online]. Available:
  \url{https://ieeexplore.ieee.org/document/9314204/}
\BIBentrySTDinterwordspacing

\bibitem{article90}
\BIBentryALTinterwordspacing
S.~Katoch, S.~S. Chauhan, and V.~Kumar, ``\BIBforeignlanguage{en}{A review on
  genetic algorithm: past, present, and future},''
  \emph{\BIBforeignlanguage{en}{Multimedia Tools and Applications}}, vol.~80,
  no.~5, p. 8091–8126, Feb. 2021. [Online]. Available:
  \url{http://link.springer.com/10.1007/s11042-020-10139-6}
\BIBentrySTDinterwordspacing

\bibitem{article29}
S.~Na, L.~Xumin, and G.~Yong, ``Research on k-means clustering algorithm: An
  improved k-means clustering algorithm,'' in \emph{2010 Third International
  Symposium on intelligent information technology and security
  informatics}.\hskip 1em plus 0.5em minus 0.4em\relax IEEE, 2010, pp. 63--67.

\bibitem{article99}
F.~S.~J. Hoekstra, H.~J. Bergveld, and M.~Donkers, ``Optimal control of active
  cell balancing: Extending the range and useful lifetime of a battery pack,''
  \emph{IEEE Transactions on Control Systems Technology}, vol.~30, no.~6, pp.
  2759--2766, 2022.

\bibitem{article102}
Y.~Li, P.~Yin, and J.~Chen, ``Active equalization of lithium-ion battery based
  on reconfigurable topology,'' \emph{Applied Sciences}, vol.~13, no.~2, p.
  1154, 2023.

\bibitem{article126}
Q.~Ouyang, Y.~Zhang, N.~Ghaeminezhad, J.~Chen, Z.~Wang, X.~Hu, and J.~Li,
  ``Module-based active equalization for battery packs: A two-layer model
  predictive control strategy,'' \emph{IEEE Transactions on Transportation
  Electrification}, vol.~8, no.~1, pp. 149--159, 2021.

\bibitem{article105}
N.~Vikhorev, A.~Kurkin, D.~Aleshin, D.~Ulyanov, M.~Konstantinov, and
  A.~Shalukho, ``Battery dynamic balancing method based on calculation of cell
  voltage reference value,'' \emph{Energies}, vol.~16, no.~9, p. 3733, 2023.

\bibitem{article27}
M.~Ya{\u{g}}c{\i} and {\"O}.~Orbeyi, ``Programmable logic controlled
  lithium-ion battery management system using passive balancing method,''
  \emph{Journal of Radiation Research and Applied Sciences}, vol.~17, no.~2, p.
  100927, 2024.

\bibitem{article85}
M.~K. Al-Smadi and J.~A.~A. Qahouq, ``Evaluation of current-mode controller for
  active battery cells balancing with peak efficiency operation,'' \emph{IEEE
  Transactions on Power Electronics}, vol.~38, no.~2, pp. 1610--1621, 2022.

\bibitem{article88}
\BIBentryALTinterwordspacing
B.~Erdoğan, M.~M. Savrun, T.~Köroğlu, M.~U. Cuma, and M.~Tümay,
  ``\BIBforeignlanguage{en}{An improved and fast balancing algorithm for
  electric heavy commercial vehicles},'' \emph{\BIBforeignlanguage{en}{Journal
  of Energy Storage}}, vol.~38, p. 102522, Jun. 2021. [Online]. Available:
  \url{https://linkinghub.elsevier.com/retrieve/pii/S2352152X2100270X}
\BIBentrySTDinterwordspacing

\bibitem{article106}
W.~Chen, Z.~Ding, J.~Liu, J.~Kan, M.~S. Nazir, and Y.~Wang, ``Half-bridge
  lithium-ion battery equalizer based on phase-shift strategy,''
  \emph{Sustainability}, vol.~15, no.~2, p. 1349, 2023.

\bibitem{article115}
W.~Yang, J.~Han, X.~Liu, and S.~Yang, ``An active cell-to-cell balancing
  circuit with planar coupled inductors for series connected batteries,'' in
  \emph{2019 IEEE 4th International Future Energy Electronics Conference
  (IFEEC)}, 2019, pp. 1--6.

\bibitem{article124}
X.~Yang, L.~Xi, Z.~Gao, Y.~Li, and J.~Wen, ``Analysis and design of a voltage
  equalizer based on boost full-bridge inverter and symmetrical voltage
  multiplier for series-connected batteries,'' \emph{IEEE Transactions on
  Vehicular Technology}, vol.~69, no.~4, pp. 3828--3840, 2020.

\bibitem{article125}
H.~Xiong, D.~Song, F.~Shi, Y.~Wei, and L.~Jinzhen, ``Novel voltage equalisation
  circuit of the lithium battery pack based on bidirectional flyback
  converter,'' \emph{IET Power Electronics}, vol.~13, no.~11, pp. 2194--2200,
  2020.

\bibitem{article127}
Y.~Yu, R.~Saasaa, A.~A. Khan, and W.~Eberle, ``A series resonant energy storage
  cell voltage balancing circuit,'' \emph{IEEE Journal of Emerging and Selected
  Topics in Power Electronics}, vol.~8, no.~3, pp. 3151--3161, 2019.

\bibitem{article128}
M.~K. Hasan, A.~A. Habib, S.~Islam, A.~T.~A. Ghani, and E.~Hossain, ``Resonant
  energy carrier base active charge-balancing algorithm,'' \emph{Electronics},
  vol.~9, no.~12, p. 2166, 2020.

\bibitem{article28}
H.~Wu, H.~Zhao, D.~Qin, J.~Yang, and J.~Chen, ``Multi-cell-to-multi-cell active
  equalization method based on k-means clustering and battery pack {SoC}
  estimation,'' \emph{International Journal of Electrochemical Science}, p.
  100588, 2024.

\bibitem{article33}
\BIBentryALTinterwordspacing
B.~Xia, Y.~Li, G.~Zhang, Q.~Cheng, and F.~Ding, ``\BIBforeignlanguage{en}{A
  double-layer ring-structured equalizer for series-connected lithium-ion
  battery pack based on model predictive control},''
  \emph{\BIBforeignlanguage{en}{Journal of Energy Storage}}, vol.~78, p.
  110047, Feb. 2024. [Online]. Available:
  \url{https://linkinghub.elsevier.com/retrieve/pii/S2352152X23034461}
\BIBentrySTDinterwordspacing

\bibitem{article34}
\BIBentryALTinterwordspacing
M.~Schwenzer, M.~Ay, T.~Bergs, and D.~Abel, ``\BIBforeignlanguage{en}{Review on
  model predictive control: an engineering perspective},''
  \emph{\BIBforeignlanguage{en}{The International Journal of Advanced
  Manufacturing Technology}}, vol. 117, no. 5–6, p. 1327–1349, Nov. 2021.
  [Online]. Available:
  \url{https://link.springer.com/10.1007/s00170-021-07682-3}
\BIBentrySTDinterwordspacing

\bibitem{article35}
M.~Kitchen and W.~Damico, \emph{Technical Report: Development of Conversion
  Factors for Heavy-Duty Bus Engines G/BHP-HR to G/Mile}, ser. National Service
  Center for Environmental Publications (NSCEP).\hskip 1em plus 0.5em minus
  0.4em\relax United States Environmental Protection Agency, 1992, no.
  EPA-AA-EVRB-92-01.

\bibitem{article78}
\BIBentryALTinterwordspacing
C.~Lu, J.~Chen, C.~Chen, Y.~Huang, and D.~Xuan, ``\BIBforeignlanguage{en}{An
  active equalization method for redundant battery based on deep reinforcement
  learning},'' \emph{\BIBforeignlanguage{en}{Measurement}}, vol. 210, p.
  112507, Mar. 2023. [Online]. Available:
  \url{https://linkinghub.elsevier.com/retrieve/pii/S0263224123000714}
\BIBentrySTDinterwordspacing

\bibitem{article79}
C.~Sun, T.~Li, and X.~Tang, ``A data-driven approach for optimizing early-stage
  electric vehicle charging station placement,'' \emph{IEEE Transactions on
  Industrial Informatics}, pp. 1--11, 2023.

\bibitem{article91}
S.~Munwaja, B.~Tanboonjit, and N.~H. Fuengwarodsakul, ``Development of cell
  balancing algorithm for {LiFePO4} battery in electric bicycles,'' in
  \emph{2012 9th International Conference on Electrical
  Engineering/Electronics, Computer, Telecommunications and Information
  Technology}, 2012, pp. 1--4.

\bibitem{article92}
\BIBentryALTinterwordspacing
F.~Hoekstra, L.~Raijmakers, M.~Donkers, and H.~Bergveld,
  ``\BIBforeignlanguage{en}{Comparison of battery electromotive-force
  measurement and modelling approaches},''
  \emph{\BIBforeignlanguage{en}{Journal of Energy Storage}}, vol.~56, p.
  105910, Dec. 2022. [Online]. Available:
  \url{https://linkinghub.elsevier.com/retrieve/pii/S2352152X22018989}
\BIBentrySTDinterwordspacing

\bibitem{article116}
A.~Devi~B and A.~N, ``Automatic cell balancing with switched capacitor for
  multi cell connectivity,'' in \emph{2021 Fourth International Conference on
  Electrical, Computer and Communication Technologies (ICECCT)}, 2021, pp.
  1--7.

\bibitem{article57}
B.~Ayadi, D.~J. Jasim, A.~E. Anqi, W.~Aich, W.~Rajhi, and M.~Marefati,
  ``Multi-criteria/comparative analysis and multi-objective optimization of a
  hybrid solar/geothermal source system integrated with a carnot battery,''
  \emph{Case Studies in Thermal Engineering}, vol.~54, p. 104031, 2024.

\bibitem{article58}
M.~M. Savrun, T.~K{\"o}ro{\u{g}}lu, E.~{\"U}nal, B.~Onur, and M.~U. Cuma,
  ``Minimization of battery pack imbalance of electric vehicles using optimized
  balancing parameters,'' in \emph{2019 Electric Vehicles International
  Conference (EV)}.\hskip 1em plus 0.5em minus 0.4em\relax IEEE, 2019, pp.
  1--5.

\bibitem{article59}
\BIBentryALTinterwordspacing
G.~Venter, \emph{\BIBforeignlanguage{en}{Review of Optimization Techniques}},
  1st~ed.\hskip 1em plus 0.5em minus 0.4em\relax Wiley, Dec. 2010. [Online].
  Available:
  \url{https://onlinelibrary.wiley.com/doi/10.1002/9780470686652.eae495}
\BIBentrySTDinterwordspacing

\bibitem{article60}
M.~A. Bagherian, K.~Mehranzamir, A.~B. Pour, S.~Rezania, E.~Taghavi,
  H.~Nabipour-Afrouzi, M.~Dalvi-Esfahani, and S.~M. Alizadeh, ``Classification
  and analysis of optimization techniques for integrated energy systems
  utilizing renewable energy sources: a review for {CHP} and {CCHP} systems,''
  \emph{Processes}, vol.~9, no.~2, p. 339, 2021.

\bibitem{article61}
P.~Sharma and R.~C. Naidu, ``Optimization techniques for grid-connected {PV}
  with retired {EV} batteries in centralized charging station with challenges
  and future possibilities: A review,'' \emph{Ain Shams Engineering Journal},
  vol.~14, no.~7, p. 101985, 2023.

\bibitem{article62}
O.~{\"O}ZKARACA, ``A review on usage of optimization methods in geothermal
  power generation,'' \emph{Mugla Journal of Science and Technology}, vol.~4,
  no.~1, pp. 130--136, 2018.

\bibitem{article64}
G.~Dong, F.~Yang, K.-L. Tsui, and C.~Zou, ``Active balancing of lithium-ion
  batteries using graph theory and a-star search algorithm,'' \emph{IEEE
  Transactions on Industrial Informatics}, vol.~17, no.~4, pp. 2587--2599,
  2020.

\bibitem{article66}
T.~Wu, C.~Zhou, Z.~Yan, H.~Peng, and L.~Wu, ``Application of pid optimization
  control strategy based on particle swarm optimization ({PSO}) for battery
  charging system,'' \emph{International Journal of Low-Carbon Technologies},
  vol.~15, no.~4, pp. 528--535, 2020.

\bibitem{article73}
Y.~Yang, J.~He, C.~Chen, and J.~Wei, ``Balancing awareness fast charging
  control for lithium-ion battery pack using deep reinforcement learning,''
  \emph{IEEE Transactions on Industrial Electronics}, 2023.

\bibitem{article67}
A.~Chandra~Shekar and S.~Anwar, ``Real-time state-of-charge estimation via
  particle swarm optimization on a lithium-ion electrochemical cell model,''
  \emph{Batteries}, vol.~5, no.~1, p.~4, 2019.

\bibitem{article65}
M.~Faisal, M.~Hannan, P.~J. Ker, M.~A. Rahman, R.~Begum, and T.~Mahlia,
  ``Particle swarm optimised fuzzy controller for charging-discharging and
  scheduling of battery energy storage system in {MG} applications,''
  \emph{Energy Reports}, vol.~6, pp. 215--228, 2020.

\bibitem{article68}
Y.-X. Wang, H.~Zhong, J.~Li, and W.~Zhang, ``Adaptive estimation-based
  hierarchical model predictive control methodology for battery active
  equalization topologies: Part i--balancing strategy,'' \emph{Journal of
  Energy Storage}, vol.~45, p. 103235, 2022.

\bibitem{article70}
S.~V.~P. Singh and P.~Agnihotri, ``Ann based modelling of optimal passive cell
  balancing,'' in \emph{2022 22nd National Power Systems Conference
  (NPSC)}.\hskip 1em plus 0.5em minus 0.4em\relax IEEE, 2022, pp. 326--331.

\bibitem{article71}
M.~Caspar, T.~Sch{\"u}rmann, M.~Anneken, and S.~Hohmann, ``Active balancing
  control for distributed battery systems based on cooperative game theory,''
  \emph{Journal of Energy Storage}, vol.~68, p. 107585, 2023.

\bibitem{article72}
C.~H. Ricardo, L.~H. Adriana, and D.~A. Nelson, ``Energy management supported
  on genetic algorithms for the equalization of battery energy storage systems
  in microgrid systems,'' \emph{Journal of Energy Storage}, vol.~72, p. 108510,
  2023.

\bibitem{article77}
\BIBentryALTinterwordspacing
C.~N. Van, T.~N. Vinh, M.-D. Ngo, and S.-J. Ahn, ``Optimal {SoC} balancing
  control for lithium-ion battery cells connected in series,'' \emph{Energies},
  vol.~14, no.~10, 2021. [Online]. Available:
  \url{https://www.mdpi.com/1996-1073/14/10/2875}
\BIBentrySTDinterwordspacing

\bibitem{article104}
Z.~Liu, X.~Liu, J.~Han, and W.~Yang, ``An optimization algorithm for
  equalization scheme of series-connected energy storage cells,'' in \emph{2017
  IEEE Transportation Electrification Conference and Expo, Asia-Pacific (ITEC
  Asia-Pacific)}, 2017, pp. 1--6.

\bibitem{article122}
M.~Luzi, M.~Paschero, A.~Rizzi, and F.~M.~F. Mascioli, ``A binary {PSO}
  approach for real time optimal balancing of electrochemical cells,'' in
  \emph{2018 International Joint Conference on Neural Networks (IJCNN)}, 2018,
  pp. 1--8.

\bibitem{article123}
M.~F. Samadi and M.~Saif, ``Nonlinear model predictive control for cell
  balancing in li-ion battery packs,'' in \emph{2014 American Control
  Conference}, 2014, pp. 2924--2929.

\bibitem{article134}
Q.~Ouyang, J.~Chen, J.~Zheng, and Y.~Hong, ``{SoC} estimation-based
  quasi-sliding mode control for cell balancing in lithium-ion battery packs,''
  \emph{IEEE Transactions on Industrial Electronics}, vol.~65, no.~4, pp.
  3427--3436, 2018.

\bibitem{article138}
I.~Rahman, P.~M. Vasant, B.~S.~M. Singh, and M.~Abdullah-Al-Wadud, ``On the
  performance of accelerated particle swarm optimization for charging plug-in
  hybrid electric vehicles,'' \emph{Alexandria Engineering Journal}, vol.~55,
  no.~1, pp. 419--426, 2016.

\bibitem{article139}
H.~Nezamabadi-pour, M.~Rostami-Shahrbabaki, and M.~Maghfoori-Farsangi, ``Binary
  particle swarm optimization: challenges and new solutions,'' \emph{CSI J
  Comput Sci Eng}, vol.~6, no.~1, pp. 21--32, 2008.

\bibitem{article140}
K.~Deb, ``Genetic algorithm in search and optimization: the technique and
  applications,'' in \emph{Proceedings of International Workshop on Soft
  Computing and Intelligent Systems, (ISI, Calcutta, India)}, 1998, pp. 58--87.

\bibitem{article141}
M.~T. Hoque, M.~Chetty, A.~Lewis, and A.~Sattar, ``Twin removal in genetic
  algorithms for protein structure prediction using low-resolution model,''
  \emph{IEEE/ACM Transactions on Computational Biology and Bioinformatics},
  vol.~8, no.~1, pp. 234--245, 2009.

\bibitem{article142}
J.~Yu, X.~You, and S.~Liu, ``A heterogeneous guided ant colony algorithm based
  on space explosion and long-short memory,'' \emph{Applied Soft Computing},
  vol. 113, p. 107991, 2021.

\bibitem{article143}
F.~Dahan, K.~El~Hindi, A.~Ghoneim, and H.~Alsalman, ``An enhanced ant colony
  optimization based algorithm to solve {QoS}-aware web service composition,''
  \emph{IEEE Access}, vol.~9, pp. 34\,098--34\,111, 2021.

\bibitem{article144}
C.~Ekaputri and A.~Syaichu-Rohman, ``Model predictive control ({MPC}) design
  and implementation using algorithm-3 on board {SPARTAN} 6 {FPGA} {SP605}
  evaluation kit,'' in \emph{2013 3rd International Conference on
  Instrumentation Control and Automation (ICA)}.\hskip 1em plus 0.5em minus
  0.4em\relax IEEE, 2013, pp. 115--120.

\bibitem{article145}
M.~Nauman, W.~Shireen, and A.~Hussain, ``Model-free predictive control and its
  applications,'' \emph{Energies}, vol.~15, no.~14, p. 5131, 2022.

\bibitem{article146}
P.~Ghamisi, F.~Sepehrband, J.~Choupan, and M.~Mortazavi, ``Binary hybrid
  {GA-PSO} based algorithm for compression of hyperspectral data,'' in
  \emph{2011 5th International Conference on Signal Processing and
  Communication Systems (ICSPCS)}.\hskip 1em plus 0.5em minus 0.4em\relax IEEE,
  2011, pp. 1--8.

\bibitem{article147}
\BIBentryALTinterwordspacing
S.~Messaoud, A.~Bradai, S.~H.~R. Bukhari, P.~T.~A. Quang, O.~B. Ahmed, and
  M.~Atri, ``\BIBforeignlanguage{en}{A survey on machine learning in internet
  of things: Algorithms, strategies, and applications},''
  \emph{\BIBforeignlanguage{en}{Internet of Things}}, vol.~12, p. 100314, Dec.
  2020. [Online]. Available:
  \url{https://linkinghub.elsevier.com/retrieve/pii/S2542660520301451}
\BIBentrySTDinterwordspacing

\bibitem{article148}
A.~Bartoszewicz, ``Discrete-time quasi-sliding-mode control strategies,''
  \emph{IEEE Transactions on Industrial Electronics}, vol.~45, no.~4, pp.
  633--637, 1998.

\bibitem{article149}
X.~Li, X.~Hu, Z.~Wang, and Z.~Du, ``Path planning based on combinaion of
  improved {A-STAR} algorithm and {DWA} algorithm,'' in \emph{2020 2nd
  International Conference on Artificial Intelligence and Advanced Manufacture
  (AIAM)}.\hskip 1em plus 0.5em minus 0.4em\relax IEEE, 2020, pp. 99--103.

\bibitem{article150}
H.~Ma, J.~Wu, and Z.~Xiong, ``Discrete-time sliding-mode control with improved
  quasi-sliding-mode domain,'' \emph{IEEE Transactions on Industrial
  Electronics}, vol.~63, no.~10, pp. 6292--6304, 2016.

\bibitem{article151}
\BIBentryALTinterwordspacing
B.~Duan, C.~Guo, and H.~Liu, ``\BIBforeignlanguage{en}{A hybrid
  genetic-particle swarm optimization algorithm for multi-constraint
  optimization problems},'' \emph{\BIBforeignlanguage{en}{Soft Computing}},
  vol.~26, no.~21, p. 11695–11711, Nov. 2022. [Online]. Available:
  \url{https://link.springer.com/10.1007/s00500-022-07489-8}
\BIBentrySTDinterwordspacing

\bibitem{article112}
Z.~Chen, W.~Liao, P.~Li, J.~Tan, and Y.~Chen, ``Simple and high-performance
  cell balancing control strategy,'' \emph{Energy Science \& Engineering},
  vol.~10, no.~9, pp. 3592--3601, 2022.

\bibitem{article137}
\BIBentryALTinterwordspacing
S.~Arendarik, \emph{Active Cell Balancing in Battery Packs}.\hskip 1em plus
  0.5em minus 0.4em\relax Rožnov pod Radhoštem, Czech Republic: Freescale
  Semiconductor, Jan. 2012, no. AN4428. [Online]. Available:
  \url{https://www.nxp.com/docs/en/application-note/AN4428.pdf}
\BIBentrySTDinterwordspacing

\bibitem{article110}
T.~Duraisamy and D.~Kaliyaperumal, ``Machine learning-based optimal cell
  balancing mechanism for electric vehicle battery management system,''
  \emph{IEEE Access}, vol.~9, pp. 132\,846--132\,861, 2021.

\bibitem{article129}
U.~K. Das, P.~Shrivastava, K.~S. Tey, M.~Y. I.~B. Idris, S.~Mekhilef, E.~Jamei,
  M.~Seyedmahmoudian, and A.~Stojcevski, ``Advancement of lithium-ion battery
  cells voltage equalization techniques: A review,'' \emph{Renewable and
  Sustainable Energy Reviews}, vol. 134, p. 110227, 2020.

\bibitem{article131}
Y.~Ye and K.~W.~E. Cheng, ``Modeling and analysis of series--parallel
  switched-capacitor voltage equalizer for battery/supercapacitor strings,''
  \emph{IEEE journal of emerging and selected topics in power electronics},
  vol.~3, no.~4, pp. 977--983, 2015.

\bibitem{article130}
T.~Ohno, T.~Suzuki, and H.~Koizumi, ``Modularized lc resonant switched
  capacitor cell voltage equalizer,'' in \emph{IECON 2014-40th Annual
  Conference of the IEEE Industrial Electronics Society}.\hskip 1em plus 0.5em
  minus 0.4em\relax IEEE, 2014, pp. 3156--3162.

\bibitem{article133}
Y.-S. Lee and G.-T. Cheng, ``Quasi-resonant zero-current-switching
  bidirectional converter for battery equalization applications,'' \emph{IEEE
  Transactions on Power electronics}, vol.~21, no.~5, pp. 1213--1224, 2006.

\bibitem{article132}
C.-H. Kim, M.-Y. Kim, and G.-W. Moon, ``A modularized charge equalizer using a
  battery monitoring {IC} for series-connected li-ion battery strings in
  electric vehicles,'' \emph{IEEE Transactions on Power Electronics}, vol.~28,
  no.~8, pp. 3779--3787, 2012.

\end{thebibliography}

\vskip -2\baselineskip plus -1fil

\begin{IEEEbiography}[{\includegraphics[width=1in,height=1.25in,clip,keepaspectratio]{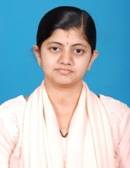}}]{Anupama R Itagi} received her M.Tech degree in VLSI and Embedded Systems from B V Bhoomraddi College of Engineering and Technology, affiliated with Visvesvaraya Technological University, in 2012. She completed her B.E. degree in Electrical and Electronics Engineering from SDM College of Engineering and Technology, Dharwad, in 2007. She has served as a Lecturer at BVBCET, Hubballi from 2008-2012. Currently, she is an Assistant Professor at EEE department, KLE Technological University, BVB campus, Hubballi. Her areas of interest include Renewable energy, Machine Learning, Embedded Systems, Electrical Vehicles, Battery Management Systems, and Engineering Education. 

\end{IEEEbiography}

\vskip -2\baselineskip plus -1fil

\begin{IEEEbiography}[{\includegraphics[width=1in,height=1.25in,clip,keepaspectratio]{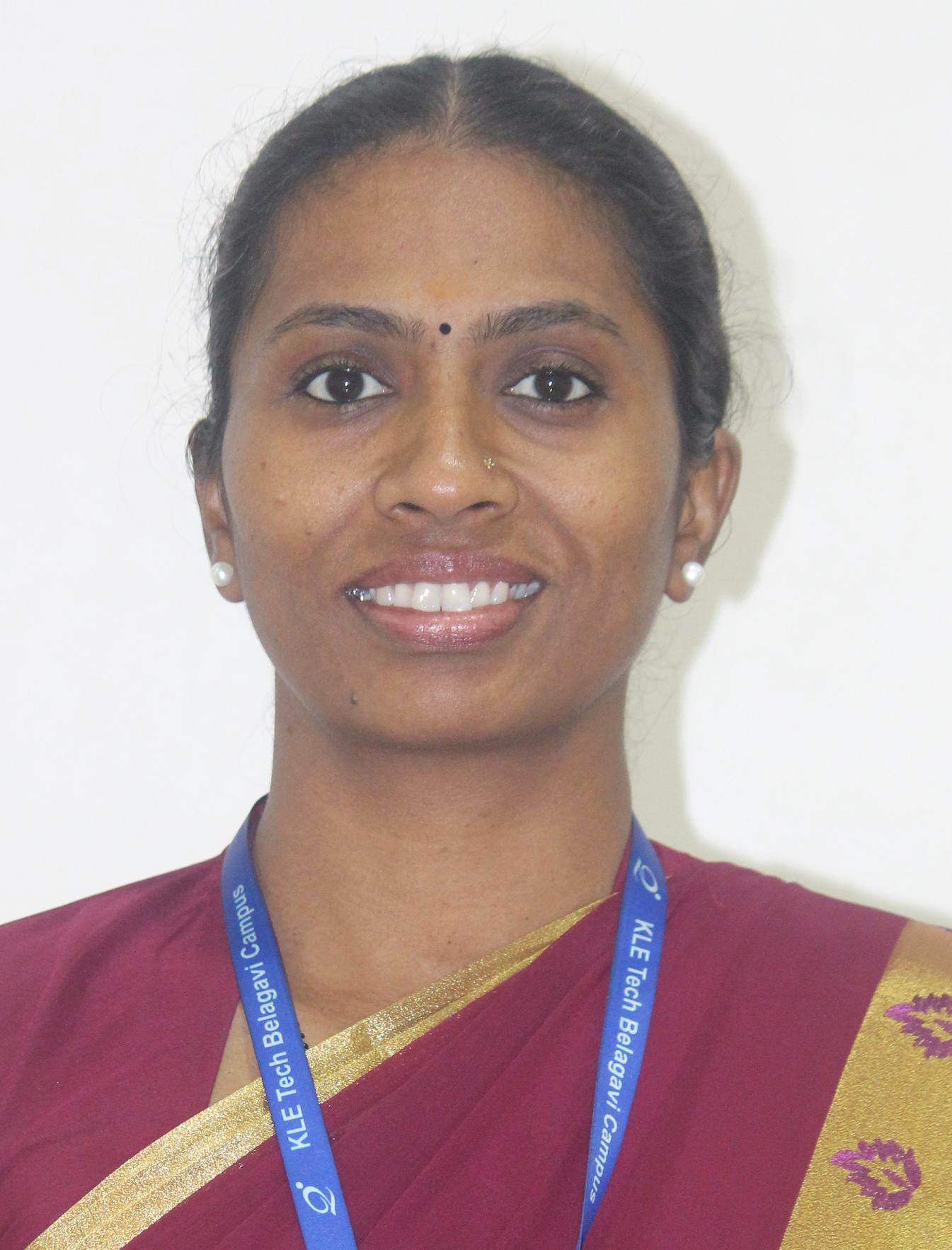}}]{Rakhee Kallimani} (Student Branch Counsellor and Member, IEEE) holds a Ph.D. from Visvesvaraya Technological University, which she received in 2023. She also holds an M.Tech degree in Embedded Systems from Calicut University, which she received in 2007, and a B.E. degree in Electrical and Electronics Engineering from B V Bhoomraddi College of Engineering and Technology, affiliated with Visvesvaraya Technological University, which she received in 2003. She has served as a Lecturer at BVBCET, Hubballi from 2003-2009 and as an Assistant Professor at KLE MSSCET, Belagavi from 2009 to 2023. Her areas of expertise include Embedded Systems, Wireless Sensor Networks, Electrical Vehicles, Battery Management Systems, and Engineering Education. Currently, she is an Associate Professor and also serves as the Head of the EEE department at KLE Technological University’s Dr M S Sheshgiri College of Engineering and Technology in Belagavi.
\end{IEEEbiography}

\vskip -2\baselineskip plus -1fil

\begin{IEEEbiography}[{\includegraphics[width=1in,height=1.25in,clip,keepaspectratio]{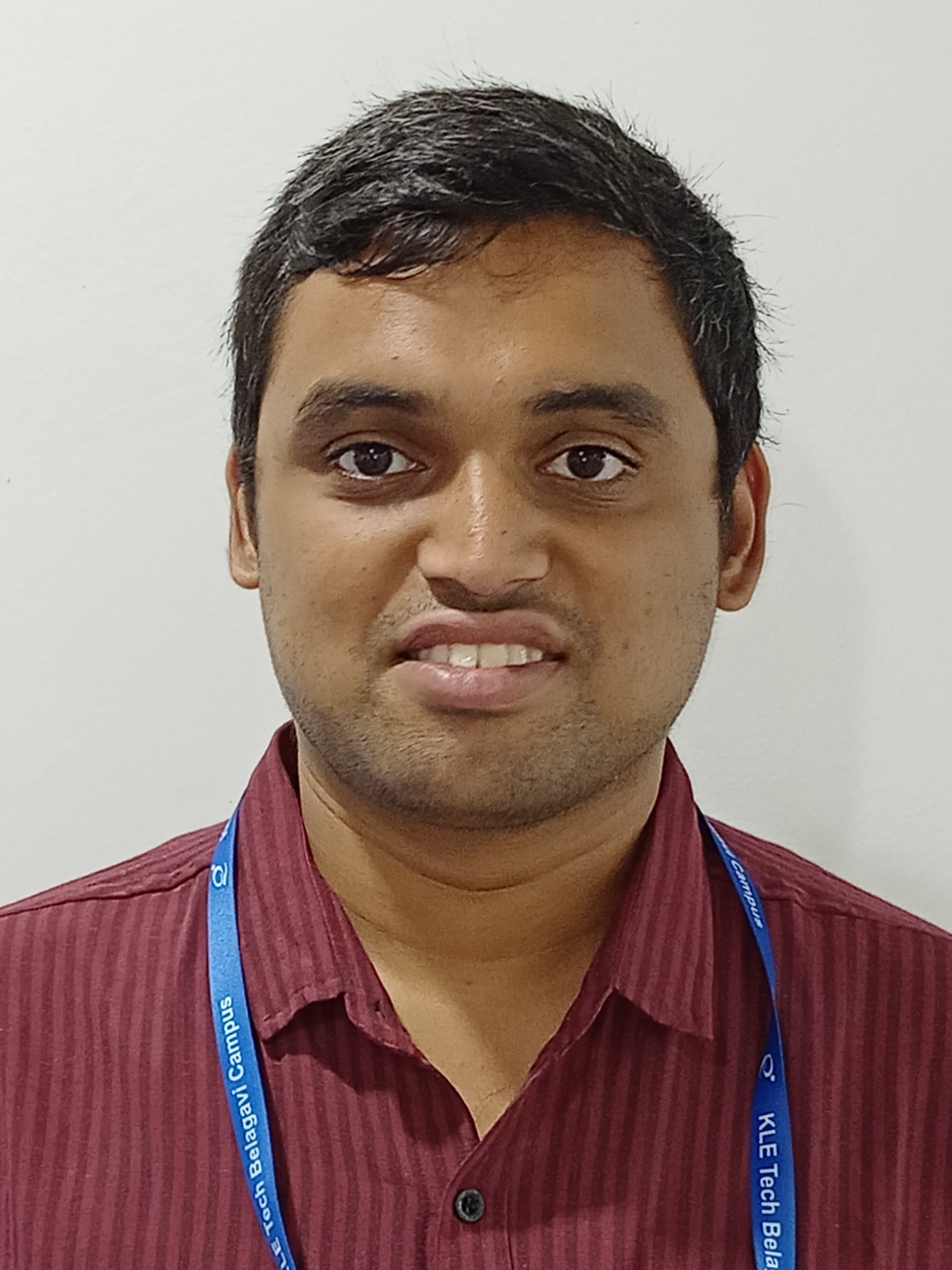}}]{Krishna Pai} received a Bachelor's degree in Electrical and Electronics Engineering from KLE Dr. M.S. Sheshgiri College of Engineering and Technology, Belagavi, India in 2021. He comes from an industrial background and works as an independent researcher in Bengaluru, India, to pursue his research interests. His research interests include Machine Learning, the Internet of Things, Electric Vehicles, Battery Management systems, Embedded Systems and Communication Theory. Currently, he is involved in multiple research projects, and his work encompasses over 16+ published articles in top-tier journals and conferences.
\end{IEEEbiography}

\vskip -2\baselineskip plus -1fil

\begin{IEEEbiography}[{\includegraphics[width=1in,height=1.25in,clip,keepaspectratio]{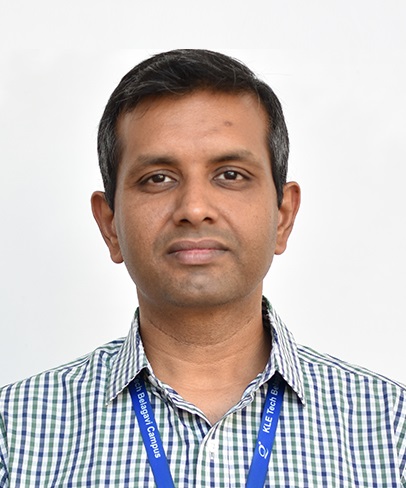}}]{Sridhar Iyer} (Senior Member, IEEE) received the M.S. degree in Electrical Engineering from New Mexico State University, U.S.A in 2008, and the Ph.D. degree from Delhi University, India in 2017. He received the young scientist award from the DST/SERB, Govt. of India in 2013, and Young Researcher Award from Institute of Scholars in 2021. He is the Recipient of the ‘Protsahan Award’ from IEEE ComSoc, Bangalore Chapter as a recognition to his contributions towards paper published/tutorial offered in recognised conferences/journals, during Jan 2020-Sep 2021, during Oct 2021-Oct 2022, and during Oct 2022-Oct 2023. He has completed two funded research projects as the Principal Investigator and is currently involved in on-going funded research projects as the Principal Investigator. His current research focus includes semantic communications and spectrum enhancement techniques for Intelligent wireless networks. Currently, he serves as a Professor at KLE Technological University, Dr MSSCET, Belagavi, Karnataka, India. 
\end{IEEEbiography}

\vskip -2\baselineskip plus -1fil

\begin{IEEEbiography}[{\includegraphics[width=1in,height=1.25in,clip,keepaspectratio]{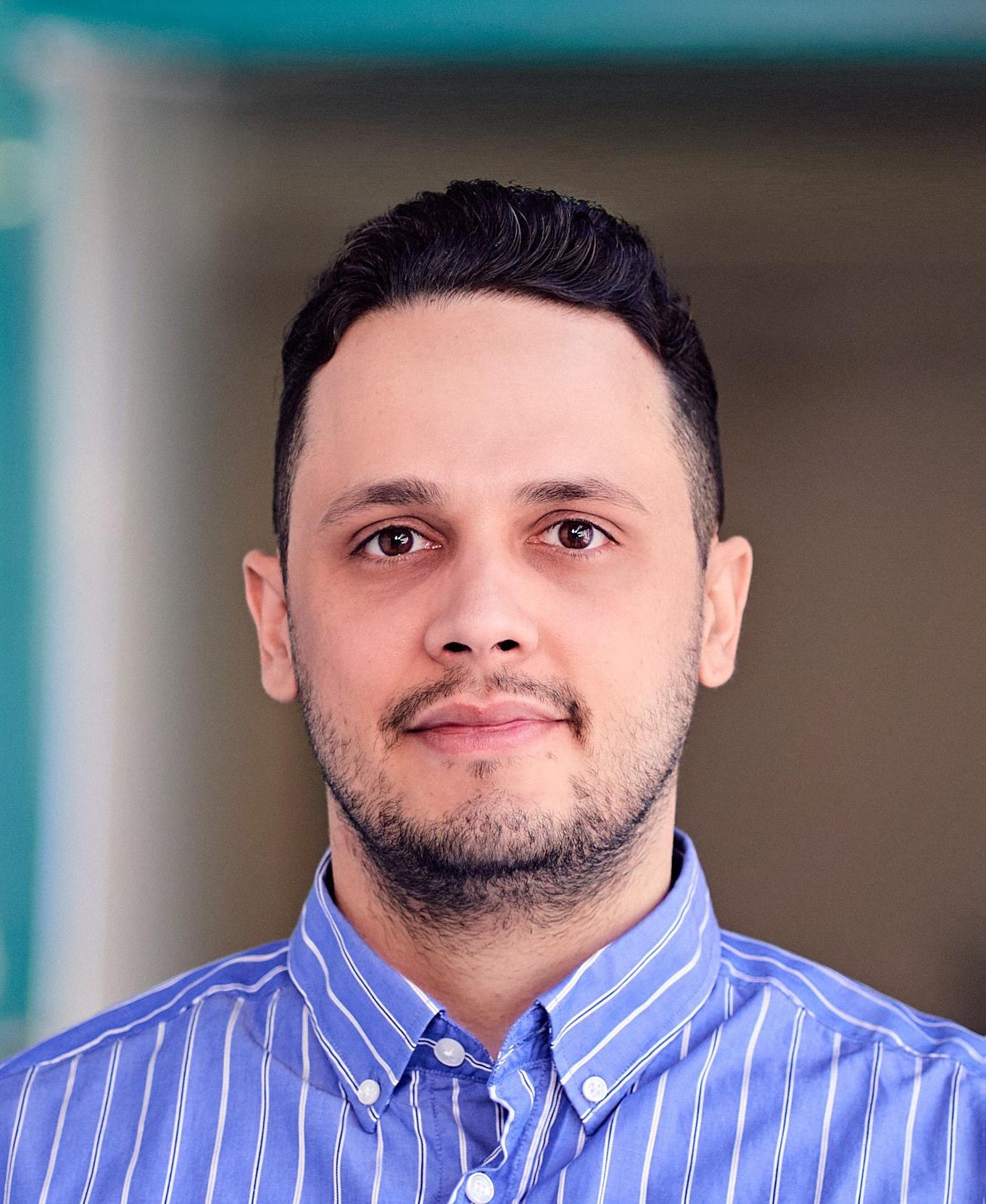}}]{Onel L. A. López} (Senior Member, IEEE) received the B.Sc. (1st class honors, 2013), M.Sc. (2017) and D.Sc. (with distinction, 2020) degree in Electrical Engineering from the Central University of Las Villas (Cuba),  the Federal University of Paraná (Brazil) and the University of Oulu (Finland), respectively. From 2013-2015 he served as a specialist in telematics at the Cuban telecommunications company (ETECSA). He is a collaborator to the 2016 Research Award given by the Cuban Academy of Sciences, a co-recipient of the 2019 and 2023 IEEE European Conference on Networks and Communications (EuCNC) Best Student Paper Award, the recipient of both the 2020 best doctoral thesis award granted by Academic Engineers and Architects in Finland TEK and Tekniska Föreningen i Finland TFiF in 2021 and the 2022 Young Researcher Award in the field of technology in Finland. He is co-author of the books entitled ``Wireless RF Energy Transfer in the massive IoT era: towards sustainable zero-energy networks'', Wiley, 2021, and ``Ultra-Reliable Low-Latency Communications: Foundations, Enablers, System Design, and Evolution Towards 6G'', Now Publishers, 2023. He is currently an Associate Professor (tenure track) in sustainable wireless communications engineering at the Centre for Wireless Communications (CWC), Oulu, Finland. His research interests include sustainable IoT, energy harvesting, wireless RF energy transfer, wireless connectivity, machine-type communications, and cellular-enabled positioning systems.
\end{IEEEbiography}

\vskip -2\baselineskip plus -1fil

\begin{IEEEbiography}[{\includegraphics[width=1in,height=1.25in,clip,keepaspectratio]{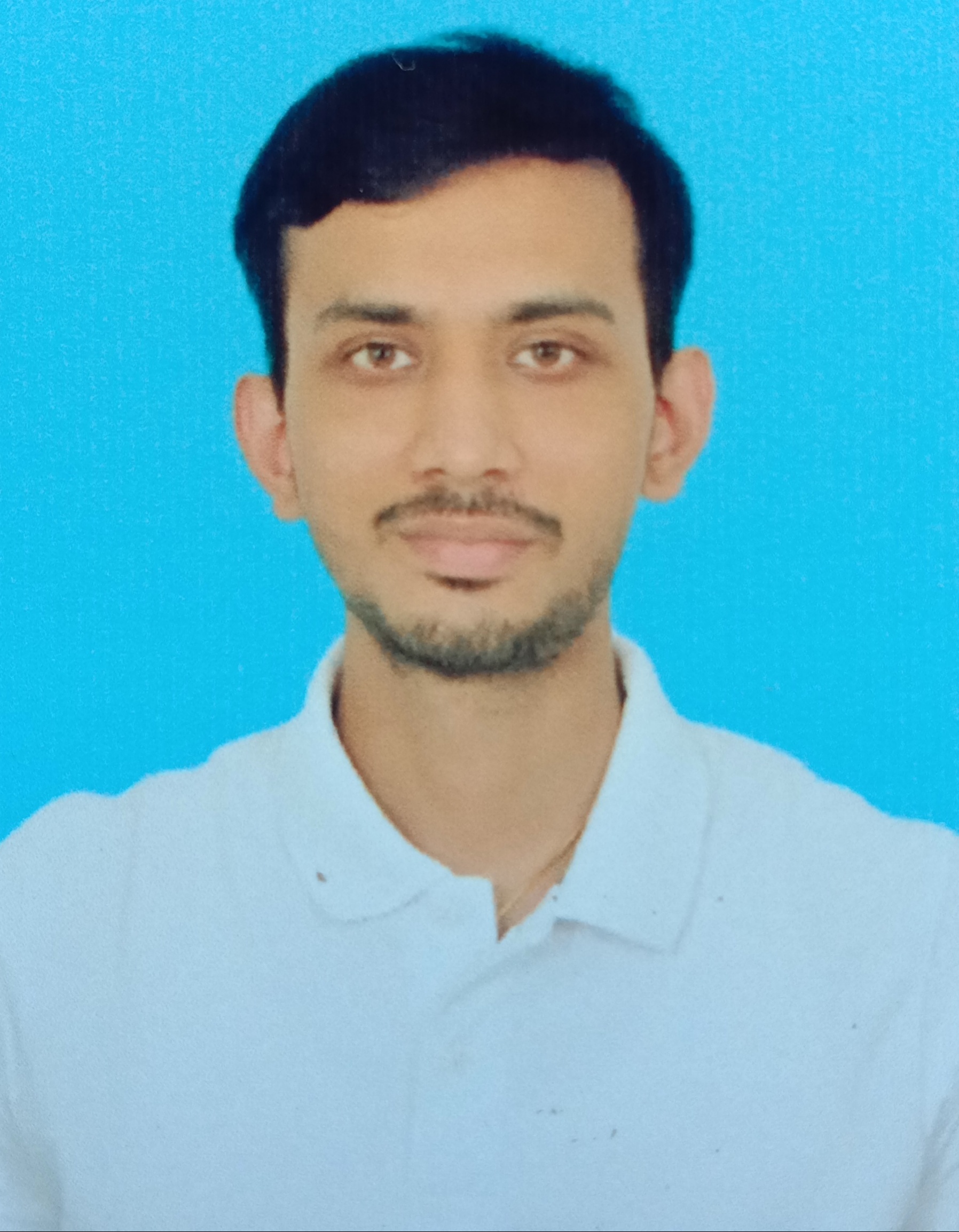}}]{Sushant Mutagekar} received his Ph.D. in Electrical Engineering from the Indian Institute of Technology Madras (IITM), India, in 2023. He has been awarded the Best Paper Award at the International Indo-Japanese Conference in 2018. He was also recognized as the Best Teaching Assistant of the IIT Madras Electrical Department. He played a pivotal role in setting up the Centre of Battery Engineering and Electric Vehicles (CBEEV) at India's first university-based Research Park at IITM. He has worked with multiple startups and led a team of young engineers in Li-ion battery testing and the development of Battery Management Systems. He was also associated with the government’s NPTEL (National Programme on Technology Enhanced Learning) programme, creating courses for engineering students on Electric Vehicles and Renewable Energy. He has published his research work in highly reputed international journals and has a patent on his name. His research in Li-ion battery technology includes developing an indigenous Li-ion Battery Tester, battery testing-modeling-characterization, Battery Management Systems, algorithm development, and data analysis.
\end{IEEEbiography}

\end{document}